\documentclass[reprint,aps,prd,superscriptaddress,nofootinbib,floats,showpacs,showkeys]{revtex4-2}

\usepackage{hyperref}
\usepackage{graphicx}
\usepackage{dcolumn}
\usepackage{subfigure}
\usepackage{mathtools}
\usepackage{amsmath,amssymb,amsfonts,amsthm,latexsym,stmaryrd,physics,mathrsfs}
\allowdisplaybreaks[4]
\usepackage{color}
\usepackage{bm}
\usepackage{float}
\usepackage{nicefrac}
\usepackage{microtype} 
\usepackage{verbatim}
\usepackage[T1]{fontenc}
\usepackage[utf8]{inputenc}
\usepackage{dsfont}
\usepackage{tikz}
\usepackage{enumerate}

\def\beq{\begin{equation}}
	\def\eeq{\end{equation}}
\newcommand{\bea}{\begin{eqnarray}}
	\newcommand{\eea}{\end{eqnarray}}
\def\bi{\begin{itemize}}
	\def\ei{\end{itemize}}
\def\bfig{\begin{figure}}
	\def\efig{\end{figure}}
\def\be{\begin{eqnarray}}
		\def\ee{\end{eqnarray}}

\newtheorem{theorem}{Theorem}[section]

\numberwithin{equation}{section}

\begin{document}

\title{Loop Quantum Kaluza-Klein Cosmology and Inflation}

\author{Shengzhi Li}
\affiliation{School of Physics and Electronic Information, Yunnan Normal University, Kunming 650500, China}
\affiliation{School of Physics and Astronomy, Key Laboratory of Multiscale Spin Physics (Ministry of Education), Beijing Normal University, Beijing 100875, China}

\author{Yongge Ma}
\thanks{Contact author}
\email{mayg@bnu.edu.cn}
\affiliation{School of Physics and Astronomy, Key Laboratory of Multiscale Spin Physics (Ministry of Education), Beijing Normal University, Beijing 100875, China}

\author{Faqiang Yuan}
\affiliation{School of Physics and Astronomy, Key Laboratory of Multiscale Spin Physics (Ministry of Education), Beijing Normal University, Beijing 100875, China}

\author{Xiangdong Zhang}
\thanks{Contact author}
\email{scxdzhang@scut.edu.cn}
\affiliation{School of Physics and Optoelectronics, South China University of Technology, Guangzhou 510641, China}

\begin{abstract}
We present the detailed analyses of five-dimensional loop quantum Kaluza-Klein cosmology based on the symmetric reduction of the connection formulation of the full theory. The previous results in a particular scenario are extended to more general cases. The effective scalar constraint for the geometric sector of the model is derived by the systematic semi-classical analysis in both the canonical and path-integral formulations, incorporating the quantum fluctuations as a subleading-order correction. The resulting effective scalar constraint not only exhibits the correct classical limit of the quantum system, but also serves as the basis for investigating the following three distinct effective scenarios through the incorporation of matter contributions: (i) vacuum, (ii) minimally coupling with a scalar field, and (iii) coupling with the dust. In all the three effective scenarios, the big bang and potential past big rip singularities in the classical model are naturally resolved by including the leading-order quantum correction of holonomies. Moreover, the visible universe undergoes a super-inflationary phase after overcoming the classical big bang singularity, during which the phenomenologically desired 55 e-folds can be achieved by appropriate initial conditions. In the case where the subleading-order quantum fluctuation term is included as a constant, the evolutions of the five-dimensional universe in all the three effective scenarios not only achieve sufficient inflation in the visible dimensions, but also exhibit re-collapse behaviors at certain large scales. Hence the cosmic inflation may originate from the interplay between compact extra dimensions and quantum geometric effects. 
\end{abstract}


\maketitle
	
\section{Introduction}\label{sec:introduction}
	
Loop quantum gravity (LQG) is a non-perturbative, background-independent quantization approach to general relativity (GR) \cite{Ro04,Th07,As04,Ma07,A.Y}. Many aspects of LQG have been explored over the past few decades. Among these, loop quantum cosmology (LQC), as the cosmological model of LQG, has become one of the most dynamic and fruitful areas of research \cite{LQC5,Boj,Ash-view,AS11,BCM}. A key achievement of LQC is the resolution of classical singularities: the big bang singularity of classical cosmology is replaced by a quantum bounce \cite{APS3,ACS,Yang23,YDM}. Furthermore, some LQC models suggest that quantum geometric corrections need not be confined to the Planck regime, but may persist at macroscopic scales \cite{DMY,YDM,Zhang:2021zfp}. Moreover, LQC has been extended to both 2+1 dimensions \cite{Zhang14} and higher dimensions \cite{Zhang16,Li:2025bzl}.

Higher-dimensional spacetime provides an appealing avenue toward the unification of gravity with the remaining fundamental interactions of matter fields. Historically, the first significant higher-dimensional theory was the well-known five-dimensional (5D) Kaluza-Klein (KK) theory, which aimed to unify 4D GR with the electromagnetic field theory \cite{Kaluza:1921tu,Klein:1926tv,bailin1987kaluza,Appelguist}. From this perspective, to study such unification mechanisms within a higher-dimensional framework of LQG is of considerable theoretical interest. Beyond its role in unification, higher-dimensional spacetime has also been shown to have profound cosmological implications. A wide class of higher-dimensional cosmological models indicates that the presence of extra dimensions provides a natural mechanism for explaining the accelerated expansion of the visible universe, a process that is often accompanied by the dynamical compactification of the extra dimensions \cite{Qiang, NM02, Panigrahi:2006wi}. Consequently, incorporating extra dimensions into LQC models and exploring this mechanism is expected to yield a richer and more diverse set of cosmological phenomena. Furthermore, several well-known theoretical paradigms, including string/M theories \cite{JP98}, AdS/CFT correspondence \cite{JM03}, and brane-world scenarios \cite{RS99a,RS99b}, intrinsically necessitate extra-dimensional formulations. In this broader context, to study higher-dimensional LQC is additionally motivated by the prospect of enabling meaningful comparisons among these different approaches to quantum gravity. Such a framework may help to establish connections and foster dialogue between distinct research programs.

By integrating the ideas of LQC and KK theory, in this paper we are going to construct the 5D loop quantum KK cosmology and study its effective dynamics. The connection formulation of classical KK cosmology is obtained through a symmetric reduction of the 5D full theory. The polymer-like quantization of the connection formulation mimicking LQG leads to the corresponding quantum model. By evaluating the expectation value of the scalar constraint operator in coherent states and analyzing the path-integral formulation of the extraction amplitude, we not only demonstrate that the quantum system possesses a correct classical limit, but also derive the effective scalar constraint for the geometric sector, which incorporates quantum fluctuations as a subleading-order correction. To explore its effective dynamics, we will study the model in the vacuum, minimally coupled to a scalar field, and coupled to the dust scenarios respectively. It turns out that, by employing the effective scalar constraint with leading-order holonomy correction, the big bang and the potential past big rip singularities appeared in the classical theory are naturally resolved by the quantum gravitational effects. Besides the singularities resolution, a common result in the three scenarios is that the visible universe undergoes a super-inflationary phase following the evolution overcoming the classical big bang singularities. Moreover, with suitable initial conditions, the e-folding number of this phase can satisfy the requirement of a sufficient inflation. It is worth recalling that in standard 4D LQC with a minimally coupled scalar field or dust, the e-folding number during the super-inflationary phase is typically too small to provide a viable alternative to the conventional inflation \cite{Ashtekar:2011ni,Ashtekar:2011rm}. In contrast, our model circumvents this limitation, offering a compelling alternative to the inflationary paradigm. Furthermore, by employing the effective scalar constraint including the subleading-order correction of a constant quantum fluctuation, the visible universe can also undergo the sufficient inflation, while the five-dimensional universe experiences re-collapses at certain large macroscopic scales.

The organization of this paper is as follows: In Sec.\ref{section2}, we perform a symmetric reduction of the connection formulation of 5D KK theory to obtain the reduced phase space for KK cosmology. In Sec.\ref{section3}, we apply the polymer-like quantization to KK cosmology and analyze the semiclassical behavior of the system to obtain the effective scalar constraint for the geometric sector. In Sec.\ref{section4}, we study the classical and effective dynamics of the model in the vacuum, minimally coupled to a scalar field, and coupled to the dust scenarios respectively. The results are summarized in the final section.

\section{Classical setting}\label{section2}
	
The phase space of the 4+1-dimensional vacuum GR in the connection formulation is described by the $\mathrm{Spin}(5)$-connection $A_{aIJ}$ and its conjugate momentum $\pi^{bKL}$ on a 4D spatial manifold $\Sigma$ with the nontrivial Poisson brackets
\begin{equation}
\{A_{aIJ}(x),\pi^{bKL}(x')\}=2\beta\kappa^{(5)}\delta^K_{[I}\delta^L_{J]}\delta_a^b\delta^{(4)}(x,x'),
\end{equation}
where $\beta$ is a positive real parameter and $\kappa^{(5)}:=16\pi G^{(5)} c^{-3}$ with $G^{(5)}$ being the 5-dimensional gravitational constant, $c$ is the speed of light, the spatial indices takes the values of $a,b,\ldots\in\{1,\ldots,4\}$, the internal indices are denoted by $I,J,\ldots\in\{1,\ldots,5\}$, and $x,x'$ are coordinates on $\Sigma$ \cite{BTTa,BTTb,BTTc,BTTd}. Note that the inverse metric density of weight 2 on $\Sigma$ is defined by $Q^{ab}:=\pi^{aIJ}\pi^{b}{}_{IJ}/2$, which is required to be nondegenerate, where the internal indices are raised and lowered through the Kronecker symbol $\delta_{IJ}$. In addition, the system is subject to the following simplicity, Gauss, vector and scalar constraints
\begin{align}
S^{ab}_{M}&:=\frac{1}{4}\epsilon_{IJKLM}\pi^{aIJ}\pi^{bKL}=0, \label{simplicity}\\
G^{IJ}&:=\frac{1}{2\beta\kappa^{(5)}}(\partial _{a}\pi ^{aIJ}+2A_{a}{}^{[I}{}_{K}\pi ^{a|K|J]})=0, \label{Gauss}\\
C_a&:=\frac{1}{2\beta\kappa^{(5)}}(\pi^{bIJ}\partial_{a}A_{bIJ}-\partial _{b}(\pi^{bIJ}A_{aIJ}))=0,\label{vectornew}\\
C_{gr}&:=-\frac{1}{\kappa^{(5)}\sqrt{q}}F_{abIJ}\pi^{aIK}\pi^{bJ}{}_{K}\nonumber\\
&+\frac{4\beta^2+1}{8\kappa^{(5)}\sqrt{q}}\pi^{[a|IJ|}\pi^{b]KL}K_{bIJ}K_{aKL}\nonumber\\
&+\frac{\beta^2}{\kappa^{(5)}\sqrt{q}}\bar{K}_{bIK}\bar{K}_{aJ}{}^{K}E^{aI}E^{bJ}=0, \label{hamiltong}
\end{align}
where $F_{abIJ}:=2\partial_{[a}A_{b]IJ}+2A_{a[I|K|}A_{b}{}^{K}{}_{J]}$ is the curvature of the connection $A_{aIJ}$, $K_{aIJ}:=(A_{aIJ}-\Gamma_{aIJ})/\beta$ with $\Gamma_{aIJ}$ being determined by the momentum $\pi^{aIJ}$, $\bar{K}_{aIJ}:=K_{aKL}\bar{\eta}^{K}_{I}\bar{\eta}^{L}_{J}$ with $\bar{\eta}^{I}_{J}:=\delta^{I}_{J}-n^{I}n_{J}$, and the spatial metric $q_{ab}$ is given by its inverse $q^{ab}:=Q^{ab}/q$ with $q:=\det(q_{ab})$. Here the unit vector $n^{I}$ and the generalized densitized 4-beins $E^{bJ}$ are initially fixed by the intrinsic decomposition $\pi^{aIJ}=2n^{[I}E^{bJ]}$ on the simplicity-constraint surface and subsequently extended arbitrarily over the phase space, and the generalized spin connection $\Gamma_{aIJ}$ is defined as an arbitrary extension on phase space of the solution to $D_{a}\pi^{bIJ}=\partial_{a}\pi^{bIJ}+\Gamma^{b}_{ac}\pi^{cIJ}+2\Gamma_{a}{}^{[I}{}_{K}\pi^{b|K|J]}-\Gamma^{c}_{ac}\pi^{bIJ}=0$ obtained on the simplicity-constraint surface, where $\Gamma^{b}_{ac}$ denotes the Christoffel symbol compatible with $q_{ab}$.

Next, we focus on the spatially homogeneous 4+1-dimensional KK cosmological model, with the spatial manifold $\Sigma=\mathbb{R}^3\times\mathbb{S}^1$ and the corresponding spatial isometry group $\mathcal{S}_{K\!K}:=(\mathbb{R}^3\rtimes\mathrm{SU}(2))\times\mathrm{U}(1)$. Here, $\mathbb{R}^3\rtimes\mathrm{SU}(2)$ is the semidirect product of the translation group $\mathbb{R}^3$ and the rotation group $\mathrm{SU}(2)$, with group multiplication $(a^{i}_{1},u_{1})(a^{i}_{2},u_{2})=(a^{i}_{1}+\chi(u_{1})^{i}_{j}a^{j}_{2},u_{1}u_{2})$, for all $a^{i}_{1},a^{i}_{2}\in\mathbb{R}^3$ and $u_{1},u_{2}\in\mathrm{SU}(2)$, where $i\in\!\{1,2,3\}$ and $\chi:\mathrm{SU}(2)\to\mathrm{SO}(3)$ is the universal covering group homomorphism. Let $(x^{i},y)$ denote the natural coordinates on $\mathbb{R}^3\times\mathbb{S}^1$, where $0\le\!y<2\pi$ and $0$ identified with $2\pi$. Then the left action of the symmetry group $\mathcal{S}_{K\!K}$ on $\Sigma$ is defined as
\begin{equation}
	\mathcal{L}:((a^j,u,e^{i\theta}),(x^j,y))\mapsto (a^j+\chi(u)^j_kx^k,y+\theta),
\end{equation}
where $e^{i\theta}\in \mathrm{U}(1)$. Thus, the $\mathrm{U}(1)$ subgroup of $\mathcal{S}_{K\!K}$ represents the homogeneity of spacetime along the extra dimension, whereas the $\mathbb{R}^3\rtimes\mathrm{SU}(2)$ subgroup encodes the spatial homogeneity and isotropy in $\mathbb{R}^3$. By imposing the condition that $A:=A_{aIJ}(x^{b})L^{IJ}dx^a/2$ and $\pi:=\pi^{aIJ}(x^{b})L_{IJ}\partial_a/2$ remain invariant under any $\mathcal{S}_{K\!K}$-symmetric transformation, up to a suitable $\mathrm{Spin}(5)$-gauge transformation, we obtain their symmetry-reduced forms as \cite{Li:2025bzl}
\begin{align}
A&=(\tilde{A}_1L_{i5}+\tilde{A}_2(\epsilon_i{}^{jk}/2)L_{jk}+\tilde{A}_3L_{i4})dx^i+\tilde{A}_yL_{45}dy,\label{symmetric A}\\
\pi&=(\tilde{\pi}^1L^{i5}+\tilde{\pi}^2(\epsilon^i{}_{jk}/2)L^{jk}+\tilde{\pi}^3L^{i4})\partial_i+\tilde{\pi}^yL^{45}\partial_y,\label{symmetric pi}
\end{align}
where the generators $L_{IJ}$ of the $\mathrm{Spin}(5)$ group can be expressed as $L_{IJ}=[\gamma_I,\gamma_J]/4$, with $\gamma_I$ being the $4\times4$ Euclidean Gamma matrix. Note that all components on the right-hand sides of Eqs.(\ref{symmetric A}) and (\ref{symmetric pi}) are independent of the coordinates $x^{b}$.

To avoid the divergences in the integration during the symmetric reduction of the full theory to the homogeneous model, we fix a fiducial cell $\Sigma^3_{(0)}:=(0,V_0^{1/3})^3\subset\mathbb{R}^3$, where $V_0$ denotes the 3-volume of the $\Sigma^3_{(0)}$ measured by the fiducial metric compatible with the chosen coordinates $x^i$, and restrict all the integrals over $\mathbb{R}^3$ to $\Sigma^3_{(0)}$ for the reduction. Then the symplectic structure $\Omega_{gr}$ of the geometric part is given by
\begin{align}
\Omega_{gr}&=\frac{1}{2\beta\kappa^{(5)}}\int_{\Sigma^3_{(0)}}d^3x\int_{\mathbb{S}^1}dy\delta\pi^{aIJ}(x^{b})\wedge\delta A_{aIJ}(x^{b})\nonumber\\
&=\frac{V_0}{\beta\kappa}[\delta\tilde{\pi}^y\wedge\delta\tilde{A}_{y}+3\delta\tilde{\pi}^{i}\wedge\delta\tilde{A}_{i}],
\end{align}
where $\kappa:=\kappa^{(5)}/(2\pi)$. By restricting the constraints to the reduced variables, we can get reduced constraints so that the vector constraints are automatically satisfied, and the simplicity and Gauss constraints are respectively equivalent to the following expressions
\begin{align}
	S^{ab}_{M}&=0\Leftrightarrow\tilde{\pi}^2=0,\\ G^{IJ}&=0\Leftrightarrow\tilde{A}_{1}\tilde{\pi}^{3}-\tilde{A}_{3}\tilde{\pi}^{1}=0.
\end{align}
Hence, at the classical level, the reduced simplicity and Gauss constraints can be solved by imposing the gauge-fixing condition $\tilde{A}_{2}=\tilde{\pi}^{3}=0$, which consequently leads to $\tilde{\pi}^2=\tilde{A}_{3}=0$. By requiring the basic variables to be invariant under the stretching coordinate transformations $x^i\to\alpha x^i$ for any given constant $\alpha$, we can absorb the coordinate volume $V_0$ into the basic variables as 
\begin{align}
	\pi^y&:=V_0\tilde{\pi}^y,&A_y&:=\tilde{A}_y,\nonumber\\
	\pi^1&:=V_0^{2/3}\tilde{\pi}^1,&A_1&:=V_0^{1/3}\tilde{A}_1.\label{XSV}
\end{align}
Therefore, the geometric sector of the phase space in KK cosmology consists of $\pi^y$, $A_y$, $\pi^1$, and $A_1$, with the following nontrivial Poisson brackets
\begin{equation}
\{A_y,\pi^{y}\}=\beta\kappa,\qquad\{A_1,\pi^{1}\}=\frac{\beta\kappa}{3}.
\end{equation} 
We consider the spatially flat KK cosmology such that the spacetime metric reads
\begin{equation}
	ds^2=-N^2(t)dt^2+a^2(t)\delta_{ij}dx^{i}dx^{j}+b^2(t)dy^2,
\end{equation}
where, $N(t)$ is the lapse function, and $a(t)$ and $b(t)$ are the scale factors of the visible and extra dimensions respectively. Here, $a$ and $b$ are related to the momentum variables as
\begin{equation}
a^2=(\frac{\pi^y}{V_0})^{2/3},\qquad b^2=\frac{(\pi^1)^2}{(\pi^y)^{4/3}}.
\end{equation}
To avoid the potential problem that the dynamic results depend on the choice of the fiducial cell or the fiducial metric, we adopt the $\bar{\mu}$ scheme in the visible dimensions, as in the 4D LQC \cite{APS3}. To quantize the scalar constraint, one needs to regularize the visible-dimensional components of the curvature $F_{abIJ}$ via the holonomy of the connection $A_{aIJ}$ along certain square loop. The idea of $\bar{\mu}$ scheme to require the physical area enclosed by the loop to be the two-dimensional area gap $\Delta:=(Ar^{(5)}_{min})^{2/3}$, where $Ar^{(5)}_{min}=2\pi\beta\kappa\hbar$ is the minimal 3D area in 5D LQG \cite{BTTc}. This choice leads to the following regularization parameter
\begin{equation}\label{mubar1}
a\bar{\mu}_{1}V_0^{1/3}=\Delta^{1/2} \Rightarrow \bar{\mu}_{1}:=\Delta^{1/2}|\pi^{y}|^{-1/3}.
\end{equation}
To adapt to this scheme, we perform the following canonical transformation
\begin{align}
	\bar{A}_1&:=\bar{\mu}_{1}A_1,&\bar{\pi}^{1}&:=(\bar{\mu}_{1})^{-1}\pi^1,\nonumber\\ \bar{A}_y&:=A_y-A_1\pi^{1}(\pi^{y})^{-1},&\bar{\pi}^y&:=\pi^{y}.\label{ZZB}
\end{align}
Since $\pi^{y}$ is unchanged under the transformation, we use the same notation for it throughout. The nontrivial Poisson brackets among the new variables are given by
\begin{equation}
\{\bar{A}_{y},\pi^{y}\}=\beta\kappa,\qquad \{\bar{A}_{1},\bar{\pi}^{1}\}=\frac{\beta\kappa}{3}.
\end{equation}
This will serve as the starting point for our quantization. The scale factors of the spacetime metric can be expressed in terms of the new variables as
\begin{equation}\label{scale factors}
a^2=(\frac{\pi^y}{V_0})^{2/3},\qquad b^2=\Delta(\frac{\bar{\pi}^1}{\pi^y})^2.
\end{equation}
Hence, $|\pi^{y}|$ represents the physical volume of the three-dimensional fiducial cell $\Sigma^3_{(0)}$, and $|\bar{\pi}^{1}|$ is proportional to the physical volume of the four-dimensional fiducial cell $\Sigma^3_{(0)}\times\mathbb{S}^1$. The smeared scalar constraint $C_{gr}[N]:=\int_{\Sigma^3_{(0)}}d^3x\int_{\mathbb{S}^1}dyNC_{gr}$ can be decomposed into the following three terms,
\begin{align}
	C_E[N]&:=\int_{\Sigma^3_{(0)}}d^3x\int_{\mathbb{S}^1}dyN(-\frac{1}{\kappa^{(5)}\sqrt{q}}F_{abIJ}\pi^{aIK}\pi^{bJ}{}_{K})\nonumber\\
	&=6N\kappa^{-1}\Delta^{-1/2}|\bar{\pi}^{1}|^{-1}[2(\bar{A}_{1}\bar{\pi}^{1})^2+\bar{A}_{1}\bar{A}_{y}\bar{\pi}^{1}\pi^{y}],\label{CE}\\
	C_L[N]&:=\int_{\Sigma^3_{(0)}}d^3x\int_{\mathbb{S}^1}dyN\frac{4\beta^2+1}{8\kappa^{(5)}\sqrt{q}}\pi^{[a|IJ|}\pi^{b]KL}K_{bIJ}K_{aKL}\nonumber\\
	&=-\frac{1}{4}(4+\beta^{-2})C_E[N],\label{CL}\\
	C_R[N]&:=\int_{\Sigma^3_{(0)}}d^3x\int_{\mathbb{S}^1}dyN\frac{\beta^2}{\kappa^{(5)}\sqrt{q}}\bar{K}_{bIK}\bar{K}_{aJ}{}^{K}E^{aI}E^{bJ}\nonumber\\
	&=0,
\end{align}
where the unit vector $n^{I}=\delta^{I}_{5}$ and the generalized spin connection $\Gamma_{aIJ}=0$ have been used. Therefore, the scalar constraint of the geometric sector of the cosmological model is given by
\begin{align}
	C_{gr}[N]&=-\frac{1}{4}\beta^{-2}C_{E}[N]\nonumber\\
	&=-\frac{3}{2}\beta^{-2}N\kappa^{-1}\Delta^{-1/2}|\bar{\pi}^{1}|^{-1}[2(\bar{A}_{1}\bar{\pi}^{1})^2\nonumber\\
	&+\bar{A}_{1}\bar{A}_{y}\bar{\pi}^{1}\pi^{y}].\label{SCKK}
\end{align}

\section{QUANTUM THEORY}\label{section3}
The classical KK cosmology typically encounters the big bang singularity or big rip singularity during its evolution (see Refs.\cite{OG,Chodos:1979vk} and Sec.\ref{section4}), where the classical description inevitably breaks down. However, in the vicinity of these classical singularities, quantum gravitational effects are expected to become dominant and to induce essential modifications to the spacetime dynamics. To check this expectation, the 5D KK cosmology described in last section will be quantized in this section by mimicking the loop quantization procedure based on the holonomy-flux algebra of the full theory. Its effective model will also be derived by semiclassical analysis.

\subsection{Kinematic Hilbert space}\label{sec:section31}
To implement the polymer-like quantization of the configuration variables, we define $\mathcal{D}_{gr}$ as the vector space generated by all possible finite linear combinations of functions of the following form
\begin{equation}
	T_{\lambda,\xi}(\bar{A}_{1},\bar{A}_{y}):=\exp(i\lambda\frac{\bar{A}_{1}}{2})\exp(i\xi\frac{\bar{A}_{y}}{2}),
\end{equation}
where $\lambda,\xi\in\mathbb{R}$. Then we employ the Dirac notation $|\lambda,\xi\rangle =T_{\lambda,\xi}$ and define the inner product on $\mathcal{D}_{gr}$ as
\begin{equation}\label{inner product}
\langle\lambda',\xi'|\lambda,\xi\rangle:=\delta_{\lambda',\lambda}\delta_{\xi',\xi},
\end{equation}
where the Kronecker-$\delta$ symbol is used. By completing $\mathcal{D}_{gr}$ with respect to the inner product (\ref{inner product}), we obtain the kinematic Hilbert space $\mathcal{H}_{gr}$. Thus, $\{\left|\lambda,\xi\right>\mid\lambda,\xi\in\mathbb{R}\}$ forms an orthonormal basis in $\mathcal{H}_{gr}$. The action of the basic operators corresponding to the momentum variables on the basis vectors are defined as
\begin{align}
\hat{\bar{\pi}}^{1}|\lambda,\xi\rangle&:=\frac{\hbar\beta\kappa}{6}\lambda|\lambda,\xi\rangle,\label{III.3}\\
\hat{\pi}^{y}|\lambda,\xi\rangle&:=\frac{\hbar\beta\kappa}{2}\xi|\lambda,\xi\rangle.\label{III.4}
\end{align}
The action of the holonomy operators corresponding to the configuration variables are defined by
\begin{align}
	\widehat{\exp(i\lambda'\frac{\bar{A}_{1}}{2})}|\lambda,\xi\rangle&:=|\lambda+\lambda',\xi\rangle,\\
	\widehat{\exp(i\xi'\frac{\bar{A}_{y}}{2})}|\lambda,\xi\rangle&:=|\lambda,\xi+\xi'\rangle.
\end{align}

Next, we need to promote the scalar constraint $C_{gr}[N]$ as an operator in the Hilbert space $\mathcal{H}_{gr}$. Since $C_{gr}[N]$ is proportional to $C_{E}[N]$ according to Eq.(\ref{SCKK}), we quantize it in the same manner as $C_{E}[N]$ in the full theory \cite{Thiemann:1996aw,BTTc,Yang:2015zda}. To this end, the holonomies of the connection $A_{aIJ}$ along the $x^{i}$ axis and the $y$ axis are introduced as
\begin{align}
	h_{i}^{(\mu_{1})}&=\mathcal{P}\exp(\int_{0}^{\mu_{1}V_{0}^{1/3}}dt\delta^a_{i}A_{aIJ}L^{IJ}/2)\nonumber\\
	&=\cos(\mu_{1}A_{1}/2)+2\sin(\mu_{1}A_{1}/2)L_{i5},\\
	h_{4}^{(\mu_{y})}&=\mathcal{P}\exp(\int_{0}^{\mu_{y}}dt\delta^a_{4}A_{aIJ}L^{IJ}/2)\nonumber\\
	&=\cos(\mu_{y}A_{y}/2)+2\sin(\mu_{y}A_{y}/2)L_{45},
\end{align}
where the regularization parameters $\mu_{1}$ and $\mu_{y}$ are sufficiently small ratios. Then, the components of the curvature $F_{abIJ}$ can be approximated by the holonomies around the square loops as follows
 \begin{align}
F_{abIJ}\delta^{a}_{i}\delta^{b}_{j}&\approx-\frac{\tr(h_{i}^{(\mu_{1})}h_{j}^{(\mu_{1})}(h_{i}^{(\mu_{1})})^{-1}(h_{j}^{(\mu_{1})})^{-1}L_{IJ})}{(\mu_{1}V_{0}^{1/3})^{2}},\label{FF1}\\
F_{abIJ}\delta^{a}_{i}\delta^{b}_{4}&\approx-\frac{\tr(h_{i}^{(\mu_{1})}h_{4}^{(\mu_{y})}(h_{i}^{(\mu_{1})})^{-1}(h_{4}^{(\mu_{y})})^{-1}L_{IJ})}{(\mu_{1}V_{0}^{1/3}\mu_{y})},\label{FF2}\\
F_{abIJ}\delta^{a}_{4}\delta^{b}_{i}&\approx-\frac{\tr(h_{4}^{(\mu_{y})}h_{i}^{(\mu_{1})}(h_{4}^{(\mu_{y})})^{-1}(h_{i}^{(\mu_{1})})^{-1}L_{IJ})}{(\mu_{1}V_{0}^{1/3}\mu_{y})}\label{FF3}.
 \end{align}
By inserting Eqs.(\ref{FF1}), (\ref{FF2}), and (\ref{FF3}) into the first line of Eq.(\ref{CE}), the scalar constraint $C_{gr}[N]$ can be regularized as
\begin{align}
	C_{gr}[N]\approx&-\frac{3}{2}N\beta^{-2}\kappa^{-1}|\pi^{y}|^{-1/3}|\pi^{1}|^{-1}\nonumber\\
	\times&[(\frac{\sin(\mu_{1}A_{1})}{\mu_{1}}\pi^{1})^{2}+\frac{\sin(\mu_{1}A_{1})}{\mu_{1}}\frac{\sin(\mu_{y}A_{y})}{\mu_{y}}\pi^{1}\pi^{y}].\label{RSCKK}
\end{align}
Since the quantization is carried out based on the new variables adapted to the $\bar{\mu}$ scheme, it is necessary to express the regularized scalar constraint (\ref{RSCKK}) in terms of these new variables. Based on the physical interpretation of $\bar{\mu}_{1}$, we replace $\mu_{1}$ with $\bar{\mu}_{1}$. For the term $\sin(\mu_{y}A_{y})/\mu_{y}$, since  $\mu_{y}$ is sufficiently small, we approximate it by performing polymerization on the inverse canonical transformation of $A_{y}$ and thus obtain
\begin{equation}
\frac{\sin(\mu_{y}A_{y})}{\mu_{y}}\approx\frac{\sin(\mu_{y}\bar{A}_{y})}{\mu_{y}}+\frac{\sin(\bar{\mu}_{1}A_{1})}{\bar{\mu}_{1}}\pi^{1}(\pi^{y})^{-1}.
\end{equation}
Collecting these results, the scalar constraint is regularized in terms of the new variables as
\begin{align}
	C_{gr}[N]\approx&-\frac{3}{2}N\beta^{-2}\Delta^{-1/2}\kappa^{-1}|\bar{\pi}^{1}|^{-1}\nonumber\\
	\times&[2\sin^{2}(\bar{A}_1)(\bar{\pi}^{1})^{2}+\sin(\bar{A}_1)\frac{\sin(\mu_{y}\bar{A}_{y})}{\mu_{y}}\bar{\pi}^{1}\pi^{y}],\label{regularizedC}
\end{align}
which can be directly promoted to the following operator
\begin{align}
	\hat{C}_{gr}[N]:=&-\frac{3}{2}N\beta^{-2}\Delta^{-1/2}\kappa^{-1}|\hat{\bar{\pi}}^{1}|^{-1/2}[2\widehat{\sin(\bar{A}_1)}{}^{2}(\hat{\bar{\pi}}^{1})^{2}\nonumber\\
	+&\widehat{\sin(\bar{A}_1)}\frac{\widehat{\sin(\mu_{y}\bar{A}_{y})}}{\mu_{y}}\hat{\bar{\pi}}^{1}\hat{\pi}^{y}]|\hat{\bar{\pi}}^{1}|^{-1/2},\label{operatorC}
\end{align}
where $\widehat{\sin(\bar{A}_1)}:=(\widehat{\exp(i\bar{A}_1)}-\widehat{\exp(-i\bar{A}_1)})/(2i)$, $\widehat{\sin(\mu_{y}\bar{A}_{y})}:=(\widehat{\exp(i\mu_{y}\bar{A}_y)}-\widehat{\exp(-i\mu_{y}\bar{A}_y)})/(2i)$ and the operator $|\hat{\bar{\pi}}^{1}|^{-1/2}$ is defined by
\begin{equation}
|\hat{\bar{\pi}}^{1}|^{-1/2}\left|\lambda,\xi\right>:=(\frac{\hbar\beta\kappa}{6})^{-1/2}\eta(\lambda)\left|\lambda,\xi\right>,\label{III.16}
\end{equation}
with 
\begin{equation}
\eta(\lambda):=\begin{cases} |\lambda|^{-1/2}, &\lambda\ne 0,\\
		0, &\lambda =0.\end{cases}
\end{equation}
For convenience, we decompose the scalar constraint operator as 
\begin{align}
\hat{C}_{gr}[N]=&4^{-2}N\hbar\beta^{-1}\Delta^{-1/2}\nonumber\\
\times&\left(\sum_{r\in\{0,4,-4\}}\hat{u}_{r}+\sum_{k,l\in\{2,-2\}}\hat{u}_{k,l}\right), \label{SCdec}
\end{align}
where the constituent operators are defined as follows
\begin{align}
	\hat{u}_{0}\left|\lambda,\xi\right>\!&:=\! u_{0}(\lambda)\left|\lambda,\xi\right>,\\
	\hat{u}_{\pm 4}\left|\lambda,\xi\right>&:=u_{\pm 4}(\lambda)\left|\lambda\pm 4,\xi\right>,\\
	\hat{u}_{\pm 2,2}\left|\lambda,\xi\right>&:=u_{\pm 2,2}(\lambda,\xi)\left|\lambda\pm 2,\xi+2\mu_{y}\right>,\\
	\hat{u}_{\pm 2,-2}\left|\lambda,\xi\right>&:=u_{\pm 2,-2}(\lambda,\xi)\left|\lambda\pm 2,\xi-2\mu_{y}\right>,
\end{align}
with 
\begin{align}
u_{0}(\lambda)&:=-4\lambda^{2}\eta^{2}(\lambda),\\
u_{\pm 4}(\lambda)&:=2\eta(\lambda\pm 4)\lambda^{2}\eta(\lambda),\\
u_{\pm 2,2}(\lambda,\xi)&:=\pm \frac{3}{\mu_{y}}\eta(\lambda\pm 2)\lambda\xi\eta(\lambda),\\
u_{\pm 2,-2}(\lambda,\xi)&:=-u_{\pm 2,2}(\lambda,\xi).
\end{align}
Moreover, through the symmetrization, the resulting operator $\hat{C}^{sym}_{gr}[N]:=(\hat{C}_{gr}[N]+\hat{C}^{\dagger}_{gr}[N])/2$ can be shown to be essentially self-adjoint in $\mathcal{H}_{gr}$ (see Appendix \ref{sec:appendixB} for the proof). In the following, we use the same notation to denote its self-adjoint extension and treat it as the quantum counterpart of the real function $C_{gr}[N]$.

\subsection{Semiclassical analysis}\label{sec:section32}
To ensure that $\hat{C}^{sym}_{gr}[N]$ is a viable quantization, we need to prove that its expectation value in an appropriate semiclassical state reduces to the classical constraint. There is a natural Gaussian coherent state in the algebraic dual space of $\mathcal{D}_{gr}$ \cite{AFJ,Taveras,DMY,YDM}, which is denoted as 
\begin{align}
 (\Psi_{\zeta}|:=&\sum_{\lambda,\xi\in\mathbb{R}}e^{-\frac{\epsilon^{2}}{2}(\lambda-\lambda_{0})^2}e^{i\frac{\bar{A}_{1}|_{0}}{2}(\lambda-\lambda_{0})}\nonumber\\
 \times&e^{-\frac{\omega^{2}}{2}(\xi-\xi_{0})^2}e^{i\frac{\bar{A}_{y}|_{0}}{2}(\xi-\xi_{0})}(\lambda,\xi|,\label{Gaussian CS1}
\end{align}
where $\zeta:=(\epsilon,\bar{A}_{1}|_{0},\bar{\pi}^{1}|_{0}\equiv\hbar\beta\kappa\lambda_{0}/6,\omega,\bar{A}_{y}|_{0},\pi^{y}|_{0}\equiv\hbar\beta\kappa\xi_{0}/2)$. Here, $\epsilon$ and $\omega$ represent the Gaussian spreads, and the subscript 0 denotes a specific point in the phase space. However, since there is no inner product defined in the algebraic dual space, we extract physical information using the shadow state obtained by projecting Eq.(\ref{Gaussian CS1}) onto a regular lattice as
\begin{align}
|\Psi_{\zeta}\rangle=&\sum_{n,m\in\mathbb{Z}}e^{-\frac{\epsilon^{2}}{2}(n-\lambda_{0})^2}e^{-i\frac{\bar{A}_{1}|_{0}}{2}(n-\lambda_{0})}\nonumber\\
\times&e^{-\frac{\bar{\omega}^{2}}{2}(m-\bar{\xi}_{0})^2}e^{-i\frac{\bar{A}_{y}|_{0}}{2}(m-\bar{\xi}_{0})}|n,m\mu_{y}\rangle,\label{Psi}
\end{align}
where $\bar{\omega}:=\mu_{y}\omega$ and $\bar{\xi}_{0}:=(\mu_{y})^{-1}\xi_{0}$. We do not expect the coherent state to exhibit semiclassical behavior for arbitrary values of $\zeta$. In fact, it is sufficient that it displays semiclassical behavior only in the late stage of the universe. At that stage, the physical volumes of the three- and four-dimensional fiducial cells are sufficiently large, the changing rates of the scale factors are sufficiently small, and the relative uncertainties of the fundamental variables can be neglected. From these considerations, we obtain the following semiclassical conditions
\begin{align}
&(\lambda_{0})^{-1}\ll\epsilon\ll|\bar{A}_{1}|_{0}|\ll1,\label{A1 condition}\\
&(\xi_{0})^{-1}\ll\omega\ll|\bar{A}_{y}|_{0}|\ll1.\label{AY condition}
\end{align}

As shown in Ref.\cite{Li:2025bzl}, under the conditions (\ref{A1 condition}) and (\ref{AY condition}), the coherent state $|\Psi_{\zeta}\rangle$ is sharply peaked at $(\bar{A}_{1}|_{0},\bar{\pi}^{1}|_{0},\bar{A}_{y}|_{0},\pi^{y}|_{0})$ with fluctuations within the specified tolerance, and the expectation value of the scalar constraint operator in the coherent state $|\Psi_{\zeta}\rangle$, retained up to the subleading-order $\epsilon^{2}$, is given by
\begin{align}\label{expectation C}
	&\langle\hat{C}^{sym}_{gr}[N]\rangle_{\zeta}\nonumber\\
	\approx&-\frac{3}{2}N\beta^{-2}\Delta^{-1/2}\kappa^{-1}|\bar{\pi}^{1}|_{0}|^{-1}[4\epsilon^2(\bar{\pi}^{1}|_{0})^{2}\nonumber\\
	+&2\sin^{2}(\bar{A}_1|_{0})(\bar{\pi}^{1}|_{0})^{2}+\sin(\bar{A}_1|_{0})\frac{\sin(\mu_{y}\bar{A}_{y}|_{0})}{\mu_{y}}\bar{\pi}^{1}|_{0}\pi^{y}|_{0}],
\end{align}
where the $\epsilon^2$ term reflects merely the fact that the term of $\bar{A}_1$ in $C_{gr}[N]$ is quadratic. It is obvious that when Eqs.(\ref{A1 condition}) and (\ref{AY condition}) are satisfied, Eq.(\ref{expectation C}) reduces to the classical scalar constraint (\ref{SCKK}). Therefore, $\hat{C}^{sym}_{gr}[N]$ is indeed a viable quantization. Moreover, we may take Eq.(\ref{expectation C}) (with the subscript 0 dropped) as the effective scalar constraint associated with the geometric sector.

The validity of the effective scalar constraint obtained in the canonical formulation can also be confirmed by calculating the extraction amplitude in the path-integral formulation with the scalar constraint operator $\hat{C}^{sym}_{gr}[N]$. Since $\hat{C}^{sym}_{gr}[N]$ is a self-adjoint operator, the physical states annihilated by $\hat{C}^{sym}_{gr}[N]$ can be constructed via the group-averaging procedure as \cite{Giulini:1998rk,Giulini:1998kf,ALMMT}
\begin{equation}
	\iota(\psi):=\lim_{\alpha_{0}\to\infty}\int_{-\alpha_{0}}^{\alpha_{0}}d\alpha\langle e^{-i\alpha\hat{C}^{sym}_{gr}[N]}\psi,\cdot\rangle,
\end{equation}
where $\psi\in\mathcal{D}_{gr}$. Additionally, the inner product can be defined on $\iota[\mathcal{D}_{gr}]$ as
\begin{align}
	\langle\iota(\psi_{1}),\iota(\psi_{2})\rangle_{phy}:=&[\iota(\psi_{1})](\psi_{2})\nonumber\\
	=&\lim_{\alpha_{0}\to\infty}\int_{-\alpha_{0}}^{\alpha_{0}}d\alpha\langle\psi_{1},e^{i\alpha\hat{C}^{sym}_{gr}[N]} \psi_{2}\rangle,
\end{align}
where $\psi_{1},\psi_{2}\in\mathcal{D}_{gr}$. In the timeless framework, the information about quantum dynamics is encoded in the extraction amplitude, which in the momentum representation is given by \cite{Ashtekar:2010gz}
\begin{align}
	A(\lambda_{f},\xi_{f};\lambda_{b},\xi_{b}):=&\langle\iota(T_{\lambda_{f},\xi_{f}}),\iota(T_{\lambda_{b},\xi_{b}}\rangle_{phy}\nonumber\\
	=&\lim_{\alpha_{0}\to\infty}\int_{-\alpha_{0}}^{\alpha_{0}}d\alpha\langle\lambda_{f},\xi_{f}|e^{i\alpha\hat{C}^{sym}_{gr}[N]}|\lambda_{b},\xi_{b}\rangle.\label{extraction amplitude1}
\end{align}

Through a careful derivation using the group-averaging method \cite{Huang:2011es,Qin:2012gaa,Zhang:2012em,Song:2020pqm}, we conclude (see Appendix \ref{sec:appendixBB} for details):
\begin{align}
	&A(\lambda_{f},\xi_{f};\lambda_{b},\xi_{b})\nonumber\\
	=&\mathfrak{c}\mathcal{D}\alpha\mathcal{D}\bar{\pi}^{1}\mathcal{D}\pi^{y}\mathcal{D}\bar{A}_{1}\mathcal{D}\bar{A}_{y}\exp\{\frac{i}{\hbar}\int_{0}^{1}dt[-\frac{3}{\beta\kappa}\bar{A}_{1}\partial_{t}\bar{\pi}^{1}\nonumber\\
	-&\frac{1}{\beta\kappa}\bar{A}_{y}\partial_{t}\pi^{y}-\alpha\hbar\frac{3}{2}N\beta^{-2}\Delta^{-1/2}\kappa^{-1}|\bar{\pi}^{1}|^{-1}[2\sin^{2}(\bar{A}_1)(\bar{\pi}^{1})^{2}\nonumber\\
	+&\sin(\bar{A}_1)\frac{\sin(\mu_{y}\bar{A}_{y})}{\mu_{y}}\bar{\pi}^{1}\pi^{y}]]\},\label{pathintegralC1}
\end{align}
where $\mathfrak{c}$ denotes an overall constant, and $t$ denotes the fictitious time variable. Thus, the effective scalar constraint (\ref{regularizedC}) has been derived from the path-integral formulation in Eq.(\ref{pathintegralC1}). Hence, it can be regarded as the effective scalar constraint with leading-order holonomy corrections for the geometric sector.

The extraction amplitude in the coherent-state representation can be computed analogously by\cite{QM1,QM2}
\begin{align}
	A(\zeta_{f};\zeta_{b}):=\lim_{\alpha_{0}\to\infty}\frac{\int_{-\alpha_{0}}^{\alpha_{0}}d\alpha\langle\Psi_{\zeta_{f}}|e^{i\alpha\hat{C}^{sym}_{gr}[N]}|\Psi_{\zeta_{b}}\rangle}{\|\Psi_{\zeta_{f}}\|\|\Psi_{\zeta_{b}}\|},\label{extraction amplitude2E}
\end{align}
where $|\Psi_{\zeta}\rangle$ denotes the coherent state defined in Eq.(\ref{Psi}) with norm $\|\Psi_{\zeta}\|:=\sqrt{\langle\Psi_{\zeta}|\Psi_{\zeta}\rangle}$, $\zeta_{\mathcal{N}}=\zeta_{f}$, $\zeta_{0}=\zeta_{b}$. After a detailed calculation based on the group-averaging method, we conclude (see Appendix \ref{sec:appendixBB} for details):
\begin{align}
	&A(\zeta_{f};\zeta_{b})\nonumber\\
	=&\mathfrak{c}'\mathcal{D}\alpha\mathcal{D}\bar{\pi}^{1}\mathcal{D}\pi^{y}\mathcal{D}\bar{A}_{1}\mathcal{D}\bar{A}_{y}\exp\{\frac{i}{\hbar}\int_{0}^{1}dt[-\frac{3}{\beta\kappa}\bar{A}_{1}\partial_{t}\bar{\pi}^{1}\nonumber\\
	-&\frac{1}{\beta\kappa}\bar{A}_{y}\partial_{t}\pi^{y}+\alpha\hbar\langle\hat{C}^{sym}_{gr}[N]\rangle_{\zeta}]\},
\end{align}
where $\mathfrak{c}'$ is an overall constant, $t$ is the fictitious time variable, and $\langle\cdot\rangle_{\zeta}$ denotes the expectation value evaluated in the coherent state $|\Psi_{\zeta}\rangle$. Thus, the expectation value of the scalar constraint operator under the coherent state, as given by Eq.(\ref{expectation C}), can be regarded as the effective scalar constraint with higher-order quantum corrections for the geometric part.

It should be noted that, although the derivation of the effective scalar constraint (\ref{expectation C}) indicates that its regime of validity is restricted to the domain in which the semiclassical conditions (\ref{A1 condition}) and (\ref{AY condition}) are satisfied, numerical studies of the quantum dynamics in various LQC models have shown that the quantum dynamics can still be well approximated by the effective dynamics even beyond this domain \cite{APS3,Martin-Benito:2009xaf,Assanioussi:2018hee,Assanioussi:2019iye}. This provides the motivation to assume that our effective model remains a good approximation even outside the domain of its validity, up to the point where the singularity occurs. To check the validity of this assumption, a tractable quantum dynamics for KK LQC, alternative to the dynamical model constructed in this section, will be studied in Appendix \ref{sec:appendixBBB}. It will be shown there that the effective dynamics is still consistent with the quantum dynamics in the vacuum case, even outside its domain of validity.

\section{Classical VS Effective Dynamics}\label{section4}
To explore the role of quantum gravitational effects in the dynamics, this section is devoted to a systematic investigation to the evolutions of the KK cosmology in both the classical and effective frameworks. Taking into account the different effects of various matter sources on spacetime, the analysis is organized into three physically representative cases of the cosmological model: vacuum, minimally coupling to a scalar field, and coupling to the dust. For convenience, we consistently choose $\bar{\pi}^{1},\pi^y>0$ and fix the lapse function to $N=1$ with $\tau$ representing the corresponding time parameter.

\subsection{The Vacuum Case}\label{sec:section41}
In the vacuum case, the equations of motion (EOM) of the classical theory generated by $C_{gr}[1]$ are given by
\begin{align}
	\partial_{\tau}\bar{A}_{1}&=\{\bar{A}_{1},C_{gr}[1]\}=-\beta^{-1}\Delta^{-1/2}(\bar{A}_{1})^2,\label{2.1}\\
	\partial_{\tau}\bar{A}_{y}&=\{\bar{A}_{y},C_{gr}[1]\}=-\frac{3}{2}\beta^{-1}\Delta^{-1/2}\bar{A}_{1}\bar{A}_{y},\label{2.2}\\
	\partial_{\tau}\bar{\pi}^{1}&=\{\bar{\pi}^{1},C_{gr}[1]\}=\frac{1}{2}\beta^{-1}\Delta^{-1/2}(4\bar{\pi}^{1}\bar{A}_{1}+\pi^{y}\bar{A}_{y}),\label{2.3}\\
	\partial_{\tau}\pi^{y}&=\{\pi^{y},C_{gr}[1]\}=\frac{3}{2}\beta^{-1}\Delta^{-1/2}\pi^{y}\bar{A}_{1}.\label{2.4}
\end{align}
Eq.(\ref{2.4}) implies that $\bar{A}_{1}$ is proportional to the Hubble parameter $\partial_{\tau}a/a$. Since observations indicate that the visible universe is currently under expanding, we focus on the region where $\bar{A}_{1}>0$. By integrating Eq.(\ref{2.1}), we obtain 
\begin{equation}\label{A1vaccl}
	\bar{A}_{1}=\beta\Delta^{1/2}\tau^{-1},
\end{equation}
where the integration constant corresponding to the time-translation freedom has been fixed by
setting $1/\bar{A}_{1}|_{\tau=0}=0$. Dividing Eq.(\ref{2.2}) by Eq.(\ref{2.1}), we obtain
\begin{equation}
	\frac{\partial_{\tau}\bar{A}_{y}}{\bar{A}_{y}}=\frac{3}{2}\frac{\partial_{\tau}\bar{A}_{1}}{\bar{A}_{1}},
\end{equation}
and hence
\begin{equation}
	\partial_{\tau}\ln|\bar{A}_{y}|=\partial_{\tau}\ln|\bar{A}_{1}|^{3/2},
\end{equation}
This implies that $O_{1}\!\equiv\!\bar{A}_{y}(\bar{A}_{1})^{-3/2}$ is also a Dirac observable, from which we obtain
\begin{equation}\label{Ayvaccl}
	\bar{A}_{y}=O_{1}(\bar{A}_{1})^{3/2}=O_{1}\beta^{3/2}\Delta^{3/4}\tau^{-3/2}.
\end{equation}
Moreover, it follows from Eq.(\ref{SCKK}) that $O_{y}\!\equiv\!\pi^{y}\bar{A}_{y}$ is a Dirac observable, and hence we obtain
\begin{equation}\label{piyvaccl}
	\pi^{y}=O_{y}(\bar{A}_{y})^{-1}=O_{y}(O_{1})^{-1}\beta^{-3/2}\Delta^{-3/4}\tau^{3/2}.
\end{equation}
By using the constraint $C_{gr}[1]=0$, we obtain
\begin{equation}\label{pibarccl}
	\bar{\pi}^{1}=-\frac{1}{2}O_{y}(\bar{A}_{1})^{-1}=-\frac{1}{2}O_{y}\beta^{-1}\Delta^{-1/2}\tau.
\end{equation}
Substituting the solutions (\ref{piyvaccl}) and (\ref{pibarccl}) into Eq.(\ref{scale factors}), the scale factors are given by
\begin{equation}
	a=a_{0}\tau^{1/2},\qquad b=b_{0}\tau^{-1/2},
\end{equation}
where
\begin{align}
	a_{0}&\equiv\![O_{y}/(V_{0}O_{1})]^{1/3}(\beta^{-1}\Delta^{-1/2})^{1/2},\nonumber\\ b_{0}&\equiv\!-\frac{1}{2}\Delta^{1/2}O_{1}(\beta^{-1}\Delta^{-1/2})^{-1/2}.
\end{align}
These results indicate the presence of a big bang (and big rip) singularity at $\tau=0$ in the vacuum KK cosmology. It should be noted that one can also solve the EOM for the region $\bar{A}_{1}<0$. Then the corresponding scale factors are given by
\begin{equation}
	a=a_{0}(-\tau)^{1/2},\qquad b=b_{0}(-\tau)^{-1/2},
\end{equation}
where the Dirac observable $O_{1}$ should be re-defined as $O_{1}\!\equiv\!\bar{A}_{y}|\bar{A}_{1}|^{-3/2}$.

We now turn to the effective dynamics of the model in the vacuum case. Let us first consider the effective scalar constraint $\mathcal{C}^{e\!f\!f}_{gr}[N]$ with first-order holonomy corrections given by Eq.(\ref{regularizedC}). The EOM generated by this effective scalar constraint read
\begin{align}
	\partial_{\tau}\bar{A}_{1}=&\{\bar{A}_{1},\mathcal{C}^{e\!f\!f}_{gr}[1]\}=-\beta^{-1}\Delta^{-1/2}\sin^{2}(\bar{A}_{1}),\label{41.1}\\
	\partial_{\tau}\bar{A}_{y}=&\{\bar{A}_{y},\mathcal{C}^{e\!f\!f}_{gr}[1]\}=-\frac{3}{2}\beta^{-1}\Delta^{-1/2}\sin(\bar{A}_{1})\frac{\sin(\mu_{y}\bar{A}_{y})}{\mu_{y}},\label{41.2}\\
	\partial_{\tau}\bar{\pi}^{1}=&\{\bar{\pi}^{1},\mathcal{C}^{e\!f\!f}_{gr}[1]\}=\frac{1}{2}\beta^{-1}\Delta^{-1/2}\cos(\bar{A}_{1})(4\bar{\pi}^{1}\sin(\bar{A}_{1})\nonumber\\
	+&\pi^{y}\frac{\sin(\mu_{y}\bar{A}_{y})}{\mu_{y}}),\label{41.3}\\
	\partial_{\tau}\pi^{y}=&\{\pi^{y},\mathcal{C}^{e\!f\!f}_{gr}[1]\}=\frac{3}{2}\beta^{-1}\Delta^{-1/2}\cos(\mu_{y}\bar{A}_{y})\pi^{y}\sin(\bar{A}_{1}).\label{41.4}
\end{align}
To compare with the classical theory, we focus on the region where $\sin(\bar{A}_{1})>0$. Integrating Eq.(\ref{41.1}), we obtain
\begin{equation}\label{Vac1}
\tan(\frac{\bar{A}_{1}}{2})=-\beta^{-1}\Delta^{-1/2}\tau+\sqrt{\beta^{-2}\Delta^{-1}\tau^{2}+1}=:f(\tau),
\end{equation}
where the integration constant associated with time-translation gauge freedom has been fixed by imposing $\tan(\bar{A}_{1}/2)|_{\tau=0}=1$. Dividing Eq.(\ref{41.2}) by Eq.(\ref{41.1}), we obtain
\begin{equation}
\frac{\mu_{y}\partial_{\tau}\bar{A}_{y}}{\sin(\mu_{y}\bar{A}_{y})}=\frac{3}{2}\frac{\partial_{\tau}\bar{A}_{1}}{\sin(\bar{A}_{1})},
\end{equation}
and hence
\begin{equation}
	\partial_{\tau}\ln\left|\tan(\frac{\mu_{y}\bar{A}_{y}}{2})\right|=\partial_{\tau}\ln\left|\tan(\frac{\bar{A}_{1}}{2})\right|^{3/2}.
\end{equation}
This indicates that $\bar{O}_{1}\!\equiv\!\tan(\mu_{y}\bar{A}_{y}/2)\tan^{-3/2}(\bar{A}_{1}/2)$ is a Dirac observable, from which we obtain
\begin{equation}\label{Ayvac}
\tan(\frac{\mu_{y}\bar{A}_{y}}{2})=\bar{O}_{1}\tan^{\frac{3}{2}}(\frac{\bar{A}_{1}}{2})=\bar{O}_{1}f^{\frac{3}{2}}(\tau).
\end{equation}
Moreover, the expression of the effective scalar constraint (\ref{regularizedC}) implies that $\bar{O}_{y}\!\equiv\!\pi^{y}\sin(\mu_{y}\bar{A}_{y})/\mu_{y}$ is also a Dirac observable, which leads to
\begin{equation}\label{piyvac}
\pi^{y}=\frac{\mu_{y}\bar{O}_{y}}{\sin(\mu_{y}\bar{A}_{y})}=\frac{\mu_{y}\bar{O}_{y}}{2}[\bar{O}_{1}f^{\frac{3}{2}}(\tau)+(\bar{O}_{1})^{-1}f^{-\frac{3}{2}}(\tau)].
\end{equation}
By using again the effective scalar constraint, we obtain
\begin{equation}\label{pi1vac}
\bar{\pi}^{1}=-\frac{\bar{O}_{y}}{2\sin(\bar{A}_{1})}=-\frac{\bar{O}_{y}}{4}[f(\tau)+f^{-1}(\tau)].
\end{equation}
Eqs.(\ref{piyvac}) and (\ref{pi1vac}) show that $\bar{\pi}^{1}$ and $\pi^{y}$ are finite and nonvanishing, thereby demonstrating that the singularity in the classical theory is naturally resolved in the effective model. Furthermore, the scale factors in the effective model are given by
\begin{align}
a&=\left(\frac{\mu_{y}\bar{O}_{y}}{2{V_0}}[\bar{O}_{1}f^{\frac{3}{2}}(\tau)+(\bar{O}_{1})^{-1}f^{-\frac{3}{2}}(\tau)]\right)^{1/3},\label{aeffvac}\\
b&=\sqrt{\Delta}\frac{-[f(\tau)+f^{-1}(\tau)]}{2\mu_{y}[\bar{O}_{1}f^{\frac{3}{2}}(\tau)+(\bar{O}_{1})^{-1}f^{-\frac{3}{2}}(\tau)]}.\label{beffvac}
\end{align}
Thus, the scale factor $a$ has a unique bounce point at $\tau_{B}\equiv\beta\Delta^{1/2}(|\bar{O}_{1}|^{2/3}-|\bar{O}_{1}|^{-2/3})/2$, and the scale factor $b$ possesses a unique collapse point. As $\tau\to\pm\infty$, the scale factors can be asymptotically approximated as
\begin{equation}
a=a_{\pm}(\pm\tau)^{1/2},\qquad b=b_{\pm}(\pm\tau)^{-1/2},
\end{equation}
where 
\begin{align}
	a_{\pm}&\equiv\![\frac{\mu_{y}\bar{O}_{y}}{2V_{0}(\bar{O}_{1})^{\pm1}}]^{\frac{1}{3}}(2\beta^{-1}\Delta^{-\frac{1}{2}})^{\frac{1}{2}},\nonumber\\
	b_{\pm}&\equiv\!-\Delta^{\frac{1}{2}}\frac{(\bar{O}_{1})^{\pm1}}{2\mu_{y}}(2\beta^{-1}\Delta^{-\frac{1}{2}})^{-\frac{1}{2}}.
\end{align}
Therefore, the effective model has the correct classical limit, and the scale factor $b$ of the extra dimension gradually contracts at late times, exhibiting the characteristic behavior of dynamical compactification. The different powers of $\bar{O}_{1}$ in $a_{\pm}$ indicate the asymmetry of the scale factor $a$ before and after the bounce. From Eq.(\ref{41.4}), the Hubble parameter is given by
\begin{equation}\label{Hvac}
H:=\frac{\partial_{\tau}a}{a}=\frac{1}{2}\beta^{-1}\Delta^{-1/2}\cos(\mu_{y}\bar{A}_{y})\sin(\bar{A}_{1}).
\end{equation}
For simplicity, we restrict our attention to the regime $|\bar{O}_{1}|\ll1$, where the Hubble parameter $H$ reaches its maximum near $\tau=0$, due to $\sin(\bar{A}_{1})|_{\tau=0}=1$ and 
\begin{equation}
\cos(\mu_{y}\bar{A}_{y})|_{\tau=0}=\left.\frac{1-\tan^{2}(\mu_{y}\bar{A}_{y}/2)}{1+\tan^{2}(\mu_{y}\bar{A}_{y}/2)}\right|_{\tau=0}=\frac{1-(\bar{O}_{1})^{2}}{1+(\bar{O}_{1})^{2}}\approx 1.
 \end{equation}
A straightforward calculation based on Eqs.(\ref{Hvac}), (\ref{41.1}) and (\ref{41.2}) gives
\begin{align}
&\partial_{\tau}H\nonumber\\
=&\frac{1}{2}\beta^{-2}\Delta^{-1}\sin^{2}(\bar{A}_{1})[\frac{3}{2}\sin^{2}(\mu_{y}\bar{A}_{y})-\cos(\mu_{y}\bar{A}_{y})\cos(\bar{A}_{1})].
\end{align}
Using Eqs.(\ref{Vac1}) and (\ref{Ayvac}), one can show that $\cos(\mu_{y}\bar{A}_{y})\cos(\bar{A}_{1})<0$ for $\tau_{B}<\tau<0$, which implies $\partial_{\tau}H>0$. Hence, the scale factor $a$ undergoes a super-inflationary phase during this interval. From Eq.(\ref{piyvac}), the e-folding number associated with this phase can be estimated as
\begin{equation}
\ln(\frac{a|_{\tau=0}}{a|_{\tau=\tau_{B}}})=\frac{1}{3}\ln(\frac{|\bar{O}_{1}|+|\bar{O}_{1}|^{-1}}{2})
\approx\frac{1}{3}\ln(\frac{|\bar{O}_{1}|^{-1}}{2}).
\end{equation}
By appropriately choosing the value of $|\bar{O}_{1}|$, the desired 55 e-foldings of inflation \cite{Planck:2018jri} can be achieved. This provides a fascinating origin of the cosmic inflation by the effects of the higher-dimensional quantum geometry.

\begin{figure*}[!htb]
	\subfigure[]{
		\label{Fig1:subfig:a}
		\includegraphics [width=0.46\textwidth]{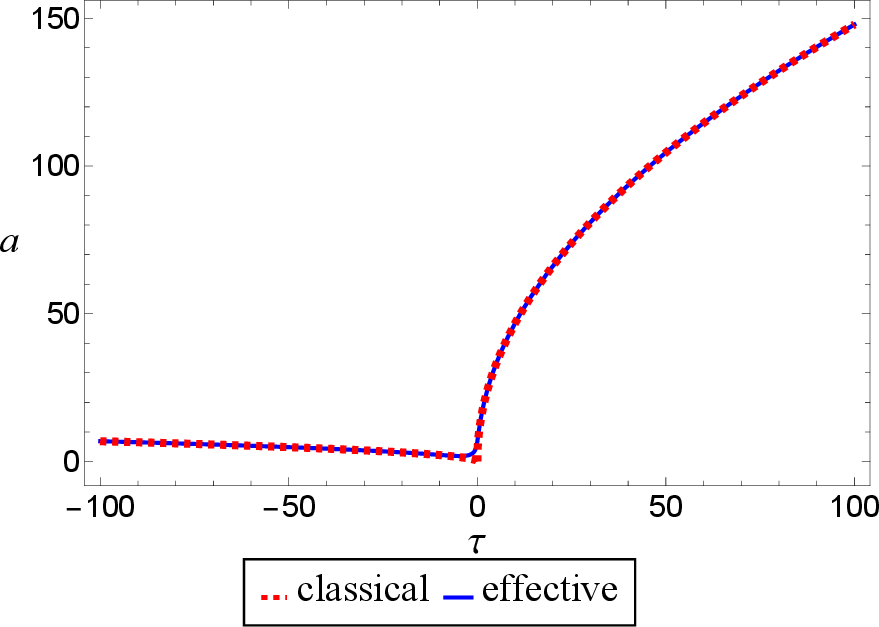}}\qquad
	\subfigure[]{
		\label{Fig:subfig:b}
		\includegraphics [width=0.46\textwidth]{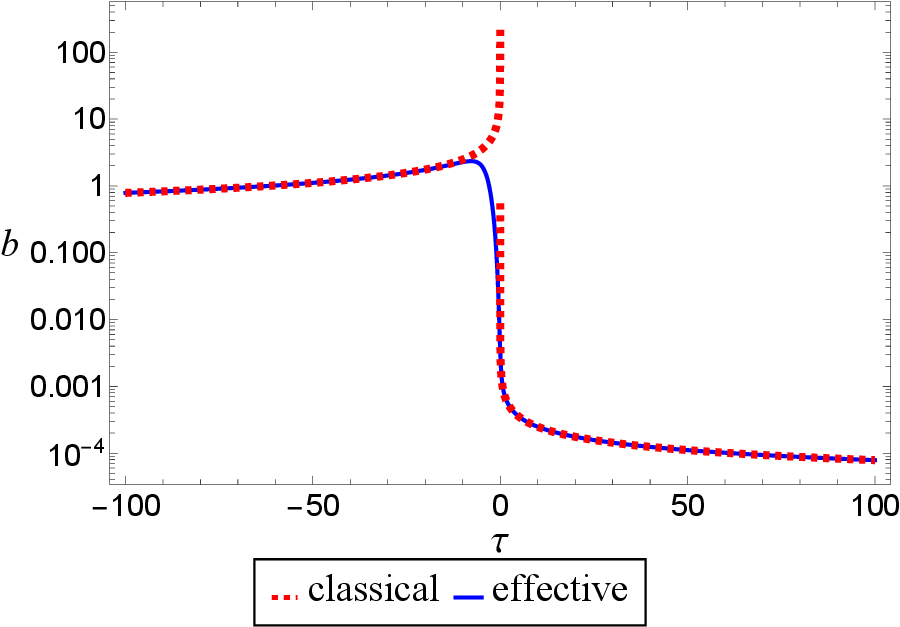}}
	\caption{(a)The evolution of the scale factor $a$ in both the classical and the effective models in the vacuum case. (b)The evolution of the scale factor $b$ in both the classical and the effective models in the vacuum case. Both figures are plotted using the parameters $G^{(5)}=c=\hbar=V_{0}=1$ and the initial values $\bar{O}_{y}=-1$ and $\bar{O}_{1}=-0.01$.}\label{Fig.1}
\end{figure*}

The evolutions of the scale factors in the effective and classical models are compared in Fig.\ref{Fig.1}. Note that we choose $\beta=0.1969227$ as indicated by the calculation of black hole entropy in 5D LQG \cite{Song:2022zit}, and $\mu_{y}=2\pi$ for simplicity. As shown in Fig.\ref{Fig.1}, the big bang singularity of scale factor $a$ and the past big rip singularity of scale factor $b$ are respectively replaced by a quantum bounce and a quantum re-collapse.

Next, let us consider the effective scalar constraint (\ref{expectation C}) with subleading-order quantum corrections. For simplicity, we only consider the case where $\epsilon$ is a constant. It turns out that, by choosing an appropriate lapse function $N$, our effective scalar constraint can be brought into the same form as the effective scalar constraint (46) in Ref.\cite{Corichi:2015xia}. Therefore, the detailed dynamical behavior in this case can be inferred by comparing with the corresponding analyses in Refs.\cite{Corichi:2015xia,Ashtekar:2018cay}. A representative result is that $\pi^{y}$ undergoes a bounce, while $\bar{\pi}^{1}$ not only bounces but also collapses to zero. However, the semiclassical condition (\ref{A1 condition}) suggests that the effective model may become invalid as $\bar{\pi}^{1}$ approaches zero. Furthermore, under the condition $\epsilon\ll1$, it is straightforward to check that one can always choose the initial conditions such that the super-inflationary phase of the visible universe following the quantum bounce provides a sufficient number of e-folds to serve as an alternative to the conventional inflation.

\subsection{The Case of Minimally Coupling to a Scalar Field}\label{sec:section42}
As a commonly used matter field in LQC\cite{LQC5,Boj,Ash-view,AS11,BCM,APS3,ACS,Zhang14,Zhang16,DMY,YDM,Zhang:2021zfp,Taveras,Huang:2011es,Qin:2012gaa,Zhang:2012em,Song:2020pqm,QM1,QM2}, the scalar field, when minimally coupled to the classical KK cosmology, gives the following scalar constraint
\begin{equation}\label{classical C2}
C_{\phi}[N]:=C_{gr}[N]+N\frac{\Delta^{-1/2}}{4\pi}|\bar{\pi}^{1}|^{-1}p_{\phi}^{2},
\end{equation}
where $\phi$ and $p_{\phi}$ represent the scalar field and its conjugate momentum respectively, with their non-trivial Poisson bracket given by $\{\phi,p_{\phi}\}=1$. The EOM for $\bar{A}_{y}$, $\bar{\pi}^{1}$, and $\pi^{y}$ generated by $C_{\phi}[1]$ are the same as those in Eqs.(\ref{2.2}), (\ref{2.3}) and (\ref{2.4}), while the EOM for the remaining variables are given by
\begin{align}
\partial_{\tau}\bar{A}_{1}=&\{\bar{A}_{1},C_{\phi}[1]\}\nonumber\\
=&-\beta^{-1}\Delta^{-1/2}(\bar{A}_{1})^{2}-\frac{\Delta^{-1/2}}{4\pi}\frac{\beta\kappa}{3}p_{\phi}^{2}(\bar{\pi}^{1})^{-2},\label{42.1}\\
\partial_{\tau}\phi=&\{\phi,C_{\phi}[1]\}=\frac{\Delta^{-1/2}}{2\pi}p_{\phi}(\bar{\pi}^{1})^{-1},\label{42.2}\\
\partial_{\tau}p_{\phi}=&\{p_{\phi},C_{\phi}[1]\}=0,\label{42.3}
\end{align}
By introducing the functions $O_{2}\equiv\bar{A}_{1}\bar{\pi}^{1}$ and $\varrho\equiv\beta^{-1}\Delta^{-1/2}+\Delta^{-1/2}\beta\kappa(12\pi)^{-1}p_{\phi}^{2}(O_{2})^{-2}$, Eqs.(\ref{42.1}) and (\ref{42.2}) can be rewritten as
\begin{align}
\partial_{\tau}\bar{A}_{1}=&-\varrho(\bar{A}_{1})^{2},\label{42.1b}\\
\partial_{\tau}\phi=&\frac{\Delta^{-1/2}}{2\pi}p_{\phi}(O_{2})^{-1}\bar{A}_{1}.\label{42.2b}
\end{align}
The expression of the scalar constraint (\ref{classical C2}) implies that $O_{y}\!\equiv\!\bar{A}_{y}\pi^{y}$, $O_{2}$ and $p_{\phi}$ are Dirac observables, and hence so is $\varrho$. We focus on the region $\bar{A}_{1}>0$ where the visible universe is expanding, as indicated by Eq.(\ref{2.4}). By integrating Eq.(\ref{42.1b}), we obtain
\begin{equation}
\bar{A}_{1}=(\varrho\tau)^{-1},
\end{equation}
where the integration constant corresponding to the time-translation freedom has been uniquely determined by the condition $1/\bar{A}_{1}|_{\tau=0}=0$. Dividing Eq.(\ref{2.2}) by Eq.(\ref{42.1b}) and Eq.(\ref{42.1b}) by Eq.(\ref{42.2b}), we obtain
\begin{align}
\frac{\partial_{\tau}\bar{A}_{y}}{\bar{A}_{y}}&=\frac{3}{2}\beta^{-1}\Delta^{-1/2}\varrho^{-1}\frac{\partial_{\tau}\bar{A}_{1}}{\bar{A}_{1}},\\
\partial_{\tau}\phi=&-\frac{\Delta^{-1/2}}{2\pi}p_{\phi}(O_{2})^{-1}\varrho^{-1}\frac{\partial_{\tau}\bar{A}_{1}}{\bar{A}_{1}},
\end{align}
and hence
\begin{align}
\partial_{\tau}\ln|\bar{A}_{y}|&=\partial_{\tau}\ln|\bar{A}_{1}|^{\frac{3}{2}\nu_{0}},\\
\partial_{\tau}\phi=&-\frac{\Delta^{-1/2}}{2\pi}p_{\phi}(O_{2}\varrho)^{-1}\partial_{\tau}\ln|\bar{A}_{1}|,
\end{align}
where $\nu_{0}\!\equiv\!\beta^{-1}\Delta^{-1/2}\varrho^{-1}$. It follows that $O_{3}\!\equiv\!\bar{A}_{y}(\bar{A}_{1})^{-3\nu_{0}/2}$ and $\phi_{0}\equiv\phi+\Delta^{-1/2}p_{\phi}(2\pi O_{2}\varrho)^{-1}\ln\bar{A}_{1}$ are also Dirac observables. Hence, one gets
\begin{align}
\bar{A}_{y}=&O_{3}(\bar{A}_{1})^{3\nu_{0}/2}=O_{3}(\varrho\tau)^{-\frac{3}{2}\nu_{0}},\\
\phi=&-\frac{\Delta^{-1/2}}{2\pi}p_{\phi}(O_{2}\varrho)^{-1}\ln\bar{A}_{1}+\phi_{0}\nonumber\\
=&-\frac{\Delta^{-1/2}}{2\pi}p_{\phi}(O_{2}\varrho)^{-1}\ln[(\varrho\tau)^{-1}]+\phi_{0}.
\end{align}
Taking into account that $O_{2}$ and $O_{y}$ are Dirac observables, we obtain
\begin{align}
\bar{\pi}^{1}=&O_{2}(\bar{A}_{1})^{-1}=O_{2}\varrho\tau,\label{pi1scal}\\ \pi^{y}=&O_{y}(\bar{A}_{y})^{-1}=O_{y}(O_{3})^{-1}(\varrho\tau)^{\frac{3}{2}\nu_{0}}.\label{piysacl}
\end{align}
Moreover, the scalar constraint implies ${O}_{y}=[(2/\nu_{0})-4]{O}_{2}$, which allows us to choose $\{{O}_{2},{O}_{3},\nu_{0},\phi_{0}\}$ as a complete set of Dirac observables for the system. Substituting Eqs.(\ref{pi1scal}) and (\ref{piysacl}) into Eq.(\ref{scale factors}), one finds the solutions of the scale factors as
\begin{equation}
a=a^{(\phi)}_{0}\tau^{\nu_{0}/2},\qquad b=b^{(\phi)}_{0}\tau^{1-3\nu_{0}/2},
\end{equation}
where
\begin{align}
	a^{(\phi)}_{0}&\equiv\!(O_{y}/(V_{0}{O}_{3}))^{1/3}(\varrho)^{\nu_{0}/2},\nonumber\\ b^{(\phi)}_{0}&\equiv\!\Delta^{1/2}(O_{2}O_{3}/O_{y})(\varrho)^{1-3\nu_{0}/2}.
\end{align}
The definition of the Dirac observable $\nu_{0}$ implies $0<\nu_{0}<1$ for $p_{\phi}\neq 0$. Thus, the scale factor $a$ expands more slowly when a scalar field is coupled than in the vacuum case. Moreover, depending on whether $\nu_{0}$ is greater than, equal to, or less than $2/3$, the scale factor $b$ respectively undergoes contraction, remains constant, or expands during the evolution. However, this spacetime exhibits a singularity at $\tau=0$. It should be noted that one can also solve the EOM for the region $\bar{A}_{1}<0$. Then the corresponding scale factors are given by 
\begin{equation}
a=a^{(\phi)}_{0}(-\tau)^{\nu_{0}/2},\qquad b=b^{(\phi)}_{0}(-\tau)^{1-3\nu_{0}/2},
\end{equation}
where the Dirac observables $O_{3}$ and $\phi_{0}$ should be respectively re-defined as $O_{3}\!\equiv\!\bar{A}_{y}|\bar{A}_{1}|^{-3\nu_{0}/2}$ and $\phi_{0}\equiv\phi+\Delta^{-1/2}p_{\phi}(2\pi O_{2}\varrho)^{-1}\ln|\bar{A}_{1}|$.

We now turn to the effective dynamics of the model coupled to the scalar field $\phi$. To obtain the contribution of the scalar field to the effective scalar constraint, one may proceed in complete parallel with the geometric part by performing a polymer-like quantization of the configuration variable $\phi$, and by considering the expectation value of the corresponding scalar constraint operator in coherent states as well as the path-integral form of its extraction amplitude. It turns out that, up to the subleading-order quantum corrections,  the contribution of the scalar field to the effective scalar constraint takes the same expression as in the classical scalar constraint (\ref{classical C2}). Similar to the geometric part, the higher-order quantum corrections of the scalar field can not be included, in order to ensure that the resulting effective scalar constraint scales by the same proportion under a scale transformation of the fiducial cell \cite{Li:2025bzl}. Thus, the effective scalar constraint for the model minimally coupled to the scalar field is given by
\begin{equation}\label{effective C2}
\mathcal{C}^{e\!f\!f}_{\phi}[N]:=\mathcal{C}^{e\!f\!f}_{gr}[N]+N\frac{\Delta^{-1/2}}{4\pi}|\bar{\pi}^{1}|^{-1}p_{\phi}^{2}.
\end{equation}

We first consider the case that the geometric part $\mathcal{C}^{e\!f\!f}_{gr}[N]$ of Eq.(\ref{effective C2}) contain only the leading-order holonomy corrections given by Eq.(\ref{regularizedC}). Then, the EOM for $\phi$, $p_{\phi}$, $\bar{A}_{y}$, $\bar{\pi}^{1}$, and $\pi^{y}$ generated by $\mathcal{C}^{e\!f\!f}_{\phi}[1]$ are the same as those in Eqs.(\ref{42.2}), (\ref{42.3}), (\ref{41.2}), (\ref{41.3}) and (\ref{41.4}), while the EOM for the remaining variable $\bar{A}_{1}$ is given by
\begin{align}
\partial_{\tau}\bar{A}_{1}=&\{\bar{A}_{1},\mathcal{C}^{e\!f\!f}_{\phi}[1]\}\nonumber\\
=&-\beta^{-1}\Delta^{-1/2}\sin^{2}(\bar{A}_{1})-\frac{\Delta^{-1/2}}{4\pi}\frac{\beta\kappa}{3}p_{\phi}^{2}(\bar{\pi}^{1})^{-2}.\label{42.4}
\end{align}
By introducing the variables $\bar{O}_{2}\!\equiv\!\sin(\bar{A}_{1})\bar{\pi}^{1}$ and $\bar{\varrho}\equiv\beta^{-1}\Delta^{-1/2}+\Delta^{-1/2}\beta\kappa(12\pi)^{-1}p_{\phi}^{2}(\bar{O}_{2})^{-2}$, Eqs.(\ref{42.4}) and (\ref{42.2}) can be rewritten as
\begin{align}
\partial_{\tau}\bar{A}_{1}=&-\bar{\varrho}\sin^{2}(\bar{A}_{1}),\label{42.4ef}\\
\partial_{\tau}\phi=&\frac{\Delta^{-1/2}}{2\pi}p_{\phi}(\bar{O}_{2})^{-1}\sin(\bar{A}_{1}).\label{42.2ef}
\end{align}
Note that the effective scalar constraint (\ref{effective C2}) implies that $\bar{O}_{y}\!\equiv\!\pi^{y}\sin(\mu_{y}\bar{A}_{y})/\mu_{y}$, $\bar{O}_{2}$ and $p_{\phi}$ are Dirac observables, and hence so is $\bar{\varrho}$. To facilitate the comparison with the classical case, we restrict our attention to the region $\sin(\bar{A}_{1})>0$. By integrating Eq.(\ref{42.4ef}), we obtain
\begin{equation}
\tan(\frac{\bar{A}_{1}}{2})=-\bar{\varrho}\tau+\sqrt{\bar{\varrho}^{2}\tau^{2}+1}=:\tilde{f}(\tau),
\end{equation}
where the integration constant corresponding to the time-translation gauge freedom is fixed by choosing the gauge condition $\tan(\bar{A}_{1}/2)|_{\tau=0}=1$. Dividing Eq.(\ref{41.2}) by Eq.(\ref{42.4ef}) and Eq.(\ref{42.4ef}) by Eq.(\ref{42.2ef}), we obtain
\begin{align}
\frac{\mu_{y}\partial_{\tau}\bar{A}_{y}}{\sin(\mu_{y}\bar{A}_{y})}=&\frac{3}{2}\beta^{-1}\Delta^{-1/2}\bar{\varrho}^{-1}\frac{\partial_{\tau}\bar{A}_{1}}{\sin(\bar{A}_{1})},\\
\partial_{\tau}\phi=&-\frac{\Delta^{-1/2}}{2\pi}p_{\phi}(\bar{O}_{2})^{-1}\bar{\varrho}^{-1}\frac{\partial_{\tau}\bar{A}_{1}}{\sin(\bar{A}_{1})},
\end{align}
and hence
\begin{align}
\partial_{\tau}\ln\left|\tan(\frac{\mu_{y}\bar{A}_{y}}{2})\right|&=\partial_{\tau}\ln\left|\tan(\frac{\bar{A}_{1}}{2})\right|^{\frac{3}{2}\bar{\nu}_{0}},\\
\partial_{\tau}\phi=&-\frac{\Delta^{-1/2}}{2\pi}p_{\phi}(\bar{O}_{2}\bar{\varrho})^{-1}\partial_{\tau}\ln\left|\tan(\frac{\bar{A}_{1}}{2})\right|,
\end{align}
where $\bar{\nu}_{0}\!\equiv\!\beta^{-1}\Delta^{-1/2}\bar{\varrho}^{-1}$. This implies that $\bar{O}_{3}\!\equiv\!\tan(\mu_{y}\bar{A}_{y}/2)\tan^{-3\bar{\nu}_{0}/2}(\bar{A}_{1}/2)$ and $\bar{\phi}_{0}\equiv\phi+\Delta^{-1/2}p_{\phi}(2\pi\bar{O}_{2}\bar{\varrho})^{-1}\ln(\tan(\bar{A}_{1}/2))$ are also Dirac observables. Hence, we obtain
\begin{align}
\tan(\frac{\mu_{y}\bar{A}_{y}}{2})=&\bar{O}_{3}\tan^{\frac{3}{2}\bar{\nu}_{0}}(\frac{\bar{A}_{1}}{2})=\bar{O}_{3}\tilde{f}^{\frac{3}{2}\bar{\nu}_{0}}(\tau),\\
\phi=&-\frac{\Delta^{-1/2}}{2\pi}p_{\phi}(\bar{O}_{2}\bar{\varrho})^{-1}\ln(\tan(\frac{\bar{A}_{1}}{2}))+\bar{\phi}_{0}\nonumber\\
=&-\frac{\Delta^{-1/2}}{2\pi}p_{\phi}(\bar{O}_{2}\bar{\varrho})^{-1}\ln(\tilde{f}(\tau))+\bar{\phi}_{0}.
\end{align}
Noting that $\bar{O}_{2}$ and $\bar{O}_{y}$ are Dirac observables, we obtain
\begin{align}
\bar{\pi}^{1}=&\frac{\bar{O}_{2}}{\sin(\bar{A}_{1})}=\frac{\bar{O}_{2}}{2}[\tilde{f}(\tau)+\tilde{f}^{-1}(\tau)],\label{pi1scl}\\ \pi^{y}=&\frac{\mu_{y}\bar{O}_{y}}{\sin(\mu_{y}\bar{A}_{y})}=\frac{\mu_{y}\bar{O}_{y}}{2}[\bar{O}_{3}\tilde{f}^{\frac{3}{2}\bar{\nu}_{0}}(\tau)+(\bar{O}_{3})^{-1}\tilde{f}^{-\frac{3}{2}\bar{\nu}_{0}}(\tau)].\label{piyscl}
\end{align}
Moreover, the effective scalar constraint implies $\bar{O}_{y}=[(2/\bar{\nu}_{0})-4]\bar{O}_{2}$, so that $\{\bar{O}_{2},\bar{O}_{3},\bar{\nu}_{0},\bar{\phi}_{0}\}$ form a complete set of Dirac observables of the system.

\begin{figure*}[!htb]
	\subfigure[]{
		\label{Figs1a}
		\includegraphics [width=0.46\textwidth]{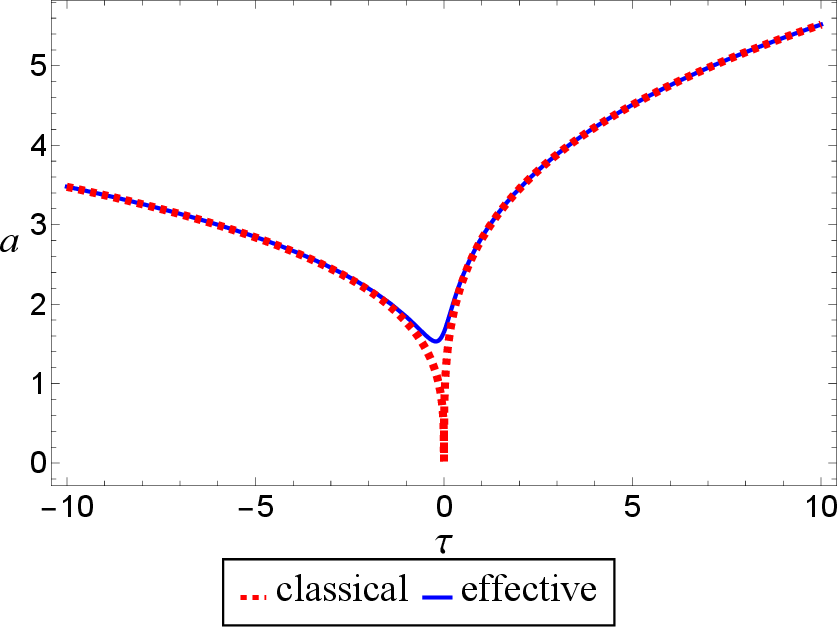}}\qquad
	\subfigure[]{
		\label{Figs1b}
		\includegraphics [width=0.46\textwidth]{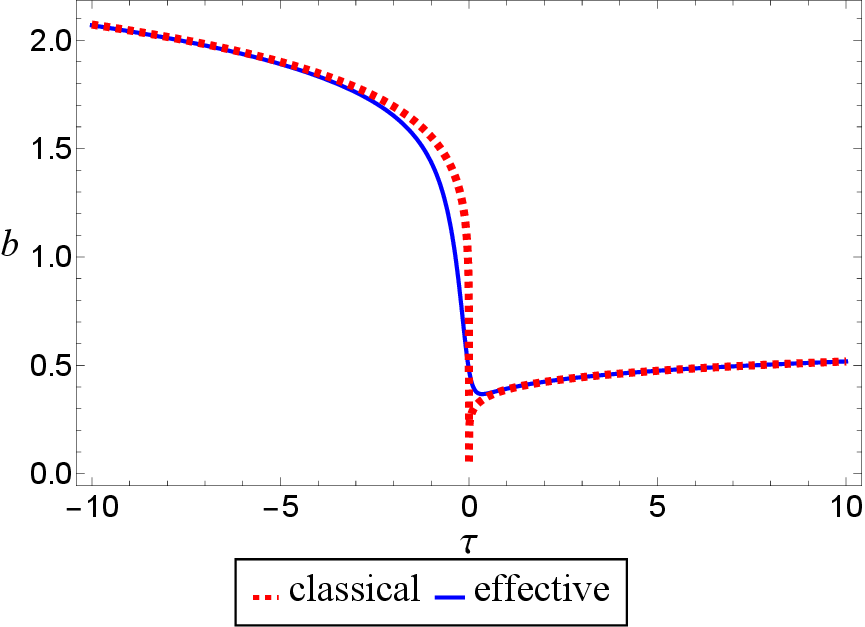}}
	\caption{(a)The evolution of the scale factor $a$ in both the classical and the effective models coupling with the scalar field in case I. (b)The evolution of the scale factor $b$ in both the classical and the effective models coupling with the scalar field in case I. Both figures are plotted using the parameters $G^{(5)}=c=\hbar=V_{0}=1$ and the initial values $\bar{\nu}_{0}=7/12$, $\bar{O}_{2}=1$ and $\bar{O}_{3}=-0.5$.}\label{Fig.s1}
\end{figure*}

\begin{figure*}[!htb]
	\subfigure[]{
		\label{Figs2a}
		\includegraphics [width=0.46\textwidth]{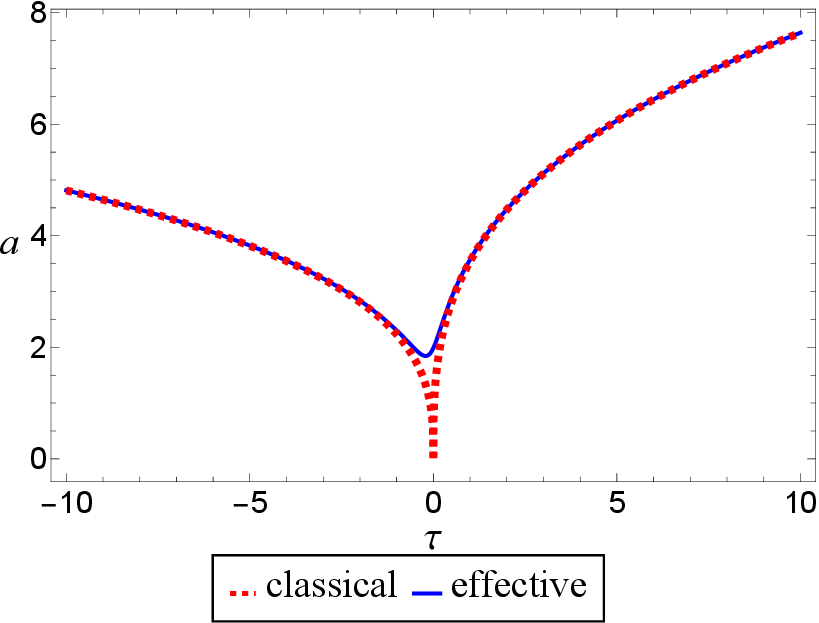}}\qquad
	\subfigure[]{
		\label{Figs2b}
		\includegraphics [width=0.46\textwidth]{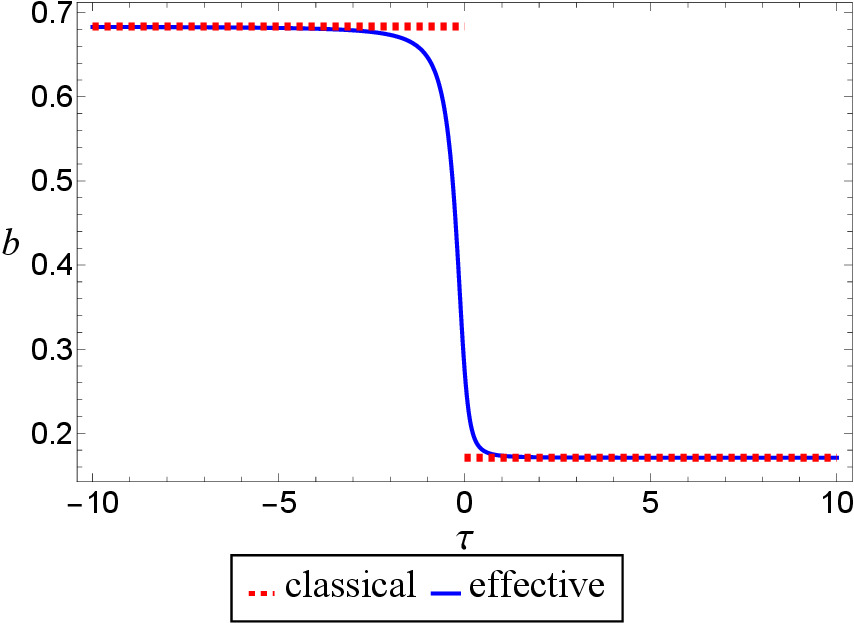}}
	\caption{(a)The evolution of the scale factor $a$ in both the classical and the effective models coupling with the scalar field in case II. (b)The evolution of the scale factor $b$ in both the classical and the effective models coupling with the scalar field in case II. Both figures are plotted using the parameters $G^{(5)}=c=\hbar=V_{0}=1$ and the initial values $\bar{\nu}_{0}=2/3$, $\bar{O}_{2}=1$ and $\bar{O}_{3}=-0.5$.}\label{Fig.s2}
\end{figure*}

\begin{figure*}[!htb]
	\subfigure[]{
		\label{Figs3a}
		\includegraphics [width=0.46\textwidth]{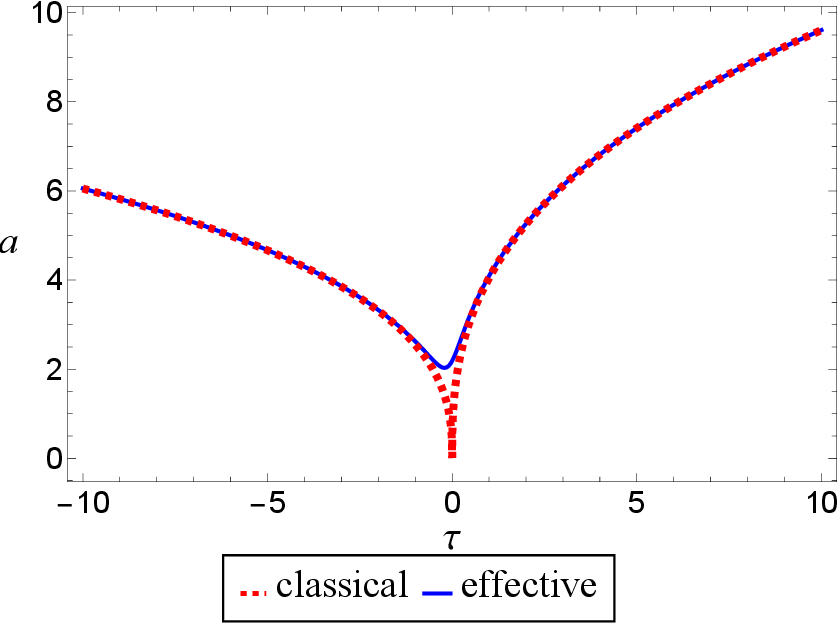}}\qquad
	\subfigure[]{
		\label{Figs3b}
		\includegraphics [width=0.46\textwidth]{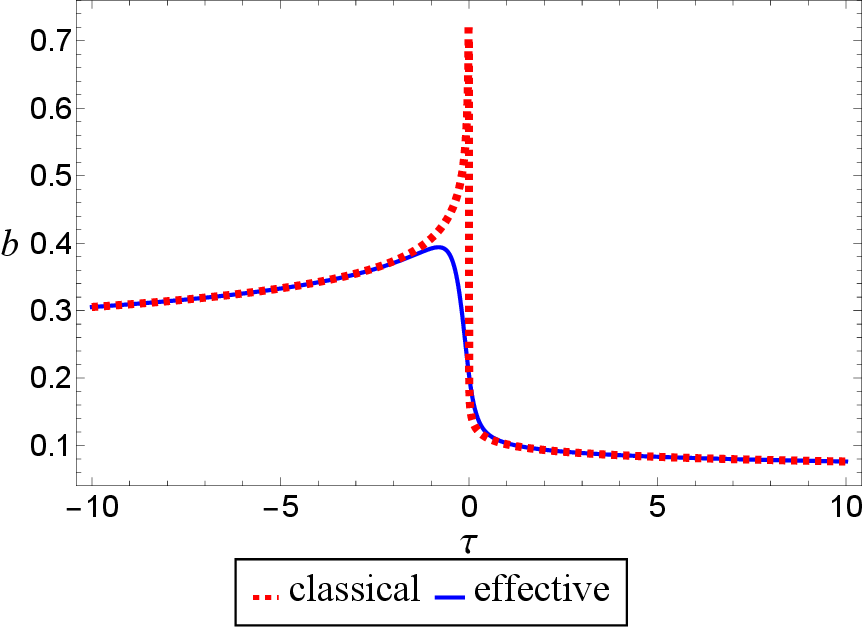}}
	\caption{(a)The evolution of the scale factor $a$ in both the classical and the effective models coupling with the scalar field in case III. (b)The evolution of the scale factor $b$ in both the classical and the effective models coupling with the scalar field in case III. Both figures are plotted using the parameters $G^{(5)}=c=\hbar=V_{0}=1$ and the initial values $\bar{\nu}_{0}=3/4$, $\bar{O}_{2}=1$ and $\bar{O}_{3}=-0.5$.}\label{Fig.s3}
\end{figure*}

For the initial conditions with $\bar{O}_{3}\ne 0$, the analysis of Sec.\ref{sec:section41} for the vacuum case can be repeated in the case of coupling to the scalar field to yield the following results: (1) The finiteness and nonvanishing of $\bar{\pi}^{1}$ and $\pi^{y}$ imply that the classical singularity is naturally resolved in the effective model. Moreover, the scale factors in the effective model read
\begin{align}
	a&=\left(\frac{\mu_{y}\bar{O}_{y}}{2V_{0}}[\bar{O}_{3}\tilde{f}^{\frac{3}{2}\bar{\nu}_{0}}(\tau)+(\bar{O}_{3})^{-1}\tilde{f}^{-\frac{3}{2}\bar{\nu}_{0}}(\tau)]\right)^{1/3},\label{aeffphi}\\
	b&=\sqrt{\Delta}\frac{\bar{O}_{2}[\tilde{f}(\tau)+\tilde{f}^{-1}(\tau)]}{\mu_{y}\bar{O}_{y}[\bar{O}_{3}\tilde{f}^{\frac{3}{2}\bar{\nu}_{0}}(\tau)+(\bar{O}_{3})^{-1}\tilde{f}^{-\frac{3}{2}\bar{\nu}_{0}}(\tau)]}.\label{beffphi}
\end{align}
Therefore, the scale factor $a$ has a unique bounce point at $\tau'_{B}\equiv(|\bar{O}_{3}|^{2/(3\bar{\nu}_{0})}-|\bar{O}_{3}|^{-2/(3\bar{\nu}_{0})})/(2\bar{\varrho})$. Note that the condition $\bar{O}_{3}\ne 0$ implies $\bar{\nu}_{0}\ne1/2$. In the cases of $\bar{\nu}_{0}>2/3$ and $1/2\ne\bar{\nu}_{0}<2/3$, the scale factor $b$ possesses a unique collapse point and a unique bounce point, respectively. In the case of $\bar{\nu}_{0}=2/3$, the scale factor $b$ decreases monotonically, remains constant, and increases monotonically with respect to $\tau$ for $|\bar{O}_{3}|<1$, $|\bar{O}_{3}|=1$, and $|\bar{O}_{3}|>1$, respectively. (2) When $|\bar{O}_{3}|\ll1$, the period $\tau'_{B}<\tau<0$ corresponds to a super-inflationary phase of the scale factor $a$, with the associated e-folding number estimated as $-\ln(2|\bar{O}_{3}|)/3$. This implies that the desired 55 e-folds of inflation in the visible dimensions can be realized by choosing an appropriate value of $|\bar{O}_{3}|$. (3) As $\tau\to\pm\infty$, the scale factors asymptotically behave as
\begin{equation}
a=a^{(\phi)}_{\pm}(\pm\tau)^{\bar{\nu}_{0}/2},\qquad b=b^{(\phi)}_{\pm}(\pm\tau)^{1-3\bar{\nu}_{0}/2},
\end{equation}
where
\begin{align}
	a^{(\phi)}_{\pm}&\equiv\![\mu_{y}\bar{O}_{y}/(2V_{0}(\bar{O}_{3})^{\pm1})]^{1/3}(2\bar{\varrho})^{\bar{\nu}_{0}/2},\nonumber\\ b^{(\phi)}_{\pm}&\equiv\!\Delta^{1/2}[\bar{O}_{2}(\bar{O}_{3})^{\pm1}/(\mu_{y}\bar{O}_{y})](2\bar{\varrho})^{1-3\bar{\nu}_{0}/2}.
\end{align}
Hence, the effective model has the correct classical limit. The evolutions of the scale factors for $1/2\ne\bar{\nu}_{0}<2/3$, $\bar{\nu}_{0}=2/3$, and $\bar{\nu}_{0}>2/3$ are shown in Fig.\ref{Fig.s1}, Fig.\ref{Fig.s2}, and Fig.\ref{Fig.s3}, respectively. It can be seen that, in all these cases, the big bang singularities of the scale factor $a$ are resolved by quantum bounces. In the cases of $1/2\ne\bar{\nu}_{0}<2/3$ and $\bar{\nu}_{0}>2/3$, the big bang singularity and the past big rip singularity of the scale factor $b$ are resolved by a quantum bounce and a quantum re-collapse, respectively. In particular, for $\bar{\nu}_{0}>2/3$, the late-time contraction of the scale factor $b$ demonstrates the occurrence of dynamical compactification. In the case of $\bar{\nu}_{0}=2/3$, the scale factor $b$ evolves from one value to another without undergoing a bounce or re-collapse.

\begin{figure*}[!htb]
	\subfigure[]{
		\label{Figs4a}
		\includegraphics [width=0.46\textwidth]{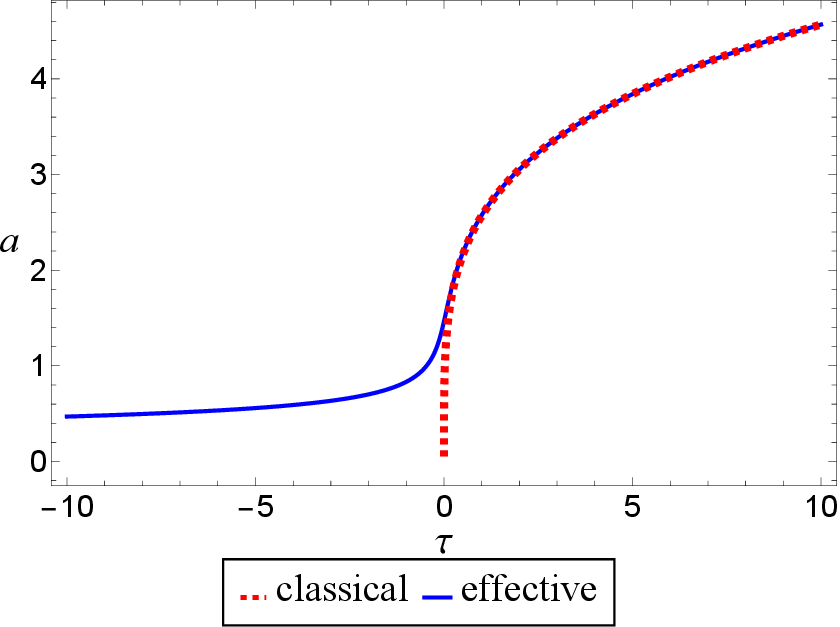}}\qquad
	\subfigure[]{
		\label{Figs4b}
		\includegraphics [width=0.46\textwidth]{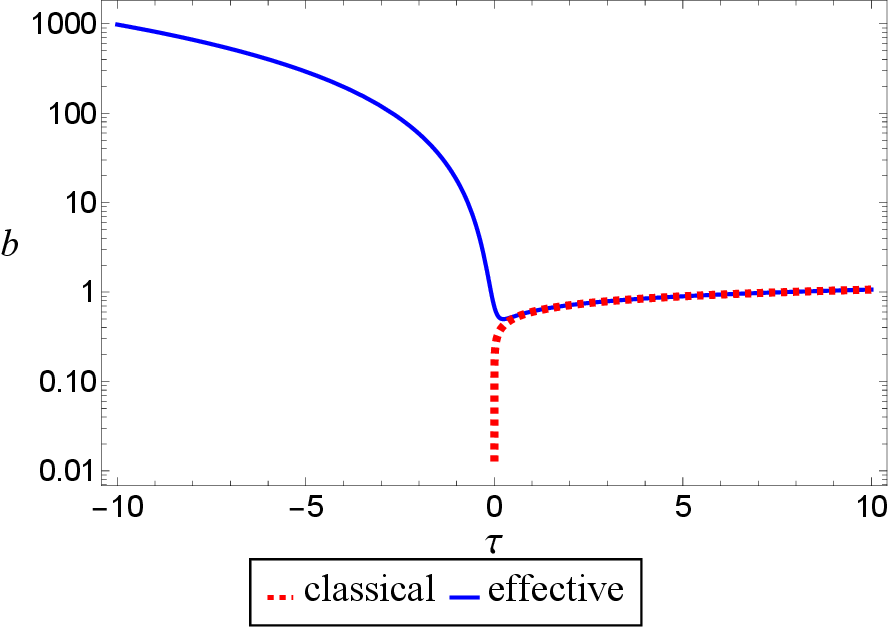}}
	\caption{(a)The evolution of the scale factor $a$ in both the classical and the effective models coupling with the scalar field in case IV. (b)The evolution of the scale factor $b$ in both the classical and the effective models coupling with the scalar field in case IV. Both figures are plotted using the parameters $G^{(5)}=c=\hbar=V_{0}=1$ and the initial values $\bar{\nu}_{0}=1/2$, $\bar{O}_{2}=1$ and $\bar{O}_{y}/\bar{O}_{3}=1$.}\label{Fig.s4}
\end{figure*}

For the initial condition with $\bar{O}_{3}=0$, one finds $\bar{O}_{y}=0$ and $\bar{\nu}_{0}=1/2$, but $\bar{O}_{y}/\bar{O}_{3}$ remains a finite Dirac observable. So Eqs.(\ref{pi1scl}) and (\ref{piyscl}) can still be used as solutions of $\bar{\pi}^1$ and $\pi^y$. Then, the corresponding scale factors are given by
\begin{align}
	a&=\left(\frac{\mu_{y}\bar{O}_{y}}{2\bar{O}_{3}V_{0}}\tilde{f}^{-\frac{3}{2}\bar{\nu}_{0}}(\tau)\right)^{1/3},\\
	b&=\sqrt{\Delta}\frac{\bar{O}_{2}[\tilde{f}(\tau)+\tilde{f}^{-1}(\tau)]}{\mu_{y}(\bar{O}_{y}/\bar{O}_{3})\tilde{f}^{-\frac{3}{2}\bar{\nu}_{0}}(\tau)}.
\end{align}
As $\tau\to\pm\infty$, the scale factors asymptotically approach
\begin{equation}
a={a'}^{(\phi)}_{\pm}(\pm\tau)^{\pm1/4},\qquad b=b'^{(\phi)}_{\pm}(\pm\tau)^{1\mp3/4},
\end{equation}
where
\begin{align}
	a'^{(\phi)}_{\pm}&\equiv\![\mu_{y}\bar{O}_{y}/(2V_{0}\bar{O}_{3})]^{1/3}(4\beta^{-1}\Delta^{-1/2})^{\pm1/4},\nonumber\\ b'^{(\phi)}_{\pm}&\equiv\!\Delta^{1/2}[\bar{O}_{2}\bar{O}_{3}/(\mu_{y}\bar{O}_{y})](4\beta^{-1}\Delta^{-1/2})^{1\mp3/4}.
\end{align}
Hence, as $\tau\to+\infty$, the scale factors in the effective model match those of the classical model with $\nu_{0}=1/2$, whereas this is no longer the case as $\tau\to-\infty$. As shown in Fig.\ref{Fig.s4}, the big bang singularities of the scale factors $a$ and $b$ are resolved by a non-bouncing mechanism and a quantum bounce respectively.


Next, let us consider the effective scalar constraint (\ref{effective C2}) in which the geometric part $\mathcal{C}^{e\!f\!f}_{gr}[N]$ contains the subleading-order quantum correction of a constant $\epsilon$. It turns out that, by choosing an appropriate lapse function $N$, the effective scalar constraint becomes formally identical to the effective scalar constraint (III.2) in Ref.\cite{Yuan:2025yjs}. Consequently, the dynamical results in this case are analogous to the those presented in Refs.\cite{Yuan:2025yjs,Zhang:2019acn}. The scale factors undergo countless bounces and re-collapses during their evolution. However, the different cyclic stages of the evolution do not have a same period, and the visible universe may expand boundlessly. Moreover, in the regime $\epsilon\ll1$, it is always possible to choose the initial values such that a super-inflationary phase following the quantum bounce can yield a sufficient e-folding number for the visible universe to replace the conventional inflationary paradigm.

\subsection{The Case of Coupling to the Dust}\label{sec:section43}
Since the present universe is matter-dominated apart from dark energy, it is also worthwhile to consider the KK cosmology coupled to a dust field. In this scenario, the classical scalar constraint reads
\begin{equation}\label{classical C3}
C_{T}[N]:=C_{gr}[N]+NP_{T},
\end{equation}
where $T$ and $P_{T}$ represent the comoving time of the dust field and its mass within the fiducial cell respectively, with the nontrivial Poisson bracket $\{T,P_{T}\}=1$. The EOM for the gravitational variables are the same as those in Eqs.(\ref{2.1}), (\ref{2.2}), (\ref{2.3}), and (\ref{2.4}), while the EOM for the dust variables read
\begin{align}
\partial_{\tau}T&=\{T,C_{T}[1]\}=1,\label{TEOM}\\
\partial_{\tau}P_{T}&=\{P_{T},C_{T}[1]\}=0.\label{PTEOM}
\end{align}
In the regime where the scale factor $a$ is expanding, so that $\bar{A}_{1}>0$, the solutions for $\bar{A}_{1}$, $\bar{A}_{y}$ and $\pi^{y}$ obtained from the EOM are still given by Eqs.(\ref{A1vaccl}), (\ref{Ayvaccl}) and (\ref{piyvaccl}) respectively, with $O_{1}\!\equiv\!\bar{A}_{y}(\bar{A}_{1})^{-3/2}$ and $O_{y}\!\equiv\!\pi^{y}\bar{A}_{y}$ remaining the Dirac observables. Moreover, Eq.(\ref{PTEOM}) shows that $P_{T}$ is a Dirac observable, which is assumed to be positive in order to match its interpretation as a mass. Dividing Eq.(\ref{TEOM}) by Eq.(\ref{2.1}), we obtain
\begin{equation}
\partial_{\tau}T=-\beta\Delta^{1/2}\frac{\partial_{\tau}\bar{A}_{1}}{(\bar{A}_{1})^2},
\end{equation}
and hence
\begin{equation}
\partial_{\tau}T=\beta\Delta^{1/2}\partial_{\tau}(\frac{1}{\bar{A}_{1}}).
\end{equation}
It then follows that $T_{0}\!\equiv\!T-\beta\Delta^{1/2}(\bar{A}_{1})^{-1}$ is a Dirac observable, which leads to
\begin{equation}
T=\beta\Delta^{1/2}(\bar{A}_{1})^{-1}+T_{0}=\tau+T_{0}.
\end{equation}
By solving the scalar constraint, we obtain
\begin{align}
\bar{\pi}^{1}=&\frac{1}{3}P_{T}\beta^{2}\kappa\Delta^{1/2}(\bar{A}_{1})^{-2}-\frac{1}{2}O_{y}(\bar{A}_{1})^{-1}\nonumber\\
=&\frac{1}{3}\kappa\Delta^{-1/2}P_{T}\tau^{2}-\frac{1}{2}O_{y}\beta^{-1}\Delta^{-1/2}\tau.
\end{align}
According to Eq.(\ref{scale factors}), this set of solutions yields the following scale factors
\begin{equation}
a=a^{(T)}_{0}\tau^{1/2},\qquad b=b^{(T)}_{0}\tau^{1/2}-b^{(T)}_{1}\tau^{-1/2},
\end{equation}
where
\begin{align}
	a^{(T)}_{0}&\equiv\!(\frac{O_{y}}{V_{0}O_{1}})^{\frac{1}{3}}(\beta^{-1}\Delta^{-1/2})^{\frac{1}{2}},\nonumber\\
	b^{(T)}_{0}&\equiv\!\frac{\kappa{P}_{T}}{3V_{0}(a^{(T)}_{0})^{3}},\nonumber\\
	b^{(T)}_{1}&\equiv\!\frac{O_{1}}{2}\beta^{\frac{1}{2}}\Delta^{\frac{3}{4}}.
\end{align}
Hence, the spacetime is singular at $\tau=0$, and that the scale factor $b$ may undergo unbounded expansion during its evolution. The latter is incompatible with the fact that there is no direct observational evidence for extra dimensions. It should be noted that, in the region $\bar{A}_{1}<0$, the corresponding scale factors given by the EOM take the form
\begin{equation}
a=a^{(T)}_{0}(-\tau)^{1/2},\qquad b=b^{(T)}_{0}(-\tau)^{1/2}-b^{(T)}_{1}(-\tau)^{-1/2},
\end{equation}
where the Dirac observable $O_{1}$ should be re-defined as $O_{1}\equiv\bar{A}_{y}|\bar{A}_{1}|^{-3/2}$.

We now turn to the effective dynamics of the model coupled to the dust. To incorporate the dust contribution into the effective scalar constraint, the polymer-like quantization of the configuration variable $T$ can be carried out in complete parallel with that of the scalar field. It turns out that in a suitable coherent state, the expectation value of $\hat{P}_{T}$ is given by $\langle\hat{P}_{T}\rangle=P_{T}|_{0}+\mathcal{O}(e^{-\pi^{2}/\sigma^2})$, where $\sigma$ is the corresponding Gaussian spread \cite{Li:2025bzl}. Hence, up to the subleading-order quantum corrections, the dust contribution to the effective scalar constraint takes the same expression as in the classical scalar constraint. A similar result can also be obtained by using the path-integral form of the extraction amplitude with the corresponding scalar constraint operator. Consequently, the effective scalar constraint of the model coupled to the dust reads
\begin{equation}\label{effective C3}
\mathcal{C}^{e\!f\!f}_{T}[N]:=\mathcal{C}^{e\!f\!f}_{gr}[N]+NP_{T}.
\end{equation}

We first consider the case that the geometric part $\mathcal{C}^{e\!f\!f}_{gr}[N]$ of Eq.(\ref{effective C3}) contain only the leading-order holonomy corrections given by Eq.(\ref{regularizedC}). Then, the EOM for the basic variables generated by $\mathcal{C}^{e\!f\!f}_{T}[1]$ are the same as those in Eqs.(\ref{41.1}), (\ref{41.2}), (\ref{41.3}), (\ref{41.4}), (\ref{TEOM}) and (\ref{PTEOM}). Therefore, in the region where $\sin(\bar{A}_{1})>0$, the solutions of $\bar{A}_{1}$, $\bar{A}_{y}$, and $\pi^{y}$ are still given by Eqs.(\ref{Vac1}), (\ref{Ayvac}) and (\ref{piyvac}), and $\bar{O}_{1}\!\equiv\!\tan(\mu_{y}\bar{A}_{y}/2)\tan^{-3/2}(\bar{A}_{1}/2)$, $\bar{O}_{y}\!\equiv\!\pi^{y}\sin(\mu_{y}\bar{A}_{y})/\mu_{y}$, and $P_T$ are the Dirac observables. Dividing Eq.(\ref{TEOM}) by Eq.(\ref{41.1}), we obtain
\begin{equation}
\partial_{\tau}T=-\beta\Delta^{1/2}\frac{\partial_{\tau}\bar{A}_{1}}{\sin^{2}(\bar{A}_{1})},
\end{equation}
and hence
\begin{equation}
\partial_{\tau}T=-\frac{\beta\Delta^{1/2}}{2}\partial_{\tau}\left(\tan(\frac{\bar{A}_{1}}{2})-\tan^{-1}\left(\frac{\bar{A}_{1}}{2}\right)\right).
\end{equation}
This implies that $\bar{T}_{0}\equiv\!T+(\beta\Delta^{1/2}/2)(\tan(\bar{A}_{1}/2)-\tan^{-1}(\bar{A}_{1}/2))$ is a Dirac observable, which leads to
\begin{equation}
T=-\frac{\beta\Delta^{1/2}}{2}\left(\tan(\frac{\bar{A}_{1}}{2})-\tan^{-1}\left(\frac{\bar{A}_{1}}{2}\right)\right)+\bar{T}_{0}=\tau+\bar{T}_{0}.
\end{equation}
It follows from the scalar constraint that
\begin{align}
\bar{\pi}^{1}=&\frac{P_{T}\beta^{2}\kappa\Delta^{1/2}}{3\sin^{2}(\bar{A}_{1})}-\frac{\bar{O}_{y}}{2\sin(\bar{A}_{1})}\nonumber\\
=&\frac{P_{T}\beta^{2}\kappa\Delta^{1/2}}{12}(f(\tau)+f^{-1}(\tau))^{2}-\frac{\bar{O}_{y}}{4}(f(\tau)+f^{-1}(\tau)).\label{pi1dust}
\end{align}

\begin{figure*}[!htb]
	\subfigure[]{
		\label{Figd4a}
		\includegraphics [width=0.46\textwidth]{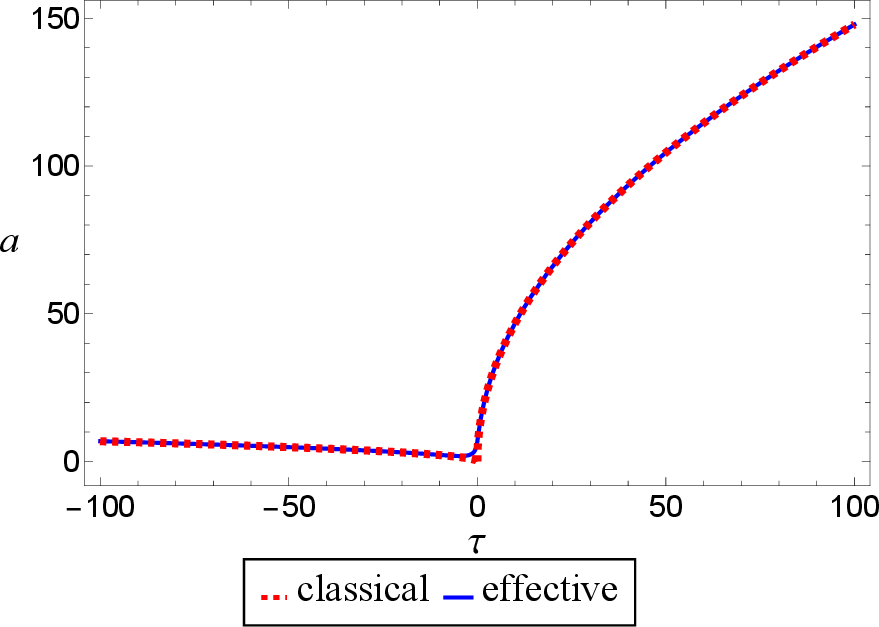}}\qquad
	\subfigure[]{
		\label{Figd4b}
		\includegraphics [width=0.46\textwidth]{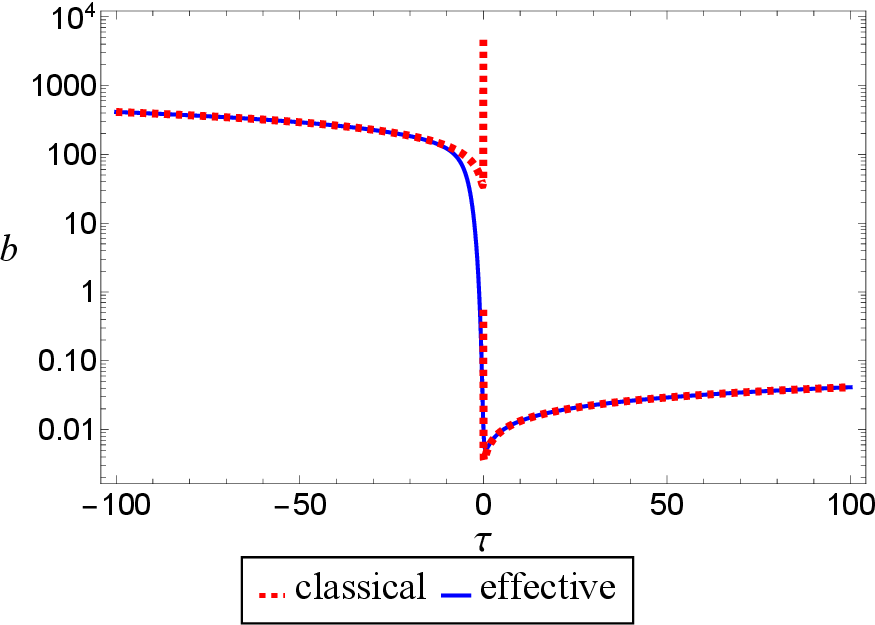}}
	\caption{(a)The evolution of the scale factor $a$ in both the classical and the effective models coupling with the dust field in case I. (b)The evolution of the scale factor $b$ in both the classical and the effective models coupling with the dust field in case I. Both figures are plotted using the parameters $G^{(5)}=c=\hbar=V_{0}=1$ and the initial values $P_{T}=5$, $\bar{O}_{y}=-1$ and $\bar{O}_{1}=-0.01$.}\label{Figd.4}
\end{figure*}

For the initial conditions with $\bar{O}_{1}\ne0$, the analysis similar to that in Sec.\ref{sec:section41} for the vacuum case directly results in the following results: (1) As long as the initial condition $2\kappa\beta^{2}\Delta^{1/2}P_{T}/3-\bar{O}_{y}>0$ is satisfied, the function $\bar{\pi}^{1}$ remains finite and non-vanishing. Combined with the finiteness and non-vanishing of $\pi^{y}$, this implies that the classical singularity is resolved in the effective model. Moreover, the scale factors in the effective model are given by
\begin{align}
a&=\left(\frac{\mu_{y}\bar{O}_{y}}{2V_0}[\bar{O}_{1}f^{\frac{3}{2}}(\tau)+(\bar{O}_{1})^{-1}f^{-\frac{3}{2}}(\tau)]\right)^{1/3},\label{aeffdust}\\
b&=\sqrt{\Delta}\frac{\frac{2P_{T}\beta^{2}\kappa\Delta^{1/2}}{3}(\frac{f(\tau)+f^{-1}(\tau)}{2})^{2}-\bar{O}_{y}(\frac{f(\tau)+f^{-1}(\tau)}{2})}{\mu_{y}\bar{O}_{y}[\bar{O}_{1}f^{\frac{3}{2}}(\tau)+(\bar{O}_{1})^{-1}f^{-\frac{3}{2}}(\tau)]}.\label{beffdust}
\end{align}
Consequently, the scale factor $a$ possesses a unique bounce point at $\tau_{B}\equiv\beta\Delta^{1/2}(|\bar{O}_{1}|^{2/3}-|\bar{O}_{1}|^{-2/3})/2$, and the scale factor $b$ has at least one bounce point. (2) In the regime $|\bar{O}_{1}|\ll1$, the scale factor $a$ undergoes a super-inflationary phase over the interval $\tau_{B}<\tau<0$, yielding an e-folding number of approximately $-\ln(2|\bar{O}_{1}|)/3$. This indicates that by suitably selecting the value of $|\bar{O}_{1}|$, one can attain the expected 55 e-folds of inflation in the visible dimensions. (3) As $\tau\to\pm\infty$, the scale factors asymptotically tend to
\begin{equation}\label{dustab}
a=a^{(T)}_{\pm}(\pm\tau)^{1/2},\qquad b=b^{(T)}_{\pm}(\pm\tau)^{1/2}-b^{(T)}_{1,\pm}(\pm\tau)^{-1/2},
\end{equation}
where 
\begin{align}
	a^{(T)}_{\pm}&\equiv\![\frac{\mu_{y}\bar{O}_{y}}{2V_{0}(\bar{O}_{1})^{\pm1}}]^{\frac{1}{3}}(2\beta^{-1}\!\Delta^{-\frac{1}{2}})^{\frac{1}{2}},\nonumber\\
	b^{(T)}_{\pm}&\equiv\!\frac{\kappa{P}_{T}}{3V_{0}(a^{(T)}_{\pm})^{3}}\nonumber\\
	b^{(T)}_{1,\pm}&\equiv\!\frac{(\bar{O}_{1})^{\pm1}}{2\mu_{y}}(2\beta^{-1}\!\Delta^{-\frac{3}{2}}\!)^{-\frac{1}{2}},
\end{align}
and hence the effective universe returns to the classical regime. It should be noted that Eq.(\ref{dustab}) indicates that the scale factor $b$ is asymptotically proportional to $G^{(5)}\sqrt{\tau}$ at late times. This implies that the size of the extra dimension can be sufficiently small in present universe as long as the five-dimensional gravitational constant $G^{(5)}$ is sufficiently small. Furthermore, by requiring the relative variation rate of the visible-dimensional scale factor, $\partial_{\tau}a/a$, to match the present Hubble parameter, we find that the relative variation rate of the extra-dimensional scale factor, $\partial_{\tau}b/b$, is also equal to the present Hubble parameter. This rate corresponds to the relative variation rate of the effective four-dimensional gravitational constant, $G^{(4)}\equiv G^{(5)}/(2\pi b)$ \cite{Qiang:2004gg}, namely $\sim 7\times10^{-11} \mathrm{yr}^{-1}$ \cite{Planck:2018vyg}. Therefore, in this sense, the extra dimension can remain sufficiently small without exceeding the observational constraint on the relative variation rate of the four-dimensional gravitational constant inferred from GW170817 \cite{Vijaykumar:2020nzc}. The evolutions of the scale factors in both the classical and the effective models are compared in Fig.\ref{Figd.4}. It is shown that the big bang singularity of the scale factor $a$ is resolved by a quantum bounce, and the past big rip singularity of the scale factor $b$ is also resolved.

\begin{figure*}[!htb]
	\subfigure[]{
		\label{Figd5a}
		\includegraphics [width=0.46\textwidth]{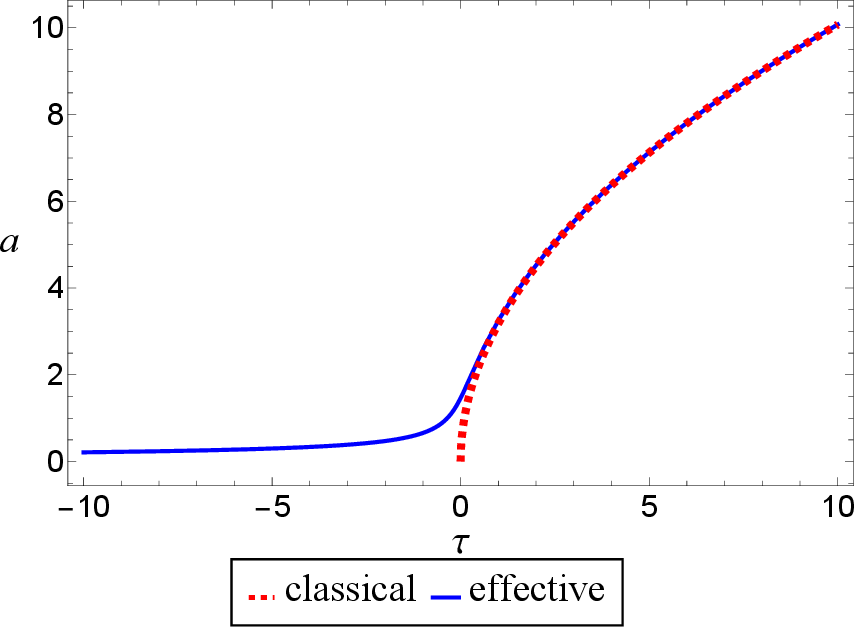}}\qquad
	\subfigure[]{
		\label{Figd5b}
		\includegraphics [width=0.46\textwidth]{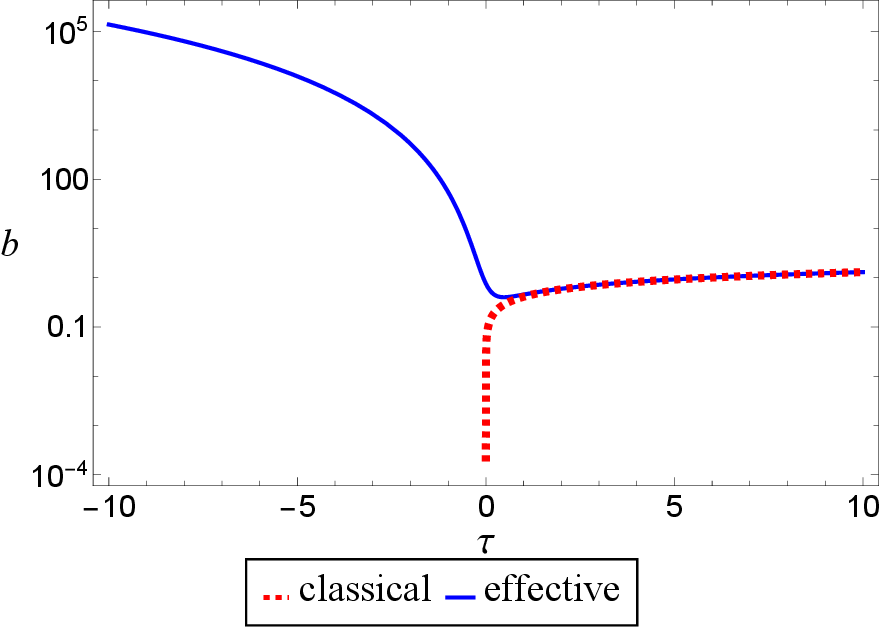}}
	\caption{(a)The evolution of the scale factor $a$ in both the classical and the effective models coupling with the dust field in case II. (b)The evolution of the scale factor $b$ in both the classical and the effective models coupling with the dust field in case II. Both figures are plotted using the parameters $G^{(5)}=c=\hbar=V_{0}=1$ and the initial values $P_{T}=5$, $\bar{O}_{1}=0$ and $\bar{O}_{y}/\bar{O}_{1}=1$.}\label{Figd.5}
\end{figure*}

For the initial conditions with $\bar{O}_{1}=0$, one finds $\bar{O}_{y}=0$, but a finite Dirac observable $\bar{O}_{y}/\bar{O}_{1}$ ensures that the solutions of $\bar{\pi}^1$ and $\pi^y$ are still given by Eqs.(\ref{pi1dust}) and (\ref{piyvac}). Then, the corresponding scale factors can be expressed as
\begin{align}
	a&=\left(\frac{\mu_{y}\bar{O}_{y}}{2\bar{O}_{1}V_0}f^{-\frac{3}{2}}(\tau)\right)^{1/3},\label{aeffdust1}\\
	b&=\sqrt{\Delta}\frac{\frac{2P_{T}\beta^{2}\kappa\Delta^{1/2}}{3}(\frac{f(\tau)+f^{-1}(\tau)}{2})^{2}}{\mu_{y}(\bar{O}_{y}/\bar{O}_{1})f^{-\frac{3}{2}}(\tau)}.\label{beffdust1}
\end{align}
As $\tau\to\pm\infty$, the scale factors asymptotically approach
\begin{equation}
a=a'^{(T)}_{\pm}(\pm\tau)^{\pm1/2},\qquad b=b'^{(T)}_{\pm}(\pm\tau)^{2\mp3/2},
\end{equation}
where
\begin{align}
	a'^{(T)}_{\pm}&\equiv\![\mu_{y}\bar{O}_{y}/(2V_{0}\bar{O}_{1})]^{1/3}(2\beta^{-1}\Delta^{-1/2})^{\pm1/2},\nonumber\\
	b'^{(T)}_{\pm}&\equiv\!(\kappa{P}_{T})/(3V_{0}(a^{(T)}_{\pm})^{3}).
\end{align}
Hence, the scale factors in the effective model converge to their classical counterparts as $\tau\to+\infty$, but deviate from them as $\tau\to-\infty$. It is shown in Fig.\ref{Figd.5} that the big bang singularities of the scale factors $a$ and $b$ are resolved via a non-bouncing process and a quantum bounce respectively.


The case that the geometric part $\mathcal{C}^{e\!f\!f}_{gr}[N]$ of the effective scalar constraint (\ref{effective C3}) includes the subleading-order quantum correction $\epsilon$ was considered in Ref.\cite{Li:2025bzl}. When $\epsilon$ is a nonzero constant, the effective model not only resolves the classical singularities through quantum bounces and re-collapses, but also leads to a cyclic universe, thereby preventing the unbounded expansion of the scale factors. Moreover, under the condition $\epsilon\ll1$, one can always choose the initial conditions such that the visible universe undergoes a super-inflationary stage with enough e-folds after each quantum bounce in every cycle, thus providing an alternative to the conventional inflation. When $\epsilon$ is chosen as a specific phase-space function, besides the resolution of the big bang and past big rip singularities by a quantum bounce and a quantum re-collapse respectively, the scale factor $a$ undergoes both the sufficient super-inflation and the late-time accelerated expansion. Meanwhile, the scale factor $b$ shrinks after the quantum re-collapse, exhibiting features of dynamical compactification.

\section{Summary and Discussion}\label{section6}
In summary, the 5D KK cosmology is introduced in Sec.\ref{section2} via the symmetric reduction of the connection formulation of the full theory. In the symmetry-reduced phase space, the vector constraint is automatically satisfied, while the simplicity and Gauss constraints are resolved via gauge fixing, leaving the scalar constraint to govern the dynamical evolution. Based on the symmetry-reduced connection formulation, the loop quantization of KK cosmology has been carried out concretely in Sec.\ref{section3}. The kinematical Hilbert space is constructed, and the classical scalar constraint is promoted to a well-defined essentially self-adjoint operator $\hat{C}^{sym}_{gr}[N]$ on it. Here, the $\bar{\mu}$ scheme is employed to express the holonomies of connection components along the visible dimensions in the quantization of the scalar constraint. By evaluating the expectation value of $\hat{C}^{sym}_{gr}[N]$ in coherent states $|\Psi_{\zeta}\rangle$, it is not only demonstrated that $\hat{C}^{sym}_{gr}[N]$ possesses the correct classical limit, but also the effective scalar constraint $\mathcal{C}^{e\!f\!f}_{gr}[N]$ for the geometric sector is derived. This effective constraint incorporates both the leading-order holonomy correction and the subleading-order quantum fluctuation term. Moreover, our further analysis on the path-integral formulation of the extraction amplitude of $\hat{C}^{sym}_{gr}[N]$ in both the momentum and coherent-state representations confirmes that $\mathcal{C}^{e\!f\!f}_{gr}[N]$ can indeed be regarded as the effective scalar constraint.

To investigate the modifications of classical dynamics induced by the quantum gravitational effects, the classical and effective dynamics of the KK cosmology have been studied in the vacuum, minimally coupled to a scalar field, and coupled to the dust cases respectively in Sec.\ref{section4}. In these three effective scenarios, by employing the effective scalar constraint with the leading-order holonomy correction the big bang and potential past big rip singularities of the classical theory are resolved. It should be noted that the classical singularities are not necessarily replaced by quantum bounces or re-collapses as shown e.g. in Fig.\ref{Figs4a} and Fig.\ref{Figd5a}. Moreover, it is worth emphasizing that, under appropriate initial conditions, the visible universe in all these effective scenarios undergo a super-inflationary phase after overcoming the big bang singularity, yielding a desirable e-folding number of about 55. This phase can serve as a compelling alternative to the conventional inflation of the universe. In contrast, in the standard 4D LQC models the e-folding number during the super-inflationary phase is typically not enough \cite{Ashtekar:2011ni,Ashtekar:2011rm}. These results suggest that the cosmic inflation may originate from quantum geometric effects associated with extra dimensions. It should be noted that the probability of sufficiently long-lasting inflation was calculated in certain LQC models by using the symplectic form on the 2-dimentioal physical phase space as the measure \cite{Ashtekar:2011ni,Ashtekar:2011rm}. However, this method cannot be applied to our scenarios of effective KK LQC model minimally coupling to a scalar field and coupling to the dust, because the physical phase spaces are now four-dimensional and hence the symplectic form cannot serve as a volume form for integrations. Thus, a more general method for constructing a suitable measure on the space of solutions is desirable in order to calculate the probability of sufficient super-inflation in our scenarios.

In the cases where the subleading-order quantum fluctuation is included into the effective scalar constraint as a non-zero constant, the visible universe is still capable of achieving sufficient inflation following the quantum bounce, while the 5D universe undergoes re-collapses at certain large scales, indicating that quantum effects could manifest themselves in large-scale cosmology. It is interesting to note that the effective dynamics of the model in the vacuum and minimally coupled to a scalar field scenarios are respectively analogous to those of the effective LQC models of the Kantowski-Sachs spacetime \cite{Corichi:2015xia,Ashtekar:2018cay} and the Janis-Newman-Winicour spacetime \cite{Yuan:2025yjs,Zhang:2019acn} in the $\mu_{0}$ scheme. Moreover, in the effective scenario of coupling with the dust, the evolutions of both the visible universe and the extra dimensional one become cyclic and thus avoid the unbounded expansion of the scale factors. It is worth noting that, in all three scenarios, the semiclassical conditions (\ref{A1 condition}) and (\ref{AY condition}) remain satisfied at late times, even if the extra-dimensional scale factor $b$ becomes very small or different initial values are chosen. This can be verified using the asymptotic expressions for the scale factors at late times given in Eqs. (IV.26), (IV.65), and (IV.88), which show that the physical volumes of both the three- and four-dimensional fiducial cells are sufficiently large and the relative rates of change of the scale factors are sufficiently small.

Our framework opens up a fascinating new direction for addressing certain most fundamental issues in cosmology within quantum cosmological models with extra dimensions. While the cosmic inflation and dark energy may both originate from the interplay between the compact extra dimension and the quantum geometric effects only in a particular scenario as shown in Ref.\cite{Li:2025bzl}, the results of this paper indicate that the sufficient inflation can also always be achieved in various scenarios by the quantum effect of the 5D KK LQC. Therefore, the cosmic inflation can be regarded as a natural result of the LQC model with a compact extra dimension. Moreover, the present paper provides the more general groundwork for further investigations to the quantum dynamics of the model. In particular, a detailed analysis of the evolution of quantum fluctuations within the quantum dynamical framework may provide deeper insights into the issue of choosing appropriate fluctuations in effective models. Studies of these topics are left for future work.

\begin{acknowledgments}
This work is supported by National Natural Science Foundation of China (NSFC) with Grants No.12275022, No.12675060 and No.12275087.
		
\end{acknowledgments}

\onecolumngrid
\appendix		
		
\section{Self-adjointness}\label{sec:appendixB}
We set the domain of the operator $\hat{C}_{gr}[N]$ to be $\mathcal{D}_{gr}$. To check whether the adjoint operator of $\hat{C}_{gr}[N]$ is densely defined, we consider the following operator with the domain $\mathcal{D}_{gr}$,
\begin{equation}
\hat{H}\left|\lambda,\xi\right>:=H(\lambda,\xi)\left|\lambda+\mathfrak{k},\xi+\mathfrak{l}\mu_{y}\right>,
\end{equation}
where $\mathfrak{k},\mathfrak{l}\in\mathbb{Z}$ and $H:\mathbb{R}^{2}\to\mathbb{C}$ is a function. By the definition of the adjoint operator, we obtain
\begin{equation}
\hat{H}^{\dagger}\left|\lambda,\xi\right>=\overline{H(\lambda-\mathfrak{k},\xi-\mathfrak{l}\mu_{y})}\left|\lambda-\mathfrak{k},\xi-\mathfrak{l}\mu_{y}\right>,
\end{equation}
where the overline denotes the complex conjugate. Thus, the domain of $\hat{H}^{\dagger}$ contains $\mathcal{D}_{gr}$, and we can define the symmetric operator $\hat{H}^{sym}:=(\hat{H}+\hat{H}^{\dagger})/2$ with the domain $\mathcal{D}_{gr}$. Since the component operators $\hat{u}_{r}$, $\hat{u}_{k,l}$, and $\hat{H}$ have the same form, and $\hat{C}_{gr}[N]$ is a finite linear combination of $\hat{u}_{r}$ and $\hat{u}_{k,l}$, we can define the symmetric operators $\hat{b}^{sym}:=(\hat{b}+\hat{b}^{\dagger})/2$ with the domain $\mathcal{D}_{gr}$, where $\hat{b}\in\{\hat{C}_{gr}[N], \hat{u}_{r}, \hat{u}_{k,l}\}$. These operators satisfy
\begin{equation} \hat{C}^{sym}_{gr}[N]=4^{-2}N\hbar\beta^{-1}\Delta^{-1/2}(\sum_{r\in\{0,4,-4\}}\hat{u}^{sym}_{r}+\sum_{k,l\in\{2,-2\}}\hat{u}^{sym}_{k,l}).
\end{equation}

Note that the kinematical Hilbert space $\mathcal{H}_{gr}$ can be decomposed into a direct sum of invariant subspaces of $\hat{C}^{sym}_{gr}[N]$,
\begin{equation} 
\mathcal{H}_{gr}=\oplus_{\gamma\in[0,2)}\mathcal{H}_{\gamma},
\end{equation}
where $\mathcal{H}_{\gamma}$ is the completion of $\mathcal{D}_{\gamma}:=span_{\mathbb{C}}\{\left|\lambda,\xi\right>\mid\lambda\in\{\gamma+2n\mid n\in\mathbb{Z}\},\xi\in\mathbb{R}\}$ with respect to the inner product Eq.(\ref{inner product}), and $\hat{C}^{sym}_{gr}[N][\mathcal{D}_{\gamma}]\subset\mathcal{D}_{\gamma}$. Thus, we can introduce the operator $\hat{C}_{\gamma}:=\hat{C}^{sym}_{gr}[N]|_{\mathcal{D}_{\gamma}}$. We first show that the operator $\hat{C}_{\gamma}$ with domain $\mathcal{D}_{\gamma}$ is essentially self-adjoint in $\mathcal{H}_{\gamma}$, and then prove that the operator $\hat{C}^{sym}_{gr}[N]$ with domain $\mathcal{D}_{gr}$ is essentially self-adjoint in $\mathcal{H}_{gr}$.

The proof of the essential self-adjointness of $\hat{C}_{\gamma}$ crucially relies on Theorem X.37 of Ref.\cite{RSR1}:
\begin{theorem}\label{theoremB1}
Let $\hat{P}$ be a self-adjoint operator with $\hat{P}\ge 1$. Let $\hat{A}$ be a symmetric operator with domain $D$ which is a core for $\hat{P}$. Suppose that: 
\begin{enumerate}[(i)]
\item For some $c$ and all $\psi\in D$, one has
\begin{equation}
\|\hat{A}\psi\|\le c \|\hat{P}\psi\|.
\end{equation}
\item For some $d$ and all $\psi\in D$, one has
\begin{equation}\label{B5}
|\langle\hat{A}\psi,\hat{P}\psi\rangle-\langle\hat{P}\psi,\hat{A}\psi\rangle|\le d \|\hat{P}^{1/2}\psi\|^{2}.
\end{equation}
\end{enumerate}
Then $\hat{A}$ is essential self-adjoint on $D$ and its closure is essentially self-adjoint on any core for $\hat{P}$.
\end{theorem}
		
In our case, $\hat{A}=\hat{C}_{\gamma}$ and $D=\mathcal{D}_{\gamma}$. To complete the proof, we introduce the following operator
\begin{equation}
\hat{P}\left|\lambda,\xi\right>:=P(\lambda,\xi)\left|\lambda,\xi\right>:=(1+|\lambda|+|\xi|)\left|\lambda,\xi\right>,
\end{equation}
with domain $\mathcal{D}_{\gamma}$. The essential self-adjointness of $\hat{P}$ follows from the following theorem\cite{RSR2}:
\begin{theorem}\label{theoremB2}
Let $\hat{T}$ be a symmetric operator on a Hilbert space. Then $\hat{T}$ is essentially self-adjoint if and only if $Ran(\hat{T}\pm i)$ are dense.
\end{theorem}
Taking $\hat{T}=\hat{P}$, it follows directly from Theorem \ref{theoremB2} that $\hat{P}$ is essentially self-adjoint, and its closure is a self-adjoint operator with the core $\mathcal{D}_{\gamma}$. Let $\left|\psi\right>\in\mathcal{D}_{\gamma}$ and define $\psi(\lambda,\xi):=\left<\lambda,\xi|\psi\right>$, then
\begin{align}
\|\hat{C}_{\gamma}\psi\|^{2}&=\sum_{\lambda',\xi'}| \sum_{\lambda,\xi}\langle\lambda',\xi'|\hat{C}_{\gamma}|\lambda,\xi\rangle\psi(\lambda,\xi) |^{2}\le 7\sum_{\lambda,\xi}\sum_{\lambda',\xi'}|\langle\lambda',\xi'|\hat{C}_{\gamma}|\lambda,\xi\rangle|^{2}|\psi(\lambda,\xi)|^{2},\label{B6}\\
\|\hat{P}\psi\|^{2}&=\sum_{\lambda,\xi}P^{2}(\lambda,\xi)|\psi(\lambda,\xi)|^{2},
\end{align}
where the factor of 7 arises from the fact that each row of the matrix of $\hat{C}_{\gamma}$ has only 7 non-vanishing entries. By direct calculation, we obtain
\begin{align}
\sum_{\lambda',\xi'}|\langle\lambda',\xi'|\hat{C}_{\gamma}|\lambda,\xi\rangle|^{2}&=
(4^{-2}N\hbar\beta^{-1}\Delta^{-1/2})^{2}[(u_{0}(\lambda))^{2}\nonumber\\
&+\frac{1}{4}(u_{4}(\lambda)+u_{-4}(\lambda+4))^{2}+\frac{1}{4}(u_{-4}(\lambda)+u_{-4}(\lambda-4))^{2}\nonumber\\
&+\frac{1}{4}(u_{2,2}(\lambda,\xi)+u_{-2,-2}(\lambda+2,\xi+2\mu_{y}))^2\nonumber\\
&+\frac{1}{4}(u_{-2,2}(\lambda,\xi)+u_{2,-2}(\lambda-2,\xi+2\mu_{y}))^2\nonumber\\
&+\frac{1}{4}(u_{-2,2}(\lambda+2,\xi-2\mu_{y})+u_{2,-2}(\lambda,\xi))^2\nonumber\\
&+\frac{1}{4}(u_{2,2}(\lambda-2,\xi-2\mu_{y})+u_{-2,-2}(\lambda,\xi))^2].
\end{align}
It is easy to verify that the functions $|u_r(\lambda+r')/P(\lambda,\xi)|,r,r'\in\{0,-4,4\}$ and $|u_{k,l}(\lambda+k',\xi+l'\mu_{y})/P(\lambda,\xi)|,k,l\in\{-2,2\},k',l'\in\{0,-2,2\}$ are bounded for all $\lambda$ and $\xi$. Consequently, there exists a constant $c$ such that 
\begin{equation}
\|\hat{C}_{\gamma}\psi\|\le c\|\hat{P}\psi\|.
\end{equation}

Since $|\langle\hat{C}_{\gamma}\psi,\hat{P}\psi\rangle-\langle\hat{P}\psi,\hat{C}_{\gamma}\psi\rangle|=|\langle\hat{P}^{1/2}\psi,\hat{a}\hat{P}^{1/2}\psi\rangle|$, where
\begin{equation}
\hat{a}:=\hat{P}^{-1/2}[\hat{C}_{\gamma},\hat{P}]\hat{P}^{-1/2},
\end{equation}
then Eq.(\ref{B5}) holds if we can prove that $\hat{a}$ is a bounded operator. Similarly to Eq.(\ref{B6}), we have
\begin{equation}
\|\hat{a}\psi\|^{2}=\sum_{\lambda',\xi'}| \sum_{\lambda,\xi}\langle\lambda',\xi'|\hat{a}|\lambda,\xi\rangle\psi(\lambda,\xi) |^{2}\le 6\sum_{\lambda,\xi}\sum_{\lambda',\xi'}|\langle\lambda',\xi'|\hat{a}|\lambda,\xi\rangle|^{2}|\psi(\lambda,\xi)|^{2},
\end{equation}
where the factor 6 comes from the fact that each row of the matrix of $\hat{a}$ has only 6 nonzero entries. By direct calculation, we obtain
\begin{align}
\sum_{\lambda',\xi'}\left|\left<\lambda',\xi'\left|\hat{a}\right|\lambda,\xi\right>\right|^{2}&=
(4^{-2}N\hbar\beta^{-1}\Delta^{-1/2})^{2}\nonumber\\
&\times[\frac{1}{4}(a_{4}(\lambda,\xi)+a'_{-4}(\lambda,\xi))^{2}+\frac{1}{4}(a_{-4}(\lambda,\xi)+a'_{4}(\lambda,\xi))^{2}\nonumber\\
&+\frac{1}{4}(a_{2,2}(\lambda,\xi)+a'_{-2,-2}(\lambda,\xi))^{2}+\frac{1}{4}(a_{-2,2}(\lambda,\xi)+a'_{2,-2}(\lambda,\xi))^{2}\nonumber\\
&+\frac{1}{4}(a_{2,-2}(\lambda,\xi)+a'_{-2,2}(\lambda,\xi))^{2}+\frac{1}{4}(a_{-2,-2}(\lambda,\xi)+a'_{2,2}(\lambda,\xi))^{2}],
\end{align}
where the corresponding coefficient functions are defined as follows
\begin{align}
a_{\pm 4}(\lambda,\xi):&=\frac{u_{\pm 4}(\lambda)}{P^{1/2}(\lambda\pm 4,\xi)P^{1/2}(\lambda,\xi)}[P(\lambda,\xi)-P(\lambda\pm 4,\xi)],\label{B13}\\		
a'_{\pm 4}(\lambda,\xi):&=\frac{u_{\pm 4}(\lambda\mp 4)}{P^{1/2}(\lambda\mp 4,\xi)P^{1/2}(\lambda,\xi)}[P(\lambda,\xi)-P(\lambda\mp 4,\xi)],\label{B14}\\		
a_{\pm 2,2}(\lambda,\xi):&=\frac{u_{\pm 2,2}(\lambda,\xi)}{P^{1/2}(\lambda\pm 2,\xi+2\mu_{y})P^{1/2}(\lambda,\xi)}[P(\lambda,\xi)-P(\lambda\pm 2,\xi+2\mu_{y})],\label{B15}\\		
a'_{\pm 2,2}(\lambda,\xi):&=\frac{u_{\pm 2,2}(\lambda\mp2,\xi-2\mu_{y})}{P^{1/2}(\lambda\mp 2,\xi-2\mu_{y})P^{1/2}(\lambda,\xi)}[P(\lambda,\xi)-P(\lambda\mp 2,\xi-2\mu_{y})],\label{B16}\\	
a_{\pm 2,-2}(\lambda,\xi):&=\frac{u_{\pm 2,-2}(\lambda,\xi)}{P^{1/2}(\lambda\pm 2,\xi-2\mu_{y})P^{1/2}(\lambda,\xi)}[P(\lambda,\xi)-P(\lambda\pm 2,\xi-2\mu_{y})],\label{B17}\\
a'_{\pm 2,-2}(\lambda,\xi):&=\frac{u_{\pm 2,-2}(\lambda\mp2,\xi+2\mu_{y})}{P^{1/2}(\lambda\mp 2,\xi+2\mu_{y})P^{1/2}(\lambda,\xi)}[P(\lambda,\xi)-P(\lambda\mp 2,\xi+2\mu_{y})].\label{B18}
\end{align}
According to the following inequality
\begin{equation}
\begin{split}
\frac{1}{P^{1/2}(\lambda+\mathfrak{k},\xi+\mathfrak{l}\mu_{y})P^{1/2}(\lambda,\xi)}&\le\max\{\frac{1}{P(\lambda+\mathfrak{k},\xi+\mathfrak{l}\mu_{y})},\frac{1}{P(\lambda,\xi)}\},\\
|P(\lambda,\xi)-P(\lambda+\mathfrak{k},\xi+\mathfrak{l}\mu_{y})|&\le|\mathfrak{k}|+|\mathfrak{l}|\mu_{y},
\end{split}
\end{equation}
where $\mathfrak{k},\mathfrak{l}\in\mathbb{Z}$, it is easy to verify that the coefficient functions in Eqs.(\ref{B13})--(\ref{B18}) are all bounded. Therefore, $\hat{a}$ is a bounded operator, which ensures the existence of a constant $d$ such that
\begin{equation}
|\langle\hat{C}_{\gamma}\psi,\hat{P}\psi\rangle-\langle\hat{P}\psi,\hat{C}_{\gamma}\psi\rangle|\le d \|\hat{P}^{1/2}\psi\|^{2}.
\end{equation}
Using \autoref{theoremB1}, we have shown that the operator $\hat{C}_{\gamma}$ with domain $\mathcal{D}_{\gamma}$ is essentially self-adjoint in $\mathcal{H}_{\gamma}$.
		
Let $\psi_{0}\in Ran(\hat{C}^{sym}_{gr}[N]\pm i)^{\perp}$, then we have
\begin{equation}
\langle(\hat{C}^{sym}_{gr}[N]\pm i)\psi,\psi_{0}\rangle=0,\qquad\forall\psi\in\mathcal{D}_{gr}.
\end{equation}
When $\psi$ ranges over $\mathcal{D}_{\gamma}$, and since $\hat{C}_{\gamma}$ is essentially self-adjoint, $Ran(\hat{C}_{\gamma}\pm i)$ is dense in $\mathcal{H}_{\gamma}$. Thus, the projection of $\psi_{0}$ onto $\mathcal{H}_{\gamma}$ is zero. When $\gamma$ ranges over $[0,2)$, it follows that $\psi_{0}=0$. Consequently, $Ran(\hat{C}^{sym}_{gr}[N]\pm i)$ is dense in $\mathcal{H}_{gr}$. By \autoref{theoremB2}, we conclude that $\hat{C}^{sym}_{gr}[N]$ is essentially self-adjoint in $\mathcal{H}_{gr}$.

\section{Calculational Details in the Path-Integral Formulation}\label{sec:appendixBB}
\subsection{Derivation of Eq. (III.35)}
By applying multiple group averaging \cite{Huang:2011es,Qin:2012gaa,Zhang:2012em,Song:2020pqm} to the right side of Eq.(\ref{extraction amplitude1}) and inserting the completeness relation $\sum_{\lambda,\xi}\left|\lambda,\xi\right>\left<\lambda,\xi\right|=\mathbb{I}$, the extraction amplitude can be written in the path-integral form as
\begin{align}
	&A(\lambda_{f},\xi_{f};\lambda_{b},\xi_{b})\nonumber\\
	=&\lim_{\tilde{\alpha}_{\mathcal{N}},\cdots,\tilde{\alpha}_{1}\to\infty}\frac{2\tilde{\alpha}_{1}}{\mathcal{N}}(\prod_{n=1}^{\mathcal{N}}\frac{1}{2\tilde{\alpha}_{n}}\int_{-\tilde{\alpha}_{n}}^{\tilde{\alpha}_{n}}d\alpha_{n})\nonumber\\
	\times&\sum_{\lambda_{\mathcal{N}-1},\cdots,\lambda_{1}}\sum_{\xi_{\mathcal{N}-1},\cdots,\xi_{1}}\prod_{n=1}^{\mathcal{N}}\langle\lambda_{n},\xi_{n}|e^{i\frac{\alpha_{n}}{\mathcal{N}}\hat{C}^{sym}_{gr}[N]}|\lambda_{n-1},\xi_{n-1}\rangle,\label{MR0}
\end{align}
where $\lambda_{\mathcal{N}}=\lambda_{f}$, $\lambda_{0}=\lambda_{b}$, $\xi_{\mathcal{N}}=\xi_{f}$ and $\xi_{0}=\xi_{b}$. Note that, as $\mathcal{N}\to\infty$, the matrix element $\langle\lambda_{n},\xi_{n}|e^{i\frac{\alpha_{n}}{\mathcal{N}}\hat{C}^{sym}_{gr}[N]}|\lambda_{n-1},\xi_{n-1}\rangle$ in Eq.(\ref{MR0}) can be expressed as
\begin{align}
	&\langle\lambda_{n},\xi_{n}|e^{i\frac{\alpha_{n}}{\mathcal{N}}\hat{C}^{sym}_{gr}[N]}|\lambda_{n-1},\xi_{n-1}\rangle\nonumber\\
	=&\delta_{\lambda_{n},\lambda_{n-1}}\delta_{\xi_{n},\xi_{n-1}}+i\frac{\alpha_{n}}{\mathcal{N}}\langle\lambda_{n},\xi_{n}|\hat{C}^{sym}_{gr}[N]|\lambda_{n-1},\xi_{n-1}\rangle\nonumber\\
	+&\mathcal{O}(\mathcal{N}^{-2}).\label{MR1}
\end{align}
The matrix element in Eq.(\ref{MR1}) can be calculated as
\begin{align}
	&\langle\lambda_{n},\xi_{n}|\hat{C}^{sym}_{gr}[N]|\lambda_{n-1},\xi_{n-1}\rangle\nonumber\\
	=&4^{-2}N\hbar\beta^{-1}\Delta^{-1/2}\nonumber\\
	\times&[\sum_{r\in\{0,4,-4\}}\frac{1}{2}\delta_{\xi_{n},\xi_{n-1}}(\delta_{\lambda_{n},\lambda_{n-1}+r}\mathfrak{u}_{r}(\lambda_{n-1},\lambda_{n})\nonumber\\
	+&\delta_{\lambda_{n}+r,\lambda_{n-1}}\mathfrak{u}_{r}(\lambda_{n},\lambda_{n-1}))\nonumber\\
	+&\sum_{k,l\in\{2,-2\}}\frac{1}{2}(\delta_{\xi_{n},\xi_{n-1}+l\mu_{y}}\delta_{\lambda_{n},\lambda_{n-1}+k}\mathfrak{u}_{k,l}(\lambda_{n},\lambda_{n-1},\xi_{n-1})\nonumber\\
	+&\delta_{\xi_{n}+l\mu_{y},\xi_{n-1}}\delta_{\lambda_{n}+k,\lambda_{n-1}}\mathfrak{u}_{k,l}(\lambda_{n},\lambda_{n-1},\xi_{n}))],\label{MR2}
\end{align}
where the coefficient functions are defined by
\begin{align}
	\mathfrak{u}_{0}(\lambda,\lambda')&:=-4\eta(\lambda')\lambda^{2}\eta(\lambda),\\
	\mathfrak{u}_{\pm 4}(\lambda,\lambda')&:=2\eta(\lambda')\lambda^{2}\eta(\lambda),\\
	\mathfrak{u}_{\pm 2,2}(\lambda,\lambda',\xi)&:=\pm \frac{3}{\mu_{y}}\eta(\lambda')\lambda\xi\eta(\lambda),\\
	\mathfrak{u}_{\pm 2,-2}(\lambda,\lambda',\xi)&:=-\mathfrak{u}_{\pm 2,2}(\lambda,\lambda',\xi).
\end{align}
By combining Eqs.(\ref{MR1}), (\ref{MR2}) and the following identities
\begin{equation}
	\begin{split}
		\delta_{\lambda_{n},\lambda_{n-1}}&=\lim_{\tilde{\beta}_{n}\to\infty}\frac{1}{2\tilde{\beta}_{n}}\int_{-\tilde{\beta}_{n}}^{\tilde{\beta}_{n}}d\bar{A}_{1,n}e^{-\frac{i}{2}(\lambda_{n}-\lambda_{n-1})\bar{A}_{1,n}},\\
		\delta_{\xi_{n},\xi_{n-1}}&=\lim_{\tilde{\gamma}_{n}\to\infty}\frac{1}{2\tilde{\gamma}_{n}}\int_{-\tilde{\gamma}_{n}}^{\tilde{\gamma}_{n}}d\bar{A}_{y,n}e^{-\frac{i}{2}(\xi_{n}-\xi_{n-1})\bar{A}_{y,n}},
	\end{split}
\end{equation}
we obtain
\begin{align}
	&\langle\lambda_{n},\xi_{n}|e^{i\frac{\alpha_{n}}{\mathcal{N}}\hat{C}^{sym}_{gr}[N]}|\lambda_{n-1},\xi_{n-1}\rangle\nonumber\\
	=&\lim_{\tilde{\beta}_{n}\to\infty}\lim_{\tilde{\gamma}_{n}\to\infty}\frac{1}{2\tilde{\beta}_{n}}\frac{1}{2\tilde{\gamma}_{n}}\int_{-\tilde{\beta}_{n}}^{\tilde{\beta}_{n}}d\bar{A}_{1,n}\int_{-\tilde{\gamma}_{n}}^{\tilde{\gamma}_{n}}d\bar{A}_{y,n}\nonumber\\
	\times&\exp\{-\frac{i}{2}(\lambda_{n}-\lambda_{n-1})\bar{A}_{1,n}-\frac{i}{2}(\xi_{n}-\xi_{n-1})\bar{A}_{y,n}\nonumber\\
	+&i\frac{\alpha_{n}}{\mathcal{N}}4^{-2}N\hbar\beta^{-1}\Delta^{-1/2}\nonumber\\
	\times&[\sum_{r\in\{0,4,-4\}}\frac{1}{2}(e^{i\frac{r}{2}\bar{A}_{1,n}}\mathfrak{u}_{r}(\lambda_{n-1},\lambda_{n})+e^{-i\frac{r}{2}\bar{A}_{1,n}}\mathfrak{u}_{r}(\lambda_{n},\lambda_{n-1}))\nonumber\\
	+&\sum_{k,l\in\{2,-2\}}\frac{1}{2}(e^{i\frac{k}{2}\bar{A}_{1,n}+i\frac{l}{2}\mu_{y}\bar{A}_{y,n}}\mathfrak{u}_{k,l}(\lambda_{n-1},\lambda_{n},\xi_{n-1})\nonumber\\
	+&e^{-i\frac{k}{2}\bar{A}_{1,n}-i\frac{l}{2}\mu_{y}\bar{A}_{y,n}}\mathfrak{u}_{k,l}(\lambda_{n},\lambda_{n-1},\xi_{n}))]\}+\mathcal{O}(\mathcal{N}^{-2}).\label{MR3}
\end{align}
By substituting Eq.(\ref{MR3}) into Eq.(\ref{MR0}) and taking the limit as $\mathcal{N}\to\infty$, we conclude
\begin{align}
	&A(\lambda_{f},\xi_{f};\lambda_{b},\xi_{b})\nonumber\\
	=&\lim_{\mathcal{N}\to\infty}\lim_{\tilde{\alpha}_{\mathcal{N}},\cdots,\tilde{\alpha}_{1}\to\infty}\frac{2\tilde{\alpha}_{1}}{\mathcal{N}}(\prod_{n=1}^{\mathcal{N}}\frac{1}{2\tilde{\alpha}_{n}}\int_{-\tilde{\alpha}_{n}}^{\tilde{\alpha}_{n}}d\alpha_{n})\nonumber\\
	\times&\lim_{\tilde{\beta}_{\mathcal{N}},\cdots,\tilde{\beta}_{1}\to\infty}(\prod_{n=1}^{\mathcal{N}}\frac{1}{2\tilde{\beta}_{n}}\int_{-\tilde{\beta}_{n}}^{\tilde{\beta}_{n}}d\bar{A}_{1,n})\nonumber\\
	\times&\lim_{\tilde{\gamma}_{\mathcal{N}},\cdots,\tilde{\gamma}_{1}\to\infty}(\prod_{n=1}^{\mathcal{N}}\frac{1}{2\tilde{\gamma}_{n}}\int_{-\tilde{\gamma}_{n}}^{\tilde{\gamma}_{n}}d\bar{A}_{y,n})\nonumber\\
	\times&\sum_{\lambda_{\mathcal{N}-1},\cdots,\lambda_{1}}\sum_{\xi_{\mathcal{N}-1},\cdots,\xi_{1}}\exp\{\frac{i}{\hbar}\mathcal{N}^{-1}\sum_{n=1}^{\mathcal{N}}[-\frac{\hbar}{2}\frac{(\lambda_{n}-\lambda_{n-1})}{\mathcal{N}^{-1}}\bar{A}_{1,n}\nonumber\\
	-&\frac{\hbar}{2}\frac{(\xi_{n}-\xi_{n-1})}{\mathcal{N}^{-1}}\bar{A}_{y,n}+\alpha_{n}\hbar4^{-2}N\hbar\beta^{-1}\Delta^{-1/2}\nonumber\\
	\times&[\sum_{r\in\{0,4,-4\}}\frac{1}{2}(e^{i\frac{r}{2}\bar{A}_{1,n}}\mathfrak{u}_{r}(\lambda_{n-1},\lambda_{n})+e^{-i\frac{r}{2}\bar{A}_{1,n}}\mathfrak{u}_{r}(\lambda_{n},\lambda_{n-1}))\nonumber\\
	+&\sum_{k,l\in\{2,-2\}}\frac{1}{2}(e^{i\frac{k}{2}\bar{A}_{1,n}+i\frac{l}{2}\mu_{y}\bar{A}_{y,n}}\mathfrak{u}_{k,l}(\lambda_{n-1},\lambda_{n},\xi_{n-1})\nonumber\\
	+&e^{-i\frac{k}{2}\bar{A}_{1,n}-i\frac{l}{2}\mu_{y}\bar{A}_{y,n}}\mathfrak{u}_{k,l}(\lambda_{n},\lambda_{n-1},\xi_{n}))]]\}\nonumber\\
	=&\mathfrak{c}\mathcal{D}\alpha\mathcal{D}\bar{\pi}^{1}\mathcal{D}\pi^{y}\mathcal{D}\bar{A}_{1}\mathcal{D}\bar{A}_{y}\exp\{\frac{i}{\hbar}\int_{0}^{1}dt[-\frac{3}{\beta\kappa}\bar{A}_{1}\partial_{t}\bar{\pi}^{1}\nonumber\\
	-&\frac{1}{\beta\kappa}\bar{A}_{y}\partial_{t}\pi^{y}-\alpha\hbar\frac{3}{2}N\beta^{-2}\Delta^{-1/2}\kappa^{-1}|\bar{\pi}^{1}|^{-1}[2\sin^{2}(\bar{A}_1)(\bar{\pi}^{1})^{2}\nonumber\\
	+&\sin(\bar{A}_1)\frac{\sin(\mu_{y}\bar{A}_{y})}{\mu_{y}}\bar{\pi}^{1}\pi^{y}]]\},
\end{align}
where $\mathfrak{c}$ denotes an overall constant, $t$ denotes the fictitious time variable, and the definitions $\bar{\pi}^{1}_{n}:=\hbar\beta\kappa\lambda_{n}/6$ and $\pi^{y}_{n}:=\hbar\beta\kappa\xi_{n}/2$ have been used.

\subsection{Derivation of Eq. (III.37)}
By applying multiple group averaging to the right side of Eq.(\ref{extraction amplitude2E}), we obtain
\begin{align}
	A(\zeta_{f};\zeta_{b})	=&\lim_{\tilde{\alpha}_{\mathcal{N}},\cdots,\tilde{\alpha}_{1}\to\infty}\frac{2\tilde{\alpha}_{1}}{\mathcal{N}}(\prod_{n=1}^{\mathcal{N}}\frac{1}{2\tilde{\alpha}_{n}}\int_{-\tilde{\alpha}_{n}}^{\tilde{\alpha}_{n}}d\alpha_{n})\nonumber\\
	\times&\frac{\langle\Psi_{\zeta_{f}}|e^{i\sum_{n=1}^{\mathcal{N}}\frac{\alpha_{n}}{\mathcal{N}}\hat{C}^{sym}_{gr}[N]}|\Psi_{\zeta_{b}}\rangle}{\|\Psi_{\zeta_{f}}\|\|\Psi_{\zeta_{b}}\|}.\label{extraction amplitude2}
\end{align}
Note that, in the invariant subspace of $\hat{C}^{sym}_{gr}[N]$ corresponding to the regular lattice, the identity operator $\mathbb{I}_{0}:=\sum_{z,m\in\mathbb{Z}}|z,m\mu_{y}\rangle\!\langle z,m\mu_{y}|$ can be expressed in terms of coherent states as
\begin{align}
	\mathbb{I}_{0}=&\lim_{\tilde{\beta}_{n},\tilde{\gamma}_{n}\to\infty}\frac{\epsilon_{n}\bar{\omega}_n}{\pi}\frac{6}{\hbar\beta\kappa}\int_{-\infty}^{\infty}d\bar{\pi}^{1}_{n}\frac{2}{\hbar\beta\kappa\mu_{y}}\int_{-\infty}^{\infty}d\pi^{y}_{n}\nonumber\\
	\times&\frac{1}{2\tilde{\beta}_{n}}\int_{-\tilde{\beta}_{n}}^{\tilde{\beta}_{n}}d\bar{A}_{1,n}\frac{1}{2\tilde{\gamma}_{n}}\int_{-\tilde{\gamma}_{n}}^{\tilde{\gamma}_{n}}d\bar{A}_{y,n}|\Psi_{\zeta}\rangle\!\langle\Psi_{\zeta}|.\label{overcompleteness}
\end{align}
By inserting this overcompleteness relation of the coherent states into the matrix element in Eq.(\ref{extraction amplitude2}), we obtain
\begin{align}
	&\frac{\langle\Psi_{\zeta_{f}}|e^{i\sum_{n=1}^{\mathcal{N}}\frac{\alpha_{n}}{\mathcal{N}}\hat{C}^{sym}_{gr}[N]}|\Psi_{\zeta_{b}}\rangle}{\|\Psi_{\zeta_{f}}\|\|\Psi_{\zeta_{b}}\|}\nonumber\\
	=&(\prod_{n=1}^{\mathcal{N}-1}\frac{6}{\hbar\beta\kappa}\int_{-\infty}^{\infty}d\bar{\pi}^{1}_{n})(\prod_{n=1}^{\mathcal{N}-1}\frac{2}{\hbar\beta\kappa\mu_{y}}\int_{-\infty}^{\infty}d\pi^{y}_{n})\nonumber\\
	\times&\lim_{\tilde{\beta}_{\mathcal{N}-1},\cdots,\tilde{\beta}_{1}\to\infty}(\prod_{n=1}^{\mathcal{N}-1}\frac{1}{2\tilde{\beta}_{n}}\int_{-\tilde{\beta}_{n}}^{\tilde{\beta}_{n}}d\bar{A}_{1,n})\nonumber\\
	\times&\lim_{\tilde{\gamma}_{\mathcal{N}-1},\cdots,\tilde{\gamma}_{1}\to\infty}(\prod_{n=1}^{\mathcal{N}-1}\frac{1}{2\tilde{\gamma}_{n}}\int_{-\tilde{\gamma}_{n}}^{\tilde{\gamma}_{n}}d\bar{A}_{y,n})\nonumber\\	
	\times&(\prod_{n=1}^{\mathcal{N}}\frac{\sqrt{\epsilon_{n}\bar{\omega}_{n}\epsilon_{n-1}\bar{\omega}_{n-1}}}{\pi}\langle\Psi_{\zeta_{n}}|e^{i\frac{\alpha_{n}}{\mathcal{N}}\hat{C}^{sym}_{gr}[N]}|\Psi_{\zeta_{n-1}}\rangle).\label{coherentmatrix}
\end{align}
In the limit $\mathcal{N}\to\infty$, the matrix element in Eq.(\ref{coherentmatrix}) can be expressed as
\begin{align}
	&\langle\Psi_{\zeta_{n}}|e^{i\frac{\alpha_{n}}{\mathcal{N}}\hat{C}^{sym}_{gr}[N]}|\Psi_{\zeta_{n-1}}\rangle\nonumber\\
	=&\langle\Psi_{\zeta_{n}}|\Psi_{\zeta_{n-1}}\rangle+i\frac{\alpha_{n}}{\mathcal{N}}\langle\Psi_{\zeta_{n}}|\hat{C}^{sym}_{gr}[N]|\Psi_{\zeta_{n-1}}\rangle+\mathcal{O}(\mathcal{N}^{-2}).\label{CH1}
\end{align}
By using the Poisson summation formula, the first term in Eq.(\ref{CH1}) can be approximated as (see Appendix \ref{sec:appendixBB} for details)
\begin{align}
	&\langle\Psi_{\zeta_{n}}|\Psi_{\zeta_{n-1}}\rangle\nonumber\\
	\approx&\frac{\pi}{\check{\epsilon}_{n}\check{\bar{\omega}}_{n}}\exp[-\frac{(\lambda_{n}-\lambda_{n-1})^{2}}{2(\epsilon^{-2}_{n}+\epsilon^{-2}_{n-1})}-\frac{(\bar{\xi}_{n}-\bar{\xi}_{n-1})^{2}}{2(\bar{\omega}^{-2}_{n}+\bar{\omega}^{-2}_{n-1})}]\nonumber\\
	\times&\exp[-i\frac{\check{\bar{A}}_{1,n}}{2}(\lambda_{n}-\lambda_{n-1})-i\mu_{y}\frac{\check{\bar{A}}_{y,n}}{2}(\bar{\xi}_{n}-\bar{\xi}_{n-1})]\nonumber\\
	\times&\exp[-\frac{1}{4\check{\epsilon}^{2}_{n}}\frac{(\bar{A}_{1,n}-\bar{A}_{1,n-1})^{2}}{4}-\frac{\mu_{y}^{2}}{4\check{\bar{\omega}}^{2}_{n}}\frac{(\bar{A}_{y,n}-\bar{A}_{y,n-1})^{2}}{4}],\label{CH2}
\end{align}
where the following variables are introduced to simplify the expression
\begin{align}
	\check{\epsilon}^{2}_{n}&:=\frac{\epsilon^{2}_{n}+\epsilon^{2}_{n-1}}{2},
	&\check{\bar{\omega}}^{2}_{n}&:=\frac{\bar{\omega}^{2}_{n}+\bar{\omega}^{2}_{n-1}}{2},\label{san481}\\
	\check{\lambda}_{n}&:=\frac{\epsilon^{2}_{n}\lambda_{n}+\epsilon^{2}_{n-1}\lambda_{n-1}}{\epsilon^{2}_{n}+\epsilon^{2}_{n-1}},
	&\check{\bar{\xi}}_{n}&:=\frac{\bar{\omega}^{2}_{n}\bar{\xi}_{n}+\bar{\omega}^{2}_{n-1}\bar{\xi}_{n-1}}{\bar{\omega}^{2}_{n}+\bar{\omega}^{2}_{n-1}},\label{san482}\\
	\check{\bar{A}}_{1,n}&:=\frac{\epsilon^{2}_{n-1}\bar{A}_{1,n}+\epsilon^{2}_{n}\bar{A}_{1,n-1}}{\epsilon^{2}_{n}+\epsilon^{2}_{n-1}},
	&\check{\bar{A}}_{y,n}&:=\frac{\bar{\omega}^{2}_{n-1}\bar{A}_{y,n}+\bar{\omega}^{2}_{n}\bar{A}_{y,n-1}}{\bar{\omega}^{2}_{n}+\bar{\omega}^{2}_{n-1}}.\label{san483}
\end{align}
To evaluate the second term in Eq.(\ref{CH1}), it suffices to evaluate the corresponding matrix elements of the constituent operators $\hat{u}^{sym}_{r}$ and $\hat{u}^{sym}_{k,l}$ in $\hat{C}^{sym}_{gr}[N]$. This motivates us to consider the action of an operator in the following form
\begin{equation}\label{Hoperator}
	\hat{H}_{\mathfrak{k},\mathfrak{l}}|\lambda,\xi\rangle:=H_{1}(\lambda)H_{y}(\xi)\left|\lambda+\mathfrak{k},\xi+\mathfrak{l}\mu_{y}\right>,
\end{equation}
where $\mathfrak{k},\mathfrak{l}\in\mathbb{Z}$, $H_{1}(\lambda)$ and $H_{y}(\xi)$ are initially assumed to be analytic functions, and the result is then extended to the non-analytic functions in our case. From Eq.(\ref{Hoperator}), it directly follows that
\begin{align}
	&\langle\Psi_{\zeta_{n}}|\hat{H}_{\mathfrak{k},\mathfrak{l}}|\Psi_{\zeta_{n-1}}\rangle\nonumber\\
	=&\sum_{r\in\mathbb{Z}}H_{1}(r)e^{-\frac{\epsilon_{n}^{2}}{2}(r+\mathfrak{k}-\lambda_{n})^2-\frac{\epsilon_{n-1}^{2}}{2}(r-\lambda_{n-1})^2+i\frac{\bar{A}_{1,n}}{2}(r+\mathfrak{k}-\lambda_{n})-i\frac{\bar{A}_{1,n-1}}{2}(r-\lambda_{n-1})}\nonumber\\
	\times&\sum_{m\in\mathbb{Z}}H_{y}(m\mu_y)e^{-\frac{\bar{\omega}_{n}^{2}}{2}(m+\mathfrak{l}-\bar{\xi}_{n})^2-\frac{\bar{\omega}_{n-1}^{2}}{2}(m-\bar{\xi}_{n-1})^2+i\frac{\bar{A}_{y,n}}{2}\mu_{y}(m+\mathfrak{l}-\bar{\xi}_{n})-i\frac{\bar{A}_{y,n-1}}{2}\mu_{y}(m-\bar{\xi}_{n-1})}\nonumber\\
	=&\exp[-\frac{(\lambda_{n}-\lambda_{n-1})^{2}}{2(\epsilon^{-2}_{n}+\epsilon^{-2}_{n-1})}-i\frac{\check{\bar{A}}_{1,n}}{2}(\lambda_{n}-\lambda_{n-1})]\nonumber\\
	\times&\exp[(\lambda_{n}-\check{\lambda}_{n})\epsilon^{2}_{n}\mathfrak{k}-\frac{i}{2}(\bar{A}_{1,n}-\bar{A}_{1,n-1})\frac{\mathfrak{k}\epsilon_{n}^{2}}{2\check{\epsilon}^{2}_{n}}]
	\exp[(\frac{\epsilon^{2}_{n}}{2\check{\epsilon}^{2}_{n}}-1)\frac{\epsilon^{2}_{n}\mathfrak{k}^{2}}{2}+i\frac{\bar{A}_{1,n}}{2}\mathfrak{k}]\nonumber\\
	\times&\sum_{r\in\mathbb{Z}}H_{1}(r)\exp[-\check{\epsilon}^{2}_{n}(r-\check{\lambda}_{n}+\frac{\mathfrak{k}\epsilon_{n}^{2}}{2\check{\epsilon}_{n}^{2}})^2+i\frac{(\bar{A}_{1,n}-\bar{A}_{1,n-1})}{2}(r-\check{\lambda}_{n}+\frac{\mathfrak{k}\epsilon_{n}^{2}}{2\check{\epsilon}_{n}^{2}})]\nonumber\\
	\times&\exp[-\frac{(\bar{\xi}_{n}-\bar{\xi}_{n-1})^{2}}{2(\bar{\omega}^{-2}_{n}+\bar{\omega}^{-2}_{n-1})}-i\mu_{y}\frac{\check{\bar{A}}_{y,n}}{2}(\bar{\xi}_{n}-\bar{\xi}_{n-1})]\nonumber\\
	\times&\exp[(\bar{\xi}_{n}-\check{\bar{\xi}}_{n})\bar{\omega}^{2}_{n}\mathfrak{l}-\frac{i}{2}\mu_{y}(\bar{A}_{y,n}-\bar{A}_{y,n-1})\frac{\mathfrak{l}\bar{\omega}_{n}^{2}}{2\check{\bar{\omega}}^{2}_{n}}]
	\exp[(\frac{\bar{\omega}^{2}_{n}}{2\check{\bar{\omega}}^{2}_{n}}-1)\frac{\bar{\omega}^{2}_{n}\mathfrak{l}^{2}}{2}+i\mu_{y}\frac{\bar{A}_{y,n}}{2}\mathfrak{l}]\nonumber\\
	\times&\sum_{m\in\mathbb{Z}}H_{y}(m\mu_{y})\exp[-\check{\bar{\omega}}^{2}_{n}(m-\check{\xi}_{n}+\frac{\mathfrak{l}\bar{\omega}_{n}^{2}}{2\check{\bar{\omega}}_{n}^{2}})^2+i\mu_{y}\frac{(\bar{A}_{1,n}-\bar{A}_{1,n-1})}{2}(m-\check{\xi}_{n}+\frac{\mathfrak{l}\bar{\omega}_{n}^{2}}{2\check{\bar{\omega}}_{n}^{2}})],\label{BBB1}
\end{align}
where Eqs.(\ref{san481}) and (\ref{san482}) have been used in the second step. Furthermore, applying the Poisson summation formula together with the method of steepest descent \cite{LQC5,Yang23,Li:2025bzl} to the two sums in the second step, we obtain
\begin{align}
	&\sum_{r\in\mathbb{Z}}H_{1}(r)\exp[-\check{\epsilon}^{2}_{n}(r-\check{\lambda}_{n}+\frac{\mathfrak{k}\epsilon_{n}^{2}}{2\check{\epsilon}_{n}^{2}})^2+i\frac{(\bar{A}_{1,n}-\bar{A}_{1,n-1})}{2}(r-\check{\lambda}_{n}+\frac{\mathfrak{k}\epsilon_{n}^{2}}{2\check{\epsilon}_{n}^{2}})]\nonumber\\
	\approx&\frac{\sqrt{\pi}}{\check{\epsilon}_{n}}\exp[-\frac{1}{4\check{\epsilon}^{2}_{n}}\frac{(\bar{A}_{1,n}-\bar{A}_{1,n-1})^{2}}{4}]\nonumber\\
	\times&[H_{1}(\check{\lambda}_{n}-\frac{\mathfrak{k}\epsilon_{n}^{2}}{2\check{\epsilon}^{2}_{n}}+\frac{i}{4\check{\epsilon}^{2}_{n}}(\bar{A}_{1,n}-\bar{A}_{1,n-1}))
	+\frac{1}{4\check{\epsilon}^{2}_{n}}\partial_{\lambda}^{2}H_{1}(\check{\lambda}_{n}-\frac{\mathfrak{k}\epsilon_{n}^{2}}{2\check{\epsilon}^{2}_{n}}+\frac{i}{4\check{\epsilon}^{2}_{n}}(\bar{A}_{1,n}-\bar{A}_{1,n-1}))],\label{BBB2}\\
	\nonumber\\
	&\sum_{m\in\mathbb{Z}}H_{y}(m\mu_{y})\exp[-\check{\bar{\omega}}^{2}_{n}(m-\check{\xi}_{n}+\frac{\mathfrak{l}\bar{\omega}_{n}^{2}}{2\check{\bar{\omega}}_{n}^{2}})^2+i\mu_{y}\frac{(\bar{A}_{1,n}-\bar{A}_{1,n-1})}{2}(m-\check{\xi}_{n}+\frac{\mathfrak{l}\bar{\omega}_{n}^{2}}{2\check{\bar{\omega}}_{n}^{2}})]\nonumber\\
	\approx&\frac{\sqrt{\pi}}{\check{\bar{\omega}}_{n}}\exp[-\frac{\mu_{y}^{2}}{4\check{\bar{\omega}}^{2}_{n}}\frac{(\bar{A}_{y,n}-\bar{A}_{y,n-1})^{2}}{4}]\nonumber\\
	\times&[H_{y}(\mu_{y}\check{\bar{\xi}}_{n}-\mu_{y}\frac{\mathfrak{l}\bar{\omega}_{n}^{2}}{2\check{\bar{\omega}}^{2}_{n}}+\frac{i\mu_{y}^{2}}{4\check{\bar{\omega}}^{2}_{n}}(\bar{A}_{y,n}-\bar{A}_{y,n-1}))
	+\frac{\mu_{y}^{2}}{4\check{\bar{\omega}}^{2}_{n}}\partial_{\xi}^{2}H_{y}(\mu_{y}\check{\bar{\xi}}_{n}-\mu_{y}\frac{\mathfrak{l}\bar{\omega}_{n}^{2}}{2\check{\bar{\omega}}^{2}_{n}}+\frac{i\mu_{y}^{2}}{4\check{\bar{\omega}}^{2}_{n}}(\bar{A}_{y,n}-\bar{A}_{y,n-1}))].\label{BBB3}
\end{align}
Collecting the above results, when $\hat{H}_{\mathfrak{k},\mathfrak{l}}$ is taken to be the identity operator, we obtain Eq.(\ref{CH2}) by combining Eqs.(\ref{BBB1}), (\ref{BBB2}), and (\ref{BBB3}). On this basis, the matrix element of the operator on Eq.(\ref{Hoperator}) can be calculated as 
\begin{align}
	&\langle\Psi_{\zeta_{n}}|\hat{H}_{\mathfrak{k},\mathfrak{l}}|\Psi_{\zeta_{n-1}}\rangle\nonumber\\
	\approx&\langle\Psi_{\zeta_{n}}|\Psi_{\zeta_{n-1}}\rangle
	\exp[(\lambda_{n}-\check{\lambda}_{n})\epsilon^{2}_{n}\mathfrak{k}-\frac{i}{2}(\bar{A}_{1,n}-\bar{A}_{1,n-1})\frac{\mathfrak{k}\epsilon_{n}^{2}}{2\check{\epsilon}^{2}_{n}}]\nonumber\\
	\times&\exp[(\frac{\epsilon^{2}_{n}}{2\check{\epsilon}^{2}_{n}}-1)\frac{\epsilon^{2}_{n}\mathfrak{k}^{2}}{2}+i\frac{\bar{A}_{1,n}}{2}\mathfrak{k}]\nonumber\\
	\times&[H_{1}(\check{\lambda}_{n}-\frac{\mathfrak{k}\epsilon_{n}^{2}}{2\check{\epsilon}^{2}_{n}}+\frac{i}{4\check{\epsilon}^{2}_{n}}(\bar{A}_{1,n}-\bar{A}_{1,n-1}))\nonumber\\
	+&\frac{1}{4\check{\epsilon}^{2}_{n}}\partial_{\lambda}^{2}H_{1}(\check{\lambda}_{n}-\frac{\mathfrak{k}\epsilon_{n}^{2}}{2\check{\epsilon}^{2}_{n}}+\frac{i}{4\check{\epsilon}^{2}_{n}}(\bar{A}_{1,n}-\bar{A}_{1,n-1}))]\nonumber\\
	\times&\exp[(\bar{\xi}_{n}-\check{\bar{\xi}}_{n})\bar{\omega}^{2}_{n}\mathfrak{l}-\frac{i}{2}\mu_{y}(\bar{A}_{y,n}-\bar{A}_{y,n-1})\frac{\mathfrak{l}\bar{\omega}_{n}^{2}}{2\check{\bar{\omega}}^{2}_{n}}]\nonumber\\
	\times&\exp[(\frac{\bar{\omega}^{2}_{n}}{2\check{\bar{\omega}}^{2}_{n}}-1)\frac{\bar{\omega}^{2}_{n}\mathfrak{l}^{2}}{2}+i\mu_{y}\frac{\bar{A}_{y,n}}{2}\mathfrak{l}]\nonumber\\
	\times&[H_{y}(\mu_{y}\check{\bar{\xi}}_{n}-\mu_{y}\frac{\mathfrak{l}\bar{\omega}_{n}^{2}}{2\check{\bar{\omega}}^{2}_{n}}+\frac{i\mu_{y}^{2}}{4\check{\bar{\omega}}^{2}_{n}}(\bar{A}_{y,n}-\bar{A}_{y,n-1}))\nonumber\\
	+&\frac{\mu_{y}^{2}}{4\check{\bar{\omega}}^{2}_{n}}\partial_{\xi}^{2}H_{y}(\mu_{y}\check{\bar{\xi}}_{n}-\mu_{y}\frac{\mathfrak{l}\bar{\omega}_{n}^{2}}{2\check{\bar{\omega}}^{2}_{n}}+\frac{i\mu_{y}^{2}}{4\check{\bar{\omega}}^{2}_{n}}(\bar{A}_{y,n}-\bar{A}_{y,n-1}))]\nonumber\\
	=:&\langle\Psi_{\zeta_{n}}|\Psi_{\zeta_{n-1}}\rangle\check{H}_{\mathfrak{k},\mathfrak{l}}(n,n-1).\label{CH3}
\end{align}
It then follows that
\begin{align}
	&\langle\Psi_{\zeta_{n}}|\hat{H}^{sym}_{\mathfrak{k},\mathfrak{l}}|\Psi_{\zeta_{n-1}}\rangle\nonumber\\
	=&\langle\Psi_{\zeta_{n}}|\Psi_{\zeta_{n-1}}\rangle\frac{1}{2}(\check{H}_{\mathfrak{k},\mathfrak{l}}(n,n-1)+\overline{\check{H}_{\mathfrak{k},\mathfrak{l}}(n-1,n)}),\label{exsym}
\end{align}
where $\hat{H}^{sym}_{\mathfrak{k},\mathfrak{l}}:=(\hat{H}_{\mathfrak{k},\mathfrak{l}}+\hat{H}^{\dagger}_{\mathfrak{k},\mathfrak{l}})/2$ and the overline denotes the complex conjugate. By combining Eqs.(\ref{exsym}), (\ref{SCdec}) and (\ref{CH1}), we obtain
\begin{align}
	&\langle\Psi_{\zeta_{n}}|e^{i\frac{\alpha_{n}}{\mathcal{N}}\hat{C}^{sym}_{gr}[N]}|\Psi_{\zeta_{n-1}}\rangle\nonumber\\
	=&\langle\Psi_{\zeta_{n}}|\Psi_{\zeta_{n-1}}\rangle\exp\{i\frac{\alpha_{n}}{\mathcal{N}}4^{-2}N\hbar\beta^{-1}\Delta^{-1/2}\nonumber\\
	\times&\frac{1}{2}(\sum_{r\in\{0,4,-4\}}\check{u}_{r}(n,n-1)+\overline{\check{u}_{r}(n-1,n)}\nonumber\\
	+&\sum_{k,l\in\{2,-2\}}\check{u}_{k,l}(n,n-1)+\overline{\check{u}_{k,l}(n-1,n)})\},\label{Ematrix}
\end{align}
where $\check{u}_{r}$ and $\check{u}_{k,l}$ are obtained by replacing $\hat{H}_{\mathfrak{k},\mathfrak{l}}$ in Eq.(\ref{exsym}) with $\hat{u}_{r}$ and $\hat{u}_{k,l}$ respectively. By combining Eqs.(\ref{extraction amplitude2}), (\ref{coherentmatrix}), and (\ref{Ematrix}) and taking the limit as $\mathcal{N}\to\infty$, we conclude
\begin{align}
	&A(\zeta_{f};\zeta_{b})\nonumber\\
	=&\lim_{\mathcal{N}\to\infty}\lim_{\tilde{\alpha}_{\mathcal{N}},\cdots,\tilde{\alpha}_{1}\to\infty}
	\frac{2\tilde{\alpha}_{1}}{\mathcal{N}}(\prod_{n=1}^{\mathcal{N}}\frac{1}{2\tilde{\alpha}_{n}}\int_{-\tilde{\alpha}_{n}}^{\tilde{\alpha}_{n}}d\alpha_{n})\nonumber\\
	\times&\lim_{\tilde{\beta}_{\mathcal{N}-1},\cdots,\tilde{\beta}_{1}\to\infty}(\prod_{n=1}^{\mathcal{N}-1}\frac{1}{2\tilde{\beta}_{n}}\int_{-\tilde{\beta}_{n}}^{\tilde{\beta}_{n}}d\bar{A}_{1,n})\nonumber\\
	\times&\lim_{\tilde{\gamma}_{\mathcal{N}-1},\cdots,\tilde{\gamma}_{1}\to\infty}(\prod_{n=1}^{\mathcal{N}-1}\frac{1}{2\tilde{\gamma}_{n}}\int_{-\tilde{\gamma}_{n}}^{\tilde{\gamma}_{n}}d\bar{A}_{y,n})\nonumber\\
	\times&(\prod_{n=1}^{\mathcal{N}-1}\frac{6}{\hbar\beta\kappa}\int_{-\infty}^{\infty}d\bar{\pi}^{1}_{n})(\prod_{n=1}^{\mathcal{N}-1}\frac{2}{\hbar\beta\kappa\mu_{y}}\int_{-\infty}^{\infty}d\pi^{y}_{n})\nonumber\\
	\times&(\prod_{n=1}^{\mathcal{N}}\frac{\sqrt{\epsilon_{n}\bar{\omega}_{n}\epsilon_{n-1}\bar{\omega}_{n-1}}}{\check{\epsilon}_{n}\check{\bar{\omega}}_{n}})\exp\{\frac{i}{\hbar}\mathcal{N}^{-1}\nonumber\\
	\times&\sum_{n=1}^{\mathcal{N}}[\frac{i\hbar(\lambda_{n}-\lambda_{n-1})^{2}}{2(\epsilon^{-2}_{n}+\epsilon^{-2}_{n-1})\mathcal{N}^{-1}}+\frac{i\hbar(\bar{\xi}_{n}-\bar{\xi}_{n-1})^{2}}{2(\bar{\omega}^{-2}_{n}+\bar{\omega}^{-2}_{n-1})\mathcal{N}^{-1}}\nonumber\\	
	-&\frac{\hbar\check{\bar{A}}_{1,n}}{2}\frac{(\lambda_{n}-\lambda_{n-1})}{\mathcal{N}^{-1}}+\frac{i\hbar}{4\check{\epsilon}^{2}_{n}}\frac{(\bar{A}_{1,n}-\bar{A}_{1,n-1})^{2}}{4\mathcal{N}^{-1}}\nonumber\\
	-&\mu_{y}\frac{\hbar\check{\bar{A}}_{y,n}}{2}\frac{(\bar{\xi}_{n}-\bar{\xi}_{n-1})}{\mathcal{N}^{-1}}+\frac{i\hbar\mu_{y}^{2}}{4\check{\bar{\omega}}^{2}_{n}}\frac{(\bar{A}_{y,n}-\bar{A}_{y,n-1})^{2}}{4\mathcal{N}^{-1}}\nonumber\\
	+&\alpha_{n}\hbar4^{-2}N\hbar\beta^{-1}\Delta^{-1/2}\frac{1}{2}\nonumber\\
	\times&(\sum_{r\in\{0,4,-4\}}\check{u}_{r}(n,n-1)+\overline{\check{u}_{r}(n-1,n)}\nonumber\\
	+&\sum_{k,l\in\{2,-2\}}\check{u}_{k,l}(n,n-1)+\overline{\check{u}_{k,l}(n-1,n)})]\}\nonumber\\
	=&\mathfrak{c}'\mathcal{D}\alpha\mathcal{D}\bar{\pi}^{1}\mathcal{D}\pi^{y}\mathcal{D}\bar{A}_{1}\mathcal{D}\bar{A}_{y}\exp\{\frac{i}{\hbar}\int_{0}^{1}dt[-\frac{3}{\beta\kappa}\bar{A}_{1}\partial_{t}\bar{\pi}^{1}\nonumber\\
	-&\frac{1}{\beta\kappa}\bar{A}_{y}\partial_{t}\pi^{y}+\alpha\hbar\langle\hat{C}^{sym}_{gr}[N]\rangle_{\zeta}]\},
\end{align}
where $\mathfrak{c}'$ is an overall constant, $t$ is the fictitious time variable, and in the second step we have used the fact that $\lim_{\mathcal{N}\to\infty}(\check{H}_{\mathfrak{k},\mathfrak{l}}(n,n-1)+\overline{\check{H}_{\mathfrak{k},\mathfrak{l}}(n-1,n)})/2=\langle\hat{H}^{sym}_{\mathfrak{k},\mathfrak{l}}\rangle_{\zeta_{n}}$ with $\langle\cdot\rangle_{\zeta}$ denoting the expectation value evaluated in the coherent state $|\Psi_{\zeta}\rangle$.

\section{An Alternative Quantum Dynamics}\label{sec:appendixBBB}
To further examine the validity of the effective scalar constraint, we consider an alternative scalar constraint operator. It differs from the scalar constraint operator used in the main text only in its factor ordering, but it leads to a solvable quantum dynamics model. For consistency, we introduce the following notation
\begin{align}
\lambda_1&:=\lambda,&\lambda_y&:=\xi,\nonumber\\
\nu_1&:=1,&\nu_y&:=\mu_y.
\end{align}
Accordingly, Eqs.(\ref{III.3}) and (\ref{III.4}) can be rewritten as
\begin{equation}
\hat{\bar{\pi}}^{1}|\lambda_1\rangle=\frac{\hbar\beta\kappa}{6}\lambda_1|\lambda_1\rangle,\qquad
\hat{\bar{\pi}}^{y}|\lambda_y\rangle=\frac{\hbar\beta\kappa}{2}\lambda_y|\lambda_y\rangle,\label{C.2}
\end{equation} 
where their respective domains are defined as $\mathcal{D}_{\mathrm{kin}}^{\alpha}:=\operatorname{span}_{\mathbb{C}}\{|\lambda_{\alpha}\rangle\mid\lambda_{\alpha}\in\mathbb{R}\}$, with $\alpha\in\{1,y\}$, and the completion of $\mathcal{D}_{\mathrm{kin}}^{\alpha}$ yields the corresponding Hilbert space $\mathcal{H}_{\mathrm{kin}}^{\alpha}$. Moreover, the regularized expression for the scalar constraint, Eq.(\ref{regularizedC}), can be rewritten as
\begin{equation}
	C_{gr}[N]\approx-\frac{3}{2}N\beta^{-2}\Delta^{-1/2}\kappa^{-1}|\bar{\pi}^{1}|^{-1/2}
	\bar{O}_1(2\bar{O}_1+\bar{O}_y)|\bar{\pi}^{1}|^{-1/2},\label{C.3}
\end{equation}
where
\begin{equation}
\bar{O}_\alpha:=\bar{\pi}^{\alpha}\frac{\sin(\nu_{\alpha}\bar{A}_{\alpha})}{\nu_{\alpha}}.
\end{equation}
Since $\bar{\pi}^{\alpha}\bar{A}_{\alpha}$ is a Dirac observable at the classical level, we are motivated to promote it to a self-adjoint operator $\hat{\bar{O}}_\alpha$ and construct the constraint operator $\hat{C}_{gr}[N]$, thereby ensuring that $[\hat{\bar{O}}_\alpha,\hat{C}_{gr}[N]]=0$ \cite{Zhang:2021wex}. To this end, we can adopt the following definition, similar to those used in Refs.\cite{Martin-Benito:2009htq,Martin-Benito:2008dfr},
\begin{equation}
\hat{\bar{O}}_\alpha:=\frac{1}{2}|\hat{\bar{\pi}}^{\alpha}|^{\frac{1}{2}}\bigg(\operatorname{sgn}(\hat{\bar{\pi}}^{\alpha})\frac{\widehat{\sin(\nu_{\alpha}\bar{A}_{\alpha})}}{\nu_{\alpha}}+\frac{\widehat{\sin(\nu_{\alpha}\bar{A}_{\alpha})}}{\nu_{\alpha}}\operatorname{sgn}(\hat{\bar{\pi}}^{\alpha})\bigg)|\hat{\bar{\pi}}^{\alpha}|^{\frac{1}{2}},
\end{equation}
where the constituent operators are defined in the same way as those in the main text. To further investigate the properties of the operator $\hat{\bar{O}}_\alpha$, we define the following dimensionless operators
\begin{equation}
\hat{\bar{O}}_1=:\frac{\hbar\beta\kappa}{3}\hat{\Omega}_1,\qquad
\hat{\bar{O}}_y=:\hbar\beta\kappa\hat{\Omega}_y,\label{C.6}
\end{equation}
where the difference in the coefficients arises from the different coefficients in Eq.(\ref{C.2}). Their actions on the basis states are given by
\begin{equation}
\hat{\Omega}_{\alpha}|\lambda_{\alpha}\rangle=\frac{1}{4\nu_{\alpha}i}
\bigg(\frac{\operatorname{sgn}(\lambda_{\alpha}+2\nu_{\alpha})+\operatorname{sgn}(\lambda_{\alpha})}{2}\sqrt{|(\lambda_{\alpha}+2\nu_{\alpha})\lambda_{\alpha}|}|\lambda_{\alpha}+2\nu_{\alpha}\rangle
-\frac{\operatorname{sgn}(\lambda_{\alpha}-2\nu_{\alpha})+\operatorname{sgn}(\lambda_{\alpha})}{2}\sqrt{|(\lambda_{\alpha}-2\nu_{\alpha})\lambda_{\alpha}|}|\lambda_{\alpha}-2\nu_{\alpha}\rangle\bigg).\label{C.7}
\end{equation}
$\mathcal{H}_{\mathrm{kin}}^{\alpha}$ can be decomposed into invariant subspaces of $\hat{\Omega}_{\alpha}$,
\begin{equation}
\mathcal{H}_{\mathrm{kin}}^{\alpha}=\mathcal{H}_{\alpha}^{0}\oplus_{0<\epsilon_{\alpha}\le1}(\mathcal{H}_{\epsilon_{\alpha}}^{+}\oplus\mathcal{H}_{\epsilon_{\alpha}}^{-}),
\end{equation}
where the Hilbert spaces $\mathcal{H}_{\alpha}^{0}$ and $\mathcal{H}_{\epsilon_{\alpha}}^{\pm}$ are respectively the completions of $\operatorname{span}_{\mathbb{C}}\{|\lambda_{\alpha}=0\rangle\}$ and $\mathcal{D}_{\epsilon_{\alpha}}^{\pm}:=\operatorname{span}_{\mathbb{C}}\{|\lambda_{\alpha}=\pm2\nu_{\alpha}(n+\epsilon_{\alpha})\rangle\mid n\in\mathbb{N}\}$. Since $[\hat{\Pi},\hat{\Omega}_{\alpha}]=0$, where the reflection operator $\hat{\Pi}$ acts as $\hat{\Pi}|\lambda_{\alpha}\rangle=|-\lambda_{\alpha}\rangle$, we can restrict $\hat{\Omega}_{\alpha}$ to the nontrivial invariant subspace $\mathcal{H}_{\epsilon_{\alpha}}^{+}$ for our analysis, and denote the resulting restriction by $\hat{\Omega}_{\epsilon_{\alpha}}\equiv\hat{\Omega}_{\alpha}|_{\mathcal{D}_{\epsilon_{\alpha}}^{+}}$. In what follows, we first investigate the properties of the operator $\hat{\Omega}_{\epsilon_{\alpha}}$ in detail, and then turn to the quantum dynamics.

In the mathematical theory of infinite tridiagonal matrices, Lemma 1 of Ref.\cite{J.Dombrowski1} implies that $\hat{\Omega}_{\epsilon_{\alpha}}$ is essentially self-adjoint on the domain $\mathcal{D}_{\epsilon_{\alpha}}^{+}$. In what follows, we use the same notation to denote its unique self-adjoint extension. Its generalized eigenvalue equation is given by
\begin{equation}
(e_{k_{\alpha}}^{\epsilon_{\alpha}}|\hat{\Omega}_{\epsilon_{\alpha}}|2\nu_{\alpha}(n+\epsilon_{\alpha})\rangle=k_{\alpha}(e_{k_{\alpha}}^{\epsilon_{\alpha}}|2\nu_{\alpha}(n+\epsilon_{\alpha})\rangle,\qquad\forall n\in\mathbb{N},\label{C.9}
\end{equation}
where $(\mathcal{D}_{\epsilon_{\alpha}}^{+})^{*}\ni (e_{k_{\alpha}}^{\epsilon_{\alpha}}|$ denotes a generalized eigenstate and $\mathbb{R}\ni k_{\alpha}$ denotes the corresponding generalized eigenvalue. Eq.(\ref{C.9}) can be explicitly rewritten as
\begin{equation}
-\frac{1}{2i}\bigg(\sqrt{(n+\epsilon_{\alpha}+1)(n+\epsilon_{\alpha})}e_{k_{\alpha}}^{\epsilon_{\alpha}}(2\nu_{\alpha}(n+\epsilon_{\alpha}+1))
-\sqrt{(n+\epsilon_{\alpha}-1)(n+\epsilon_{\alpha})}e_{k_{\alpha}}^{\epsilon_{\alpha}}(2\nu_{\alpha}(n+\epsilon_{\alpha}-1))\bigg)=k_{\alpha}e_{k_{\alpha}}^{\epsilon_{\alpha}}(2\nu_{\alpha}(n+\epsilon_{\alpha})),\label{C.10}
\end{equation}
where $e_{k_{\alpha}}^{\epsilon_{\alpha}}(2\nu_{\alpha}(n+\epsilon_{\alpha})):=\langle2\nu_{\alpha}(n+\epsilon_{\alpha})|e_{k_{\alpha}}^{\epsilon_{\alpha}}):=\overline{(e_{k_{\alpha}}^{\epsilon_{\alpha}}|2\nu_{\alpha}(n+\epsilon_{\alpha})\rangle}$ and $e_{k_{\alpha}}^{\epsilon_{\alpha}}(2\nu_{\alpha}(\epsilon_{\alpha}-1)):=0$. To establish the orthogonality and completeness of the generalized eigenstates, the fact that $|2\nu_{\alpha}\epsilon_{\alpha}\rangle$ is a cyclic vector of the operator $\hat{\Omega}_{\epsilon_{\alpha}}$, together with Theorem VIII.6 of Ref.\cite{RSR2}, implies the existence of a unitary isomorphism $\hat{U}_{\alpha}:\operatorname{L}^2(\mathbb{R},d\mu({k_{\alpha}}))\to\mathcal{H}_{\epsilon_{\alpha}}^{+}$, where the measure $\mu({k_{\alpha}}):=\langle2\nu_{\alpha}\epsilon_{\alpha}|\hat{E}_{k_{\alpha}}|2\nu_{\alpha}\epsilon_{\alpha}\rangle$, with $\hat{E}_{k_{\alpha}}$ being the projection-valued measure of the self-adjoint operator $\hat{\Omega}_{\epsilon_{\alpha}}$. Its explicit action is given by
\begin{equation}
\hat{U}_{\alpha}:f\mapsto f(\hat{\Omega}_{\epsilon_{\alpha}})|2\nu_{\alpha}\epsilon_{\alpha}\rangle.
\end{equation}
It is straightforward to see that $\hat{U}_{\alpha}^{-1}\hat{\Omega}_{\epsilon_{\alpha}}\hat{U}_{\alpha}$ becomes a multiplication operator $\hat{k}_{\alpha}$ in the Hilbert space $\operatorname{L}^2(\mathbb{R},d\mu({k_{\alpha}}))$. Furthermore, by defining $\varphi_{n}^{\epsilon_{\alpha}}:=\hat{U}_{\alpha}^{-1}|2\nu_{\alpha}(n+\epsilon_{\alpha})\rangle$, Eq.(\ref{C.7}) in the Hilbert space $\mathcal{H}_{\epsilon_{\alpha}}^{+}$ is mapped by $\hat{U}_{\alpha}^{-1}$ to
\begin{equation}
k_{\alpha}\varphi_{n}^{\epsilon_{\alpha}}(k_{\alpha})=\frac{1}{2i}\bigg(\sqrt{(n+\epsilon_{\alpha}+1)(n+\epsilon_{\alpha})}\varphi_{n+1}^{\epsilon_{\alpha}}(k_{\alpha})
-\sqrt{(n+\epsilon_{\alpha}-1)(n+\epsilon_{\alpha})}\varphi_{n-1}^{\epsilon_{\alpha}}(k_{\alpha})\bigg),
\end{equation}
where $\varphi_{0}^{\epsilon_{\alpha}}(k_{\alpha})=1$ and $\varphi_{-1}^{\epsilon_{\alpha}}(k_{\alpha}):=0$. By comparing with the generalized eigenvalue equation \ref{C.10}, it can be seen that $\tilde{e}_{k_{\alpha}}^{\epsilon_{\alpha}}(2\nu_{\alpha}(n+\epsilon_{\alpha})):=\langle2\nu_{\alpha}(n+\epsilon_{\alpha})|\tilde{e}_{k_{\alpha}}^{\epsilon_{\alpha}}):=\overline{\varphi_{n}^{\epsilon_{\alpha}}(k_{\alpha})}$ is a solution of the generalized eigenvalue equation. Therefore, for any state $\psi\in\mathcal{H}_{\epsilon_{\alpha}}^{+}$, we obtain
\begin{align}
[\hat{U}_{\alpha}^{-1}\cdot\psi](k_{\alpha})=&\sum\limits_{n\in\mathbb{N}}\varphi_{n}^{\epsilon_{\alpha}}(k_{\alpha})\langle\varphi_{n}^{\epsilon_{\alpha}},\hat{U}_{\alpha}^{-1}\cdot\psi\rangle_{\operatorname{L}^2}\nonumber\\
=&\sum\limits_{n\in\mathbb{N}}(\tilde{e}_{k_{\alpha}}^{\epsilon_{\alpha}}|2\nu_{\alpha}(n+\epsilon_{\alpha})\rangle\langle2\nu_{\alpha}(n+\epsilon_{\alpha})|\psi\rangle\nonumber\\
=:&(\tilde{e}_{k_{\alpha}}^{\epsilon_{\alpha}}|\psi\rangle,\label{C.13}
\end{align}
where the first equality follows from the fact that $\{\varphi_{n}^{\epsilon_{\alpha}},n\in\mathbb{N}\}$ form a basis of $\operatorname{L}^2(\mathbb{R},d\mu({k_{\alpha}}))$, and the second equality follows from the unitarity of the isomorphism $\hat{U}_{\alpha}$. On this basis, for any two states $\psi,\psi'\in\mathcal{H}_{\epsilon_{\alpha}}^{+}$, we further obtain
\begin{align}
\langle\psi|\psi'\rangle=&\int_{\mathbb{R}}d\mu(k_{\alpha})\overline{[\hat{U}_{\alpha}^{-1}\cdot\psi](k_{\alpha})}[\hat{U}_{\alpha}^{-1}\cdot\psi'](k_{\alpha})\nonumber\\
=&\int_{\mathbb{R}}d\mu(k_{\alpha})\langle\psi|\tilde{e}_{k_{\alpha}}^{\epsilon_{\alpha}})(\tilde{e}_{k_{\alpha}}^{\epsilon_{\alpha}}|\psi'\rangle.\label{C.14}
\end{align}
This is precisely the completeness relation for the generalized eigenstates $\{|\tilde{e}_{k_{\alpha}}^{\epsilon_{\alpha}}),k_{\alpha}\in\mathbb{R}\}$. A further analysis of the orthogonality of the generalized eigenstates relies on Theorem 2 of Ref.\cite{A.I.Aptekarev} and Theorem 1.3(i) of Ref.\cite{J.Sahbani}, which imply that the measure $\mu({k_{\alpha}})$ is absolutely continuous and that the absolutely continuous spectrum of the operator $\hat{\Omega}_{\epsilon_{\alpha}}$ is $\mathbb{R}$. Consequently, there exists a nonnegative function $\rho_{\epsilon_{\alpha}}$ on $\mathbb{R}$ such that $d\mu(k_{\alpha})=\rho_{\epsilon_{\alpha}}(k_{\alpha})dk_{\alpha}$. We thus obtain a new set of generalized eigenstates $\{|e_{k_{\alpha}}^{\epsilon_{\alpha}}):=\sqrt{\rho_{\epsilon_{\alpha}}(k_{\alpha})}|\tilde{e}_{k_{\alpha}}^{\epsilon_{\alpha}}),k_{\alpha}\in\mathbb{R}\}$. In terms of these states, Eq.(\ref{C.14}) can be rewritten as 
\begin{equation}
	\langle\psi|\psi'\rangle=\int_{\mathbb{R}}dk_{\alpha}\langle\psi|e_{k_{\alpha}}^{\epsilon_{\alpha}})(e_{k_{\alpha}}^{\epsilon_{\alpha}}|\psi'\rangle,\label{C.15}
\end{equation}
and Eq.(\ref{C.13}) takes the form
\begin{align}
(e_{k_{\alpha}}^{\epsilon_{\alpha}}|\psi\rangle:=&\sum\limits_{n\in\mathbb{N}}(e_{k_{\alpha}}^{\epsilon_{\alpha}}|2\nu_{\alpha}(n+\epsilon_{\alpha})\rangle\langle2\nu_{\alpha}(n+\epsilon_{\alpha})|\psi\rangle\nonumber\\
=&\sum\limits_{n\in\mathbb{N}}(e_{k_{\alpha}}^{\epsilon_{\alpha}}|2\nu_{\alpha}(n+\epsilon_{\alpha})\rangle\int_{\mathbb{R}}dk'_{\alpha}\langle2\nu_{\alpha}(n+\epsilon_{\alpha})|e_{k'_{\alpha}}^{\epsilon_{\alpha}})(e_{k'_{\alpha}}^{\epsilon_{\alpha}}|\psi\rangle,\label{C.16}
\end{align}
where the second equality is obtained by inserting the completeness relation (\ref{C.15}). It follows from Eq.(\ref{C.16}) that the generalized eigenstates $\{|e_{k_{\alpha}}^{\epsilon_{\alpha}}),k_{\alpha}\in\mathbb{R}\}$ satisfy the following orthogonality relation
\begin{equation}
(e_{k_{\alpha}}^{\epsilon_{\alpha}}|e_{k'_{\alpha}}^{\epsilon_{\alpha}}):=\sum\limits_{n\in\mathbb{N}}(e_{k_{\alpha}}^{\epsilon_{\alpha}}|2\nu_{\alpha}(n+\epsilon_{\alpha})\rangle\langle2\nu_{\alpha}(n+\epsilon_{\alpha})|e_{k'_{\alpha}}^{\epsilon_{\alpha}})=\delta(k_{\alpha},k'_{\alpha}).\label{C.17}
\end{equation}

For convenience, we focus on the case $\epsilon_{\alpha}=1$ to derive the explicit form of $e_{k_{\alpha}}^{\epsilon_{\alpha}}(2\nu_{\alpha}(n+\epsilon_{\alpha}))$. The other cases can be handled similarly. In this case, the corresponding notation is introduced as follows
\begin{align}
	(e_{k_{\alpha}}|&:=(e_{k_{\alpha}}^{\epsilon_{\alpha}=1}|,&\hat{\tilde{\Omega}}_{\alpha}&:=\hat{\Omega}_{\epsilon_{\alpha}=1},\nonumber\\
	\tilde{\mathcal{D}}_\alpha&:=\mathcal{D}_{\epsilon_{\alpha}=1}^{+},&\tilde{\mathcal{H}}_\alpha&:=\mathcal{H}_{\epsilon_{\alpha}=1}^{+}.
\end{align}
With this notation, Eq.(\ref{C.10}) can be expressed as
\begin{equation}
-\frac{1}{2i}\bigg(\sqrt{(n+1)n}e_{k_{\alpha}}(2\nu_{\alpha}(n+1))-\sqrt{(n-1)n}e_{k_{\alpha}}(2\nu_{\alpha}(n-1))\bigg)=k_{\alpha}e_{k_{\alpha}}(2\nu_{\alpha}n),\qquad\forall n\in\mathbb{Z}^{+},\label{C.19}
\end{equation}
where $n$ has been reparametrized such that its range changes from $\mathbb{N}$ to $\mathbb{Z}^{+}$. To apply the generating function method to the recurrence relation (\ref{C.19}), we introduce $\check{e}(2\nu_{\alpha}n):=e_{k_{\alpha}}(2\nu_{\alpha}n)/\sqrt{n}$ to remove the square root in Eq.(\ref{C.19}),
\begin{equation}
-\frac{1}{2i}\bigg((n+1)\check{e}_{k_{\alpha}}(2\nu_{\alpha}(n+1))-(n-1)\check{e}_{k_{\alpha}}(2\nu_{\alpha}(n-1))\bigg)=k_{\alpha}\check{e}_{k_{\alpha}}(2\nu_{\alpha}n),\qquad\forall n\in\mathbb{Z}^{+}.\label{C.20}
\end{equation}
By multiplying both sides of Eq.(\ref{C.20}) by $t^n$ and summing over $n$ from 0 to infinity, we obtain
\begin{equation}
(1-t^2)\partial_{t}G(t,k_{\alpha})=-2ik_{\alpha}G(t,k_{\alpha}),\label{C.21}
\end{equation}
where $G(t,k_{\alpha}):=\sum_{n=0}^{\infty}\check{e}_{k_{\alpha}}(2\nu_{\alpha}n)t^n$, the range of $n$ is naturally extended to $\mathbb{Z}$, and $\check{e}_{k_{\alpha}}(-2\nu_{\alpha})\equiv0$ has been introduced to simplify the expression. By regarding the initial value $\check{e}_{k_{\alpha}}(0)$ as an undetermined normalization coefficient $A(k_{\alpha})$ associated with the Dirac delta, and solving Eq.(\ref{C.21}), we obtain
\begin{equation}
G(t,k_{\alpha})=A(k_{\alpha})(1-t)^{ik_{\alpha}}(1+t)^{ik_{\alpha}}.\label{C.22}
\end{equation}
The general expression for $\check{e}_{k_{\alpha}}(2\nu_{\alpha}n)$ can be obtained by expanding Eq.(\ref{C.22}) using the generalized binomial theorem. However, a more useful approach is to express $\check{e}_{k_{\alpha}}(2\nu_{\alpha}n)$ in terms of the Cauchy integral formula as follows,
\begin{equation}
\check{e}_{k_{\alpha}}(2\nu_{\alpha}n)=\frac{1}{n!}\frac{\partial^n}{\partial t^n}\bigg|_{t=0}G(t,k_{\alpha})=\frac{1}{2\pi i}\oint_{|z|=1}dz\frac{G(z,k_{\alpha})}{z^{n+1}}.\label{C.23}
\end{equation}
By parametrizing the unit circle as $z=e^{i\theta_{\alpha}}$, where $-\pi<\theta_{\alpha}<\pi$, Eq.(\ref{C.23}) can be regarded as defining the Fourier coefficients of the function  
\begin{equation}
\mathcal{G}(\theta_{\alpha},k_{\alpha}):=G(e^{i\theta_{\alpha}},k_{\alpha})=A(k_{\alpha})e^{\operatorname{sgn}(\theta_{\alpha})\pi k_{\alpha}/2}e^{ik_{\alpha}\ln(\tan(\theta_{\alpha}/2))}.
\end{equation}
Consequently, $\mathcal{G}(\theta_{\alpha},k_{\alpha})$ admits the following Fourier expansion
\begin{equation}
\mathcal{G}(\theta_{\alpha},k_{\alpha})=\sum\limits_{n\in\mathbb{N}}\check{e}_{k_{\alpha}}(2\nu_{\alpha}n)e^{in\theta_{\alpha}}.\label{C.25}
\end{equation}
It follows from the definition of the inner product between generalized eigenstates in Eq.(\ref{C.17}) that
\begin{equation}
\begin{aligned}
	(e_{k_\alpha}|e_{k'_\alpha})
	&=\sum_{n\in\mathbb{Z}^{+}}
	\overline{e_{k_\alpha}(2\nu_\alpha n)}
	e_{k'_\alpha}(2\nu_\alpha n)\\
	&=\sum_{n\in\mathbb{Z}^{+}}
	\overline{\sqrt{n}\check{e}_{k_\alpha}(2\nu_\alpha n)}
	\sqrt{n}\check{e}_{k'_\alpha}(2\nu_\alpha n)\\
	&=\frac{1}{2\pi i}\int_{-\pi}^{\pi}d\theta_\alpha\,
	\mathcal{G}(\theta_\alpha,k_\alpha)
	\partial_{\theta_\alpha}
	\overline{\mathcal{G}(\theta_\alpha,k'_\alpha)}\\
	&=\frac{\overline{A(k_\alpha)}e^{-\pi k_\alpha/2}
		A(k'_\alpha)e^{-\pi k'_\alpha/2}}{2\pi i}
	\int_{-\pi}^{0}d\theta_\alpha\,
	e^{-ik_\alpha\ln(-\tan(\theta_\alpha/2))}
	\partial_{\theta_\alpha}
	e^{ik'_\alpha\ln(-\tan(\theta_\alpha/2))}\\
	&\quad+
	\frac{\overline{A(k_\alpha)}e^{\pi k_\alpha/2}
		A(k'_\alpha)e^{\pi k'_\alpha/2}}{2\pi i}
	\int_{0}^{\pi}d\theta_\alpha\,
	e^{-ik_\alpha\ln(\tan(\theta_\alpha/2))}
	\partial_{\theta_\alpha}
	e^{ik'_\alpha\ln(\tan(\theta_\alpha/2))}\\
	&=-\frac{\overline{A(k_\alpha)}e^{-\pi k_\alpha/2}
		A(k'_\alpha)e^{-\pi k'_\alpha/2}}{2\pi i}
	\int_{-\infty}^{\infty}dx\,
	e^{-ik_\alpha x}\partial_x e^{ik'_\alpha x}\\
	&\quad+
	\frac{\overline{A(k_\alpha)}e^{\pi k_\alpha/2}
		A(k'_\alpha)e^{\pi k'_\alpha/2}}{2\pi i}
	\int_{-\infty}^{\infty}dx\,
	e^{-ik_\alpha x}\partial_x e^{ik'_\alpha x}\\
	&=|A(k_\alpha)|^2\,2k_\alpha
	\sinh(\pi k_\alpha)\,
	\delta(k_\alpha,k'_\alpha),
\end{aligned}
\end{equation}
where Eq.(\ref{C.25}) is used in the third step, and the substitution $x=\ln|\tan(\theta_{\alpha}/2)|$ is made in the fifth step. Therefore, the coefficient $A(k_\alpha)$ can be chosen as $1/\sqrt{2k_{\alpha}\sinh(\pi k_\alpha)}$. It can be seen that all the information about the generalized eigenstates is encoded in the function $\mathcal{G}(\theta_{\alpha},k_{\alpha})$. We can further exploit this property to express the inner product in the Hilbert space $\tilde{\mathcal{H}}_\alpha$ and the action of the operator $\hat{\bar{\pi}}^{\alpha}$, both of which will play a crucial role in the subsequent calculation of expectation values. To this end, for any $\psi\in\tilde{\mathcal{H}}_\alpha$, we define
\begin{equation}
\check{\psi}(2\nu_\alpha n) := \frac{\psi(2\nu_\alpha n)}{\sqrt{n}} := \frac{\langle 2\nu_\alpha n | \psi \rangle}{\sqrt{n}} = \frac{1}{\sqrt{n}} \int_{\mathbb{R}} dk_\alpha \langle 2\nu_\alpha n | e_{k_\alpha}) (e_{k_\alpha} | \psi \rangle = \int_{\mathbb{R}} dk_\alpha \check{e}_{k_\alpha}(2\nu_\alpha n) \tilde{\psi}(k_\alpha), \quad \forall n\in\mathbb{Z}^+,\label{C.27}
\end{equation}
where $\tilde{\psi}(k_\alpha):=(e_{k_\alpha} | \psi \rangle$. Then, we further define
\begin{equation}
\chi_{\psi}(\theta_{\alpha}):= \sum_{n=1}^{\infty}\check{\psi}(2\nu_{\alpha}n)e^{in\theta_{\alpha}}
= \int_{\mathbb{R}} dk_{\alpha}\,
(\mathcal{G}(\theta_{\alpha},k_{\alpha})-A(k_{\alpha}))\tilde{\psi}(k_{\alpha}),
\end{equation}
where Eqs.(\ref{C.25}) and (\ref{C.27}) are used in the second step. Collecting the above results, the inner product between any two states $\psi,\psi'\in\tilde{\mathcal{H}}_\alpha$ can be expressed as
\begin{equation}
\langle \psi | \psi' \rangle = \sum_{n \in \mathbb{Z}^+} \overline{\psi(2\nu_\alpha n)} \psi'(2\nu_\alpha n) = \sum_{n \in \mathbb{Z}^+} n \check{\psi}(2\nu_\alpha n) \check{\psi}'(2\nu_\alpha n) = \frac{1}{2\pi i} \int_{-\pi}^{\pi} d\theta_\alpha \overline{\chi_\psi(\theta_\alpha)} \partial_{\theta_\alpha} \chi_{\psi'}(\theta_\alpha).
\end{equation}
The action of the operator $\hat{\bar{\pi}}^{y}$ can be expressed as
\begin{equation}
\hat{\bar{\pi}}^{y} \cdot \chi_\psi(\theta_{y}) := \chi_{\hat{\bar{\pi}}^{y} \cdot \psi}(\theta_{y}) = \hbar\beta\kappa\sum_{n=1}^\infty \nu_{y} n \check{\psi}(2\nu_{y} n) e^{in\theta_{y}} = -i\hbar\beta\kappa \nu_{y} \partial_{\theta_{y}} \chi_\psi(\theta_{y}).
\end{equation}
The action of the operator $\hat{\bar{\pi}}^{1}$ takes a similar form and is omitted here.

Next, we focus on the vacuum case, while the cases involving coupling to dust and a scalar field can be treated in a similar manner. On the one hand, the classical constraints $C_{gr}[1]=0$ and $C_{gr}[|\bar{\pi}^1|]=0$ are equivalent. On the other hand, according to Eq.(\ref{C.3}), $C_{gr}[1]=0$ can be promoted to the following operator
\begin{equation}
\widehat{C_{gr}[1]}:=|\hat{\bar{\pi}}^{1}|^{-1/2}\widehat{C_{gr}[|\bar{\pi}^1|]}|\hat{\bar{\pi}}^{1}|^{-1/2},
\end{equation}
where the operator $|\hat{\bar{\pi}}^{1}|^{-1/2}$ is defined as in Eq.(\ref{III.16}), and the operator corresponding to $C_{gr}[|\bar{\pi}^1|]$ is given by
\begin{equation}
\widehat{C_{gr}[|\bar{\pi}^1|]}=-\frac{3}{2}\beta^{-2}\Delta^{-1/2}\kappa^{-1}\hat{\bar{O}}_1(2\hat{\bar{O}}_1+\hat{\bar{O}}_y).
\end{equation}
Consequently, it is easy to verify that the generalized kernel spaces $\operatorname{Gker}(\widehat{C_{gr}[1]})$ and $\operatorname{Gker}(\widehat{C_{gr}[|\bar{\pi}^1|]})$ are isomorphic to each other, where the generalized kernel space of the operator $\hat{H}$ is defined as follows
\begin{equation}
\operatorname{Gker}(\hat{H}):=\{(\psi|\in\mathcal{D}_{\mathrm{kin}}^{*}\mid(\psi|\hat{H}|\phi\rangle=0,\forall|\phi\rangle\in\mathcal{D}_{\mathrm{kin}}\},
\end{equation}
and $\mathcal{D}_{\mathrm{kin}}^*$ denotes the algebraic dual of $\mathcal{D}_{\mathrm{kin}}:=\mathcal{D}_{\mathrm{kin}}^{1}\otimes\mathcal{D}_{\mathrm{kin}}^{y}$. Therefore, for simplicity, we focus on the more tractable constraint operator $\widehat{C_{gr}[|\bar{\pi}^1|]}$ and restrict our discussion to its invariant subspace $\tilde{\mathcal{H}}_1\otimes\tilde{\mathcal{H}}_y$. Using Eq.(\ref{C.6}), we can define the following dimensionless constraint operator
\begin{equation}
\hat{\bar{O}}_1(2\hat{\bar{O}}_1+\hat{\bar{O}}_y)|_{\tilde{\mathcal{D}}_{1}\otimes\tilde{\mathcal{D}}_{y}}=\frac{2}{9}(\hbar\beta\kappa)^2\hat{\tilde{\Omega}}_{1}(\hat{\tilde{\Omega}}_{1}+\frac{3}{2}\hat{\tilde{\Omega}}_{y})=:\frac{2}{9}(\hbar\beta\kappa)^2\hat{\mathcal{C}}_{0}.
\end{equation}
As discussed on p.259 of Ref.\cite{RSR2}, $\hat{\mathcal{C}}_{0}$ can be shown to be self-adjoint. We can therefore use the group-averaging map $\mathcal{P}_{\hat{\mathcal{C}}_{0}}:\tilde{\mathcal{D}}_{1}\otimes\tilde{\mathcal{D}}_{y}\to(\tilde{\mathcal{D}}_{1}\otimes\tilde{\mathcal{D}}_{y})^{*}$ to construct the physical Hilbert space consisting of states annihilated by $\hat{\mathcal{C}}_{0}$. The explicit action of $\mathcal{P}_{\hat{\mathcal{C}}_{0}}$ is given by	
\begin{equation}
\mathcal{P}_{\hat{\mathcal{C}}_{0}} (\psi) := \int_{-\infty}^{\infty} d\xi \langle e^{i\xi\hat{\mathcal{C}}_{0}} \cdot \psi, \cdot \rangle, \qquad \forall \psi \in \tilde{\mathcal{D}}_1 \otimes \tilde{\mathcal{D}}_y. 
\end{equation}
A natural inner product can be defined on the range $\mathcal{P}_{\hat{\mathcal{C}}_{0}}[\tilde{\mathcal{D}}_{1}\otimes\tilde{\mathcal{D}}_{y}]$ of $\mathcal{P}_{\hat{\mathcal{C}}_{0}}$ as follows
\begin{align}
	\langle \mathcal{P}_{\hat{\mathcal{C}}_{0}}(\psi), \mathcal{P}_{\hat{\mathcal{C}}_{0}}(\psi') \rangle_0 &:= \int_{-\infty}^{\infty} d\xi \langle e^{i\xi\hat{\mathcal{C}}_{0}} \cdot \psi, \psi' \rangle = \int_{-\infty}^{\infty} d\xi \int_{-\infty}^{\infty} dk_1 \int_{-\infty}^{\infty} dk_y \langle \psi | e^{-i\xi\hat{\mathcal{C}}_{0}} | e_{k_1} ) | e_{k_y} ) (e_{k_y} | (e_{k_1} | \psi' \rangle \nonumber\\
	&= \int_{-\infty}^{\infty} dk_1 \int_{-\infty}^{\infty} dk_y \int_{-\infty}^{\infty} d\xi e^{-i\xi((k_1)^2+\frac{3}{2}k_1 k_y)} \overline{\tilde{\psi}(k_1, k_y)} \tilde{\psi}'(k_1, k_y) \nonumber\\
	&= \int_{-\infty}^{\infty} dk_1 \int_{-\infty}^{\infty} dk_y \frac{4\pi}{3|k_y|} \bigg(\delta(k_1) + \delta(k_1 + \frac{3}{2}k_y)\bigg) \overline{\tilde{\psi}(k_1, k_y)} \tilde{\psi}'(k_1, k_y),\label{C.36}
\end{align}
where $\tilde{\psi}(k_{1},k_{y}):=(e_{k_1}|(e_{k_y}|\psi \rangle$. By completing $\mathcal{P}_{\hat{\mathcal{C}}_{0}}[\tilde{\mathcal{D}}_{1}\otimes\tilde{\mathcal{D}}_{y}]$ with respect to the inner product (\ref{C.36}), we obtain the physical Hilbert space $\tilde{\mathcal{H}}^{0}_{phy} := \langle\mathcal{P}_{\hat{\mathcal{C}}_{0}}[\tilde{\mathcal{D}}_1 \otimes\tilde{\mathcal{D}}_y]\rangle_0$. Following an entirely analogous construction, we can define the physical Hilbert spaces $\tilde{\mathcal{H}}^{1}_{phy} := \langle\mathcal{P}_{\hat{\mathcal{C}}_{1}}[\tilde{\mathcal{D}}_1 \otimes\tilde{\mathcal{D}}_y]\rangle_1$ and $\tilde{\mathcal{H}}^{2}_{phy} := \langle\mathcal{P}_{\hat{\mathcal{C}}_{2}}[\tilde{\mathcal{D}}_1 \otimes\tilde{\mathcal{D}}_y]\rangle_2$ associated with the constraint operators $\hat{\mathcal{C}}_{1}:=\hat{\tilde{\Omega}}_{1}$ and $\hat{\mathcal{C}}_{2}:=\hat{\tilde{\Omega}}_{1}+\frac{3}{2}\hat{\tilde{\Omega}}_{y}$, respectively. It should be noted that the common zero-eigenvalue generalized eigenstate $(e_{k_{1}=0}|(e_{k_{y}=0}|$ of $\hat{\mathcal{C}}_{1}$ and $\hat{\mathcal{C}}_{2}$ does not belong to any of the spaces $\tilde{\mathcal{H}}^{0}_{phy}$, $\tilde{\mathcal{H}}^{1}_{phy}$, or $\tilde{\mathcal{H}}^{2}_{phy}$. Consequently, $\tilde{\mathcal{H}}^{1}_{phy}\cap\tilde{\mathcal{H}}^{2}_{phy}=\{0\}$. It then follows from Eq.(\ref{C.36}) that $\tilde{\mathcal{H}}^{0}_{phy}$ is unitarily isomorphic to $\tilde{\mathcal{H}}^{1}_{phy}\oplus\tilde{\mathcal{H}}^{2}_{phy}$, where $\tilde{\mathcal{H}}^{1}_{phy}$ and $\tilde{\mathcal{H}}^{2}_{phy}$ are orthogonal to each other.

To introduce the notion of evolution in $\tilde{\mathcal{H}}^{\iota}_{phy}$, where $\iota\in\{1,2\}$, we can promote the family of Dirac observables obtained through deparametrization in the classical theory to a family of operators in $\tilde{\mathcal{H}}^{\iota}_{phy}$, thereby establishing a notion of operator evolution analogous to that in the Heisenberg picture. To this end, we first introduce the operator $\hat{T}_{\alpha}:=i\partial_{k_{\alpha}}$ in $\operatorname{L}^2(\mathbb{R},dk_{\alpha})$, together with its standard generalized eigenstates
\begin{equation}
e_{T_{\alpha}}(k_{\alpha}):=(e_{k_{\alpha}}|e_{T_{\alpha}}):=\frac{1}{\sqrt{2\pi}}e^{-iT_{\alpha}k_{\alpha}}.
\end{equation}
At the quantum level, the operators satisfy $[\hat{T}_{\alpha},\hat{\tilde{\Omega}}_{\alpha}]=i$. Meanwhile, at the classical level, $(\ln|\tan(\nu_{1}\bar{A}_{1}/2)|,\bar{O}_{1}/(\frac{\beta\kappa}{3}))$ and $(\ln|\tan(\nu_{y}\bar{A}_{y}/2)|,\bar{O}_{y}/(\beta\kappa))$ constitute a set of fundamental canonical variables, with the following nonvanishing Poisson brackets:
\begin{equation}
\{\ln|\tan(\nu_{1}\bar{A}_{1}/2)|,\bar{O}_{1}/(\frac{\beta\kappa}{3})\}=\{\ln|\tan(\nu_{y}\bar{A}_{y}/2)|,\bar{O}_{y}/(\beta\kappa)\}=1.
\end{equation}
Thus, based on Eq.(\ref{C.6}) and the correspondence principle of canonical quantization, $T_{\alpha}:=\ln|\tan(\nu_{\alpha}\bar{A}_{\alpha}/2)|$ may be identified as the classical counterpart of $\hat{T}_{\alpha}$. Using the completeness relation $\int_{-\infty}^{\infty} dT_{1}|e_{T_1})(e_{T_1}|=\mathbb{I}$, a physical state in $\tilde{\mathcal{H}}^{\iota}_{phy}$ can be further expressed as
\begin{align}
	\mathcal{P}_{\hat{\mathcal{C}}_\iota}(\psi) &= \int_{-\infty}^{\infty} d\xi \langle e^{i\xi \hat{\mathcal{C}}_\iota} \cdot \psi, \cdot \rangle = \int_{-\infty}^{\infty} dT_1 \int_{-\infty}^{\infty} dk_y \int_{-\infty}^{\infty} dk_1 \int_{-\infty}^{\infty} d\xi \langle e^{i\xi \hat{\mathcal{C}}_\iota} \cdot\psi | e_{k_1} ) | e_{k_y} ) (e_{k_1} | e_{T_1} ) (e_{T_1} | (e_{k_y} | \nonumber\\
	&= \int_{-\infty}^{\infty} dT_1 \int_{-\infty}^{\infty} dk_y \overline{\widetilde{\psi}(h_\iota (k_y), k_y)} \sqrt{2\pi} e^{-iT_1 h_\iota (k_y)} (e_{T_1} | (e_{k_y} | \nonumber\\
	&=: \int_{-\infty}^{\infty} dT_1 \int_{-\infty}^{\infty} dk_y \overline{\Psi_\iota (T_1, k_y)} (e_{T_1} | (e_{k_y} |,\label{C.39}
\end{align}
where $h_1 (k_y):=0$ and $h_2 (k_y):=-3k_{y}/2$. Therefore, Eq.(\ref{C.39}) establishes a one-to-one correspondence between physical states $\mathcal{P}_{\hat{\mathcal{C}}_\iota}(\psi)$ and functions $\Psi_\iota$ satisfying the following conditions
\begin{equation}\label{C.40}
	\begin{aligned}
		\text{(i)}\quad & \Psi_{\iota} (0,\cdot)\in\tilde{\mathcal{D}}_y,\\
		\text{(ii)}\quad & \Psi_{\iota} (T_1,\cdot)=e^{iT_1 \hat{H}_\iota}\cdot\Psi_{\iota} (0,\cdot),
		\quad \forall T_1\in\mathbb{R},
	\end{aligned}
\end{equation}
where $\hat{H}_1 :=0$ and $\hat{H}_2 :=-3\hat{\tilde{\Omega}}_{y}/2$. Furthermore, the inner product between two physical states $\mathcal{P}_{\hat{\mathcal{C}}_\iota}(\psi)$ and $\mathcal{P}_{\hat{\mathcal{C}}_\iota}(\psi')$ can also be expressed in terms of $\Psi_\iota$ and $\Psi'_\iota$,
\begin{align}
	\langle \mathcal{P}_{\hat{\mathcal{C}}_\iota}(\psi), \mathcal{P}_{\hat{\mathcal{C}}_\iota}(\psi') \rangle_t &:= \int_{-\infty}^{\infty} d\xi \langle e^{i\xi \hat{\mathcal{C}}_\iota} \cdot \psi, \psi' \rangle \nonumber\\
	&= \int_{-\infty}^{\infty} dk_y 2\pi \overline{\tilde{\psi}(h_{\iota}(k_y), k_y)} \tilde{\psi}(h_{\iota}(k_y), k_y) \nonumber\\
	&= \int_{-\infty}^{\infty} dk_y \overline{\Psi_{\iota}(0, k_y)} \Psi_{\iota}'(0, k_y)=: \langle \Psi_{\iota}, \Psi_{\iota}' \rangle_{phy}.\label{C.41}
\end{align}
Therefore, the physical Hilbert space $\tilde{\mathcal{H}}^{\iota}_{\mathrm{phy}}$ can be identified with the completion of $\operatorname{span}_{\mathbb{C}}\{\Psi_{\iota}\mid \Psi_{\iota}\text{ satisfies condition (\ref{C.40})}\}$ with respect to the inner product (\ref{C.41}). With this identification, the Dirac observable $O_{t}(L)$, corresponding to the value of the phase-space function $L=L(\bar{A}_{y},\bar{\pi}^y)$ at $T_1=t$, can be promoted to the following operator in the physical Hilbert space $\tilde{\mathcal{H}}^{\iota}_{\mathrm{phy}}$ \cite{Domagala:2010bm},
\begin{align}
\widehat{O_t (L)}_{\iota} \cdot \Psi_{\iota} (T_1, \cdot) := e^{i(T_1-t) \hat{H}_{\iota}}  \hat{L} e^{-i(T_1-t) \hat{H}_{\iota}} \cdot \Psi_{\iota} (T_1, \cdot),\quad \forall T_1\in\mathbb{R},
\end{align}
where $\hat{L}$ denotes the operator in $\tilde{\mathcal{H}}_y$ corresponding to $L$. The expectation value of the operator $\widehat{O_t (L)}_{\iota}$ with respect to the physical state $\Psi_{\iota}$ is given by,
\begin{align}
\langle \Psi_{\iota}, \widehat{O_t(L)}_{\iota} \cdot \Psi_{\iota} \rangle_{phy} = \langle e^{it\hat{H}_{\iota}} \cdot \psi, \hat{L} e^{it\hat{H}_{\iota}} \cdot \psi \rangle,
\end{align}
where $\psi\equiv\Psi_{\iota}(0,\cdot)$. In the case of $\iota=1$, the expectation value $\langle \Psi_{1}, \widehat{O_t(L)}_{1} \cdot \Psi_{1} \rangle_{phy}$ does not evolve with $t$, corresponding to a static universe that is inconsistent with observations. Hence, we focus on the case $\iota=2$ and omit the subscript $\iota=2$ for simplicity. In this case, the expectation value of the Dirac observable operator corresponding to the physical volume of the three-dimensional fiducial cell at $T_1=t$, evaluated in a general physical state $\Psi$, is given by
\begin{align}
		\langle \widehat{O_{t}(\bar{\pi}^y)}\rangle&=\langle
		e^{it\hat{H}_2}\cdot\psi,\,
		\hat{\bar{\pi}}^{y}e^{it\hat{H}_2}\cdot\psi\rangle \nonumber\\
		&=\frac{\mu_y\hbar\beta\kappa}{2\pi}
		\int_{-\pi}^{\pi}d\theta_y\,\partial_{\theta_y}
		\overline{
			\int_{\mathbb{R}}dk'_y\,
			\mathcal{G}(\theta_y,k'_y)e^{-1.5itk'_y}
			\tilde{\psi}(k'_y)
		}
		\partial_{\theta_y}
		\int_{\mathbb{R}}dk_y\,
		\mathcal{G}(\theta_y,k_y)e^{-1.5itk_y}
		\tilde{\psi}(k_y) \nonumber\\
		&=\frac{\mu_y\hbar\beta\kappa}{2\pi}
		\Bigg[
		\int_{-\pi}^{0}d\theta_y\,\partial_{\theta_y}
		\overline{
			\int_{\mathbb{R}}dk'_y\,A(k'_y)
			e^{-\pi k'_y/2}
			e^{ik'_y\ln(-\tan(\theta_y/2))}
			e^{-1.5itk'_y}
			\tilde{\psi}(k'_y)	}\nonumber\\
		&\times
		\partial_{\theta_y}
		\int_{\mathbb{R}}dk_y\,A(k_y)
		e^{-\pi k_y/2}
		e^{ik_y\ln(-\tan(\theta_y/2))}
		e^{-1.5itk_y}
		\tilde{\psi}(k_y) \nonumber\\
		&+\int_{0}^{\pi}d\theta_y\,\partial_{\theta_y}
		\overline{
			\int_{\mathbb{R}}dk'_y\,A(k'_y)
			e^{\pi k'_y/2}
			e^{ik'_y\ln(\tan(\theta_y/2))}
			e^{-1.5itk'_y}
			\tilde{\psi}(k'_y)}\nonumber\\
		&\times\partial_{\theta_y}
		\int_{\mathbb{R}}dk_y\,A(k_y)
		e^{\pi k_y/2}
		e^{ik_y\ln(\tan(\theta_y/2))}
		e^{-1.5itk_y}
		\tilde{\psi}(k_y)
		\Bigg]\nonumber\\
		&=\frac{\mu_y\hbar\beta\kappa}{2\pi}
		\int_{-\infty}^{\infty}dx\,
		\Big[
		\partial_x\overline{M_-(x-1.5t)}
		\cosh(x)\,
		\partial_x M_-(x-1.5t)\nonumber\\
		&+\partial_x\overline{M_+(x-1.5t)}
		\cosh(x)\,
		\partial_x M_+(x-1.5t)
		\Big]\nonumber\\
		&=V_+e^{1.5t}+V_-e^{-1.5t}\nonumber\\
		&=V_+\left|\tan\left(\frac{\bar{A}_1}{2}\right)\right|^{\frac{3}{2}}
		+V_-\left|\tan\left(\frac{\bar{A}_1}{2}\right)\right|^{-\frac{3}{2}}.\label{C.44}
\end{align}
where the fourth step follows from $x=\ln|\tan(\theta_{y}/2)|$, and the last step follows from $t=\ln|\tan(\bar{A}_{1}/2)|$. In addition, we have introduced the definitions 
\begin{align}
M_{\pm}(x) &:= \int_{\mathbb{R}} dk_y A(k_y) e^{\pm \pi k_y / 2} e^{i k_y x} \widetilde{\psi}(k_y),\\
V_{\pm} &:= \frac{\mu_y \hbar \beta \kappa}{2\pi} \int_{-\infty}^{\infty} dx \frac{e^{\pm x}}{2} \left[ \overline{\partial_x M_-(x)} \partial_x M_-(x) + \overline{\partial_x M_+(x)} \partial_x M_+(x) \right].\label{C.46}
\end{align}
Eq.(\ref{C.44}) implies that the classical singularity in the visible universe is avoided by a quantum bounce. Moreover, we can calculate the expectation values of the basic operators $\widehat{O_{t}(\tilde{\Omega}_y)}$ and $\widehat{O_{t}(T_y)}$ as follows,
\begin{align}
	\langle \widehat{O_{t}(\tilde{\Omega}_y)}\rangle&=\langle \widehat{O_{0}(\tilde{\Omega}_y)}\rangle,\label{C.47}\\
	\langle \widehat{O_{t}(T_y)}\rangle&=\frac{3}{2}t+\langle \widehat{O_{0}(T_y)}\rangle.\label{C.48}
\end{align}
To further demonstrate the consistency between the quantum dynamics and the effective model, we introduce the following coherent states in the physical Hilbert space $\tilde{\mathcal{H}}^{2}_{\mathrm{phy}}$,
\begin{align}
\Psi_{\zeta}(T_1,k_y) = e^{-iT_1 1.5k_y}\frac{1}{\sqrt{\pi \sigma_0}} e^{-(k_y - k_0)^2 / (2\sigma_0^2)} e^{-i(q_0 - 1.5t_0)k_y},\label{C.49}
\end{align}
where $\zeta:=(k_0,\sigma_0,q_0,t_0)$. A straightforward calculation yields the expectation values and uncertainties of the operators $\widehat{O_{t_0}(\tilde{\Omega}_y)}$ and $\widehat{O_{t_0}(T_y)}$ with respect to the coherent state $\Psi_\zeta$, as follows
\begin{align}
	\langle \widehat{O_{t_0}(\tilde{\Omega}_y)} \rangle = k_0, \qquad 
	\Delta \widehat{O_{t_0}(\widetilde{\Omega}_y)} = \frac{\sigma_0}{\sqrt{2}}, \qquad \langle \widehat{O_{t_0}(T_y)} \rangle = q_0, \qquad 
	\Delta \widehat{O_{t_0}(T_y)} = \frac{1}{\sqrt{2}\sigma_0}.\label{C.50}
\end{align}
Here, $t_0$ denotes a time in the late stage of cosmic evolution. Therefore, the parameters in Eq.(\ref{C.50}) must satisfy semiclassical conditions similar to those in Eqs.(\ref{A1 condition}) and (\ref{AY condition}), 
\begin{align}
	|k_0|^{-1} \ll \sigma_0^{-1} \ll |q_0| \ll 1,
\end{align}
ensuring that the coherent state $\Psi_\zeta$ is sharply peaked at $(k_0,q_0)$, with fluctuations within the specified tolerance. Substituting the coherent state $\Psi_\zeta$ into Eqs.(\ref{C.47}) and (\ref{C.48}), we obtain
\begin{align}
	\langle \widehat{O_{t}(\tilde{\Omega}_y)}\rangle&=\langle \widehat{O_{0}(\tilde{\Omega}_y)}\rangle=k_{0},\label{C.52}\\
	\langle \widehat{O_{t}(T_y)}\rangle&=\frac{3}{2}t+\langle \widehat{O_{0}(T_y)}\rangle=\frac{3}{2}t+q_0-\frac{3}{2}t_0.\label{C.53}
\end{align}
With the identification $t=\ln|\tan(\bar{A}_{1}/2)|$, Eq.(\ref{C.53}) can be matched with Eq.(\ref{Ayvac}) in the main text. On this basis, we further evaluate Eq.(\ref{C.46}) using the coherent state $\Psi_\zeta$. By employing the steepest-descent approximation, we obtain
\begin{align}
	V_{\pm} \approx \frac{\mu_y \hbar \beta \kappa}{2} e^{\pm(q_0 - 1.5t_0) + 1/(4\sigma_0^2)} \sqrt{k_0^2 + \frac{1}{4}} \approx \frac{\mu_y}{2} |\langle \widehat{O_0(\bar{O}_y)} \rangle| e^{\pm \langle \widehat{O_0(T_y)} \rangle}.
\end{align}
Hence, Eq.(\ref{C.44}) is consistent with Eq.(\ref{piyvac}) in the main text.

In summary, in the quantum dynamics of the vacuum case, we deparameterize the system by using $T_{1}=\ln|\tan(\bar{A}_{1}/2)|$, the conjugate variable of $\bar{\pi}^{1}\sin(\bar{A}_{1})/(\frac{\beta\kappa}{3})$, as the physical time. By calculating the relational evolution of the expectation values of the operators $\widehat{O_{t}(T_y)}$ and $\widehat{O_{t}(\bar{\pi}^y)}$, corresponding to the remaining basic variables $T_y:=\ln|\tan(\mu_y\bar{A}_y/2)|$ and $\bar{\pi}^y$, respectively, with respect to the coherent state (\ref{C.49}), we reproduce their evolution with respect to $T_1$ obtained from the effective model presented in the main text. In this sense, the quantum-dynamical results are consistent with those of the effective dynamics. This demonstrates that the validity of the effective scalar constraint presented in the main text extends beyond the regime in which the semiclassical conditions (\ref{A1 condition}) and (\ref{AY condition}) are satisfied.

\twocolumngrid


\begin{thebibliography}{99}
		
\bibitem{Ro04} 
C. Rovelli, 
Quantum Gravity, 
Cambridge University Press, 2004.
		
\bibitem{Th07} 
T. Thiemann, 
Modern Canonical Quantum General Relativity, 
Cambridge University Press, 2007.
		
\bibitem{As04}
A. Ashtekar and J. Lewandowski, 
Background independent quantum gravity: A Status report, 
Class. Quant. Grav. \textbf{21}, R53 (2004).
		
\bibitem{Ma07} 
M. Han, W. Huang, and Y. Ma, 
Fundamental structure of loop quantum gravity, 
Int. J. Mod. Phys. D {\bf16}, 1397 ,(2007).

\bibitem{A.Y}
A. Ashtekar and Y. Ma, 
Loop Quantum Gravity and Spinfoams,
in Handbook of Quantum Gravity (Eds. C. Bambi et al.), pp. 10-13. Springer Singapore, 2024.
		
\bibitem{LQC5}
A. Ashtekar, M. Bojowald, and J. Lewandowski, 
Mathematical structure of loop quantum cosmology, 
Adv. Theor. Math. Phys. \textbf{7}, 233 (2003).



\bibitem{Boj}
M. Bojowald, 
Loop quantum cosmology, 
Living Rev. Relativity \textbf{8}, 11 (2005).
		
\bibitem{Ash-view}
A. Ashtekar, 
Loop quantum cosmology: An overview, 
Gen. Rel. Grav. {\bf41}, 707 (2009).
		
\bibitem{AS11} 
A. Ashtekar, P. Singh, 
Loop quantum cosmology: A status report,  
Class. Quant. Grav. {\bf28}, 213001 (2011).



\bibitem{BCM} 
K. Banerjee, G. Calcagni, M. Mart¨ªn-Benito, 
Introduction to Loop Quantum Cosmology, 
SIGMA {\bf 8}, 016 (2012).
		
\bibitem{APS3} 
A. Ashtekar, T. Pawlowski, P. Singh,  
Quantum nature of the big bang: Improved dynamics,  
Phys. Rev. D {\bf 74}, 084003 (2006).
		
\bibitem{ACS}
A. Ashtekar, A. Corichi, and P. Singh, 
Robustness of key features of loop quantum cosmology, 
Phys. Rev. D  {\bf 77}, 024046 (2008).

\bibitem{Yang23}
J. Yang, C. Zhang, X. Zhang, 
Alternative $k=-1$ loop quantum cosmology, 
Phys. Rev. D  {\bf 107} 4, 046012 (2023).

\bibitem{YDM}
J. Yang, Y. Ding and Y. Ma, 
Alternative quantization of the Hamiltonian in loop quantum cosmology, 
Phys. Lett. B \textbf{682}, 1 (2009).


\bibitem{DMY}
Y. Ding, Y. Ma and J. Yang, 
Effective scenario of loop quantum cosmology, 
Phys. Rev. Lett. \textbf{102}, 051301 (2009).



\bibitem{Zhang:2021zfp}
X.~Zhang, G.~Long and Y.~Ma,
Loop quantum gravity and cosmological constant,
Phys. Lett. B \textbf{823}, 136770 (2021).




\bibitem{Zhang14}
X. Zhang, 
Loop quantum cosmology in 2+1 dimension, 
Phys. Rev. D  {\bf 90}, 124018 (2014).
		
\bibitem{Zhang16}
X. Zhang, 
Higher dimensional Loop Quantum Cosmology, 
Eur. Phys. J. C \textbf{76}, no.7, 395 (2016).

\bibitem{Li:2025bzl}
S.~Li, Y.~Ma, F.~Yuan and X.~Zhang,
Effective dynamics of loop quantum Kaluza-Klein cosmology,
Phys. Rev. D \textbf{113}, no.4, L041503 (2026).



\bibitem{Kaluza:1921tu}
T.~Kaluza,
Zum Unit\"atsproblem der Physik,
Sitzungsber. Preuss. Akad. Wiss. Berlin (Math. Phys. ) \textbf{1921}, 966-972 (1921).

\bibitem{Klein:1926tv}
O.~Klein,
Quantum Theory and Five-Dimensional Theory of Relativity. (In German and English),
Z. Phys. \textbf{37}, 895-906 (1926).

\bibitem{bailin1987kaluza}
David Bailin and Alex Love.
\newblock Kaluza-klein theories.
\newblock {\em Reports on Progress in Physics}, 50(9):1087, 1987.

\bibitem{Appelguist}
T. Appelquist, A. Chodos, and P. G. O. Freund (ed.), 
Modern Kaluza-Klein Theories, 
Frontiers in Physics Series Vol.\textbf{65}(Addison-Wesley, Reading, MA, 1986).


\bibitem{NM02}
N. Mohammedi, 
Dynamical compactification, standard cosmology, and the accelerating universe, 
Phys. Rev. D \textbf{65}, 104018(2002).

\bibitem{Qiang}
L. Qiang, Y. Ma, M. Han and D. Yu, 
5-dimensional Brans-Dicke theory and cosmic acceleration, 
Phys. Rev. D {\bf71},  061501(R) (2005).

\bibitem{Panigrahi:2006wi}
D. Panigrahi, Y. Z. Zhang and S. Chatterjee, 
Accelerating universe as window for extra dimensions, 
Int. J. Mod. Phys. A \textbf{21}, 6491-6512 (2006).


		
\bibitem{JP98}
J. Polchinski, 
String Theory, 
Vol 1 and Vol 2, (Cambridge University Press, 1998).
		
\bibitem{JM03}
J. M. Maldacena, 
TASI 2003 Lectures on AdS/CFT, 
arXiv:hep-th/0309246.
		
\bibitem{RS99a}
L. Randall and R. Sundrum, 
Large Mass Hierarchy from a Small Extra Dimension, 
Phys. Rev. Lett. \textbf{83}, 3370(1999).
		
\bibitem{RS99b}
L. Randall and R. Sundrum, 
An Alternative to Compactification, 
Phys. Rev. Lett. \textbf{83}, 4690 (1999).
		

		


\bibitem{Ashtekar:2011ni}
A.~Ashtekar and P.~Singh,
Loop Quantum Cosmology: A Status Report,
Class. Quant. Grav. \textbf{28}, 213001 (2011).

\bibitem{Ashtekar:2011rm}
A.~Ashtekar and D.~Sloan,
Probability of Inflation in Loop Quantum Cosmology,
Gen. Rel. Grav. \textbf{43}, 3619-3655 (2011).




\bibitem{BTTa} 
N. Bodendorfer,  T. Thiemann and A. Thurn, 
New variables for classical and quantum gravity in all dimensions I. Hamiltonian analysis,
Class. Quant. Grav, {\bf30}, 045001 (2013).
		
\bibitem{BTTb} 
N. Bodendorfer,  T. Thiemann and A. Thurn, 
New variables for classical and quantum gravity in all dimensions II. Lagrangian analysis,
Class. Quant. Grav. {\bf30}, 045002 (2013).
		
\bibitem{BTTc} 
N. Bodendorfer,  T. Thiemann and A. Thurn, 
New variables for classical and quantum gravity in all dimensions III. Quantum theory,
Class. Quant. Grav. {\bf30}, 045003 (2013).
		
\bibitem{BTTd} 
N. Bodendorfer,  T. Thiemann and A. Thurn, 
New variables for classical and quantum gravity in all dimensions IV. Matter Coupling,
Class. Quant. Grav. {\bf30}, 045004 (2013).

\bibitem{OG}
O. Gr\o n,
Inflationary cosmology according to Wesson's gravitational theory, 
Astronomy Ashtrophysics \textbf{193}, 1 (1988).
	
\bibitem{Chodos:1979vk}
A.~Chodos and S.~L.~Detweiler,
Where Has the Fifth-Dimension Gone?,
Phys. Rev. D \textbf{21}, 2167 (1980).




\bibitem{Thiemann:1996aw}
T.~Thiemann,
Quantum spin dynamics (QSD),
Class. Quant. Grav. \textbf{15}, 839-873 (1998).
		
\bibitem{Yang:2015zda} 
J. Yang and Y. Ma, 
New Hamiltonian constraint operator for loop quantum gravity, 
Phys. Lett. B \textbf{751}, 343-347 (2015).		
		
\bibitem{AFJ}
A. Ashtekar, S. Fairhurst and J. L. Willis, 
Quantum gravity, shadow states, and quantum mechanics, 
Class. Quant. Grav. \textbf{20}, 1031-1062 (2003).
		
\bibitem{Taveras}
V. Taveras, 
Corrections to the Friedmann equations from loop quantum gravity for a universe with a free scalar field, 
Phys. Rev. D \textbf{78}, 064072 (2008).




\bibitem{Giulini:1998rk}
D. Giulini and D. Marolf, 
On the generality of refined algebraic quantization, 
Class. Quant. Grav. \textbf{16}, 2479-2488 (1999).
		
\bibitem{Giulini:1998kf}
D. Giulini and D. Marolf, 
A Uniqueness theorem for constraint quantization, 
Class. Quant. Grav. \textbf{16}, 2489-2505 (1999).
		
\bibitem{ALMMT}
A. Ashtekar, J. Lewandowski, D. Marolf, J. Mourao and T. Thiemann, 
Quantization of diffeomorphism invariant theories of connections with local degrees of freedom, 
J. Math. Phys. \textbf{36}, 6456-6493 (1995).
	
\bibitem{Ashtekar:2010gz}
A. Ashtekar, M. Campiglia and A. Henderson, 
Path Integrals and the WKB approximation in Loop Quantum Cosmology, 
Phys. Rev. D \textbf{82}, 124043 (2010).
		
\bibitem{Huang:2011es}
H. Huang, Y. Ma and L. Qin, 
Path Integral and Effective Hamiltonian in Loop Quantum Cosmology, 
Gen. Rel. Grav. \textbf{45}, 1191-1210 (2013).
		
\bibitem{Qin:2012gaa}
L. Qin, G. Deng and Y. Ma, 
Path integrals and alternative effective dynamics in loop quantum cosmology, 
Commun. Theor. Phys. \textbf{57}, 326-332 (2012).

\bibitem{Zhang:2012em}
X.~Zhang, Y.~Ma and M.~Artymowski,
Loop quantum Brans-Dicke cosmology,
Phys. Rev. D \textbf{87}, no.8, 084024 (2013).

\bibitem{Song:2020pqm}
S.~Song, C.~Zhang and Y.~Ma,
Alternative dynamics in loop quantum Brans-Dicke cosmology,
Phys. Rev. D \textbf{102}, no.2, 024024 (2020).





		
\bibitem{QM1}
L. Qin and Y. Ma, 
Coherent state functional integrals in quantum cosmology, 
Phys. Rev. D \textbf{85}, 063515 (2012).
		
\bibitem{QM2}
L. Qin and Y. Ma, 
Coherent state functional integral in loop quantum cosmology: Alternative dynamics, 
Mod. Phys. Lett. \textbf{27}, 1250078 (2012).


\bibitem{Martin-Benito:2009xaf}
M.~Martin-Benito, G.~A.~M.~Marugan and T.~Pawlowski,
Physical evolution in Loop Quantum Cosmology: The Example of vacuum Bianchi I,
Phys. Rev. D \textbf{80}, 084038 (2009).



\bibitem{Assanioussi:2018hee}
M.~Assanioussi, A.~Dapor, K.~Liegener and T.~Paw{\l}owski,
Emergent de Sitter Epoch of the Quantum Cosmos from Loop Quantum Cosmology,
Phys. Rev. Lett. \textbf{121}, no.8, 081303 (2018).


\bibitem{Assanioussi:2019iye}
M.~Assanioussi, A.~Dapor, K.~Liegener and T.~Paw{\l}owski,
Emergent de Sitter epoch of the Loop Quantum Cosmos: a detailed analysis,
Phys. Rev. D \textbf{100}, no.8, 084003 (2019).












\bibitem{Planck:2018jri}
Y.~Akrami \textit{et al.} [Planck],
Planck 2018 results. X. Constraints on inflation,
Astron. Astrophys. \textbf{641}, A10 (2020).

		
		
\bibitem{Song:2022zit}
S. Song, G. Long, C. Zhang and X. Zhang, 
Thermodynamics of isolated horizons in loop quantum gravity, 
Phys. Rev. D \textbf{106}, no.12, 126007 (2022).


\bibitem{Corichi:2015xia}
A.~Corichi and P.~Singh,
Loop quantization of the Schwarzschild interior revisited,
Class. Quant. Grav. \textbf{33}, no.5, 055006 (2016).


		
\bibitem{Ashtekar:2018cay}
A. Ashtekar, J. Olmedo and P. Singh, 
Quantum extension of the Kruskal spacetime, 
Phys. Rev. D \textbf{98}, no.12, 126003 (2018).







\bibitem{Yuan:2025yjs}
F.~Yuan, S.~Li, Z.~Li and Y.~Ma,
Effective dynamics of Janis-Newman-Winicour spacetime,
Phys. Rev. D \textbf{113}, no.8, 084063 (2026).




\bibitem{Zhang:2019acn}
C.~Zhang and X.~Zhang,
Quantum geometry and effective dynamics of Janis-Newman-Winicour singularities,
Phys. Rev. D \textbf{101}, no.8, 086002 (2020).







\bibitem{Qiang:2004gg}
L.~e.~Qiang, Y.~g.~Ma, M.~x.~Han and D.~Yu,
5-dimensional Brans-Dicke theory and cosmic acceleration,
Phys. Rev. D \textbf{71}, 061501 (2005).


\bibitem{Planck:2018vyg}
N.~Aghanim \textit{et al.} [Planck],
Planck 2018 results. VI. Cosmological parameters,
Astron. Astrophys. \textbf{641}, A6 (2020)
[erratum: Astron. Astrophys. \textbf{652}, C4 (2021)].


\bibitem{Vijaykumar:2020nzc}
A.~Vijaykumar, S.~J.~Kapadia and P.~Ajith,
Constraints on the time variation of the gravitational constant using gravitational-wave observations of binary neutron stars,
Phys. Rev. Lett. \textbf{126}, no.14, 141104 (2021).





\bibitem{RSR1} 
M. Reed and B. Simon. 
Methods of Modern Mathematical Physics: Fourier Analysis, Self-adjointness. 
Elsevier, 2003.

\bibitem{RSR2} 
M. Reed and B. Simon. 
Methods of modern mathematical physics: Functional analysis. 
Elsevier, 2003.		




\bibitem{Zhang:2021wex}
C.~Zhang, Y.~Ma, S.~Song and X.~Zhang,
Loop quantum deparametrized Schwarzschild interior and discrete black hole mass,
Phys. Rev. D \textbf{105}, no.2, 024069 (2022).



\bibitem{Martin-Benito:2009htq}
M.~Martin-Benito, G.~A.~M.~Marugan and J.~Olmedo,
Further Improvements in the Understanding of Isotropic Loop Quantum Cosmology,
Phys. Rev. D \textbf{80}, 104015 (2009).



\bibitem{Martin-Benito:2008dfr}
M.~Martin-Benito, G.~A.~Mena Marugan and T.~Pawlowski,
Loop Quantization of Vacuum Bianchi I Cosmology,
Phys. Rev. D \textbf{78}, 064008 (2008).




\bibitem{J.Dombrowski1}
J.~Dombrowski,
Cyclic operators, commutators, and absolutely continuous measures.
Proceedings of the American Mathematical Society 100, no.3 (1987): 457-463.





\bibitem{A.I.Aptekarev}
A.~I.~Aptekarev and J.~S.~Geronimo, 
Measures for orthogonal polynomials with unbounded recurrence coefficients. 
Journal of Approximation Theory 207 (2016): 339-347.



\bibitem{J.Sahbani}
J.~Sahbani,
Spectral theory of certain unbounded Jacobi matrices. 
Journal of mathematical analysis and applications 342, no. 1 (2008): 663-681.








\bibitem{Domagala:2010bm}
M.~Domagala, K.~Giesel, W.~Kaminski and J.~Lewandowski,
Gravity quantized: Loop Quantum Gravity with a Scalar Field,
Phys. Rev. D \textbf{82}, 104038 (2010).






































		
\end{thebibliography}
\end{document}